\documentclass[10pt]{article}
\usepackage[margin=1in]{geometry}                
\usepackage{graphicx}
\usepackage{amssymb}
\usepackage{epstopdf}
\usepackage{amsmath}
\usepackage{hyperref}
\usepackage{mathrsfs}
\usepackage{varioref}
\usepackage{textcomp}
\usepackage{enumerate}
\usepackage{paralist}

\hypersetup{
    bookmarks=false,         
    pdfstartview={FitH},    
    colorlinks=true,       
    linkcolor=black,          
    citecolor=black,        
    filecolor=blue,      
    urlcolor=blue           
}

\DeclareFontFamily{OT1}{pzc}{}
\DeclareFontShape{OT1}{pzc}{m}{it}%
             {<-> s * [0.900] pzcmi7t}{}
\DeclareMathAlphabet{\mathscr}{OT1}{pzc}%
                                 {m}{it}
\labelformat{section}{Section #1} 
\labelformat{subsection}{Section #1} 
\labelformat{subsubsection}{Section #1}
\labelformat{subsubsubsection}{Section #1}
\labelformat{equation}{Eq.~(#1)} 
\labelformat{figure}{Fig.~#1} 
\labelformat{subfigure}{Fig.~\thefigure#1} 
\labelformat{table}{Tab.~#1} 
\newcommand{\be}{\begin{equation}}
\newcommand{\ee}{\end{equation}}
\newcommand{\bea}{\begin{eqnarray}}
\newcommand{\eea}{\end{eqnarray}}
\newcommand{\nn}{\nonumber}

\newcommand{\sqg}{\sqrt{-g}}
\newcommand{\sqh}{\sqrt{|h|}}

\newcommand{\df}{\delta}

\newcommand{\vbar}{\left\vert\vphantom{\int}\right.}
\newcommand{\eqrtoz}{:=}
\newcommand{\eqH}{:=}
\newcommand{\eqscs}{\overset{*}=}
\newcommand{\intdx}{\int_\mathcal{\partial V} d^3x}
\newcommand{\intlt}{\int_\mathcal{\partial V} d\lambda d^2z_{\perp}}
\newcommand{\STn}{\left[\df \mathcal{A}_{_{\partial  \mathcal \tiny V}}\right]_{null}}
\newcommand{\STnn}{\left[\df \mathcal{A}_{_{\partial  \mathcal \tiny V}}\right]_{non-null}}
\newcommand{\un}{n_{\perp}}
\newcommand{\ul}{\ell_{\perp}}

\newcommand{\uA}{A_{\perp}}
\newcommand{\unt}{\tilde{n}_{\perp}}

\begin{document}
\title{A Boundary Term for the Gravitational Action with Null Boundaries}
\author{ 
	{\bf {\normalsize Krishnamohan Parattu$^a$,}$
		$\thanks{E-mail: krishna@iucaa.ernet.in, mailofkrishnamohan@gmail.com}} \,
	{\bf {\normalsize Sumanta Chakraborty$^a$,}$
		$\thanks{E-mail: sumanta@iucaa.in}} \, 
	{\bf {\normalsize Bibhas Ranjan Majhi$^{a,b}$}$
		$\thanks{E-mail: bibhas.majhi@iitg.ernet.in}} \,\\\\ and
	{\bf {\normalsize T. Padmanabhan$^a$}$
		$\thanks{E-mail: paddy@iucaa.in}}\\\\
	{\normalsize $^a$IUCAA, Post Bag 4, Ganeshkhind,}
	\\{\normalsize Pune University Campus, Pune 411 007, India}\\\\
	{\normalsize $^b$Department of Physics, Indian Institute of Technology Guwahati,}
	\\{\normalsize Guwahati 781039, Assam, India}
	\\[0.3cm]
}\date{\today}
\maketitle

\begin{abstract}
Constructing a well-posed variational principle is a non-trivial issue in general relativity. For spacelike and timelike boundaries, one knows that the addition of the Gibbons-Hawking-York (GHY) counter-term will make the variational principle well-defined. This result, however, does not directly generalize to null boundaries on which the 3-metric becomes degenerate. In this work, we address the following question: What is the counter-term that may be added on a null boundary to make the variational principle well-defined? We propose the boundary integral of $2 \sqrt{-g} \left( \Theta+\kappa \right)$ as an appropriate counter-term for a null boundary. We also conduct a preliminary analysis of the variations of the metric on the null boundary and conclude that isolating the degrees of freedom that may be fixed for a well-posed variational principle requires a deeper investigation.
\end{abstract}



\section{Introduction and Summary}\label{NullIntro}

Just like any other field theory, the dynamics of gravity can be obtained from an action, the Einstein-Hilbert action.  On varying the action with respect to the metric, we obtain an equations-of-motion term and a boundary term generated by integration by parts. The equations of motion turn out to be second order in the derivatives of the metric. But the boundary term is unusual, as it contains variations both of the metric and its first derivatives, both tangential and normal to the boundary \cite{gravitation}. Thus, setting the variation of the action to zero will lead to the equations of motion on the bulk only if  we fix the metric \textit{and} its normal derivatives on the boundary. (The tangential derivatives get fixed when the metric is fixed on the boundary.) The problem with such a structure is that it makes the variational principle ill-defined \cite{Dyer:2008hb}. In general, the equations of motion and the boundary conditions will turn out to be inconsistent \cite{Dyer:2008hb,Parattu:2013gwb}.

There is a widely accepted prescription for resolving this issue. One adds an extra boundary term to the action, called a counter-term, such that the surface term in the variation of the new action contains only variations of the metric and does not involve the variations of the normal derivatives. (We shall be using the terms surface term, boundary term and counter-term interchangeably. Note that the use of the word ``counter-term" here, unlike in many places in the literature, does not mean that our boundary term is supposed to cancel off a divergent part of the action.) Thus, we need to fix only the metric on the boundary and the variational principle becomes well-posed. The most commonly used counter-term is the Gibbons-Hawking-York (GHY) counter-term \cite{York:1972sj, Gibbons:1976ue}, although there are other counter-terms available in the literature \cite{Charap:1982kn}. In cases where the boundary is taken to infinity, other boundary terms in addition to the Gibbons-Hawking-York term are added to make the action finite on classical solutions \cite{Poisson,Mann:2005yr}. But we shall not be discussing this issue in the current work.

The GHY prescription has been around for a long time and is now textbook material (see, e.g., \cite{gravitation, Poisson}). If the spacetime region $\mathcal{V}$ under consideration has a boundary surface denoted by $\partial \mathcal{V}$, the counter-term is an integral over $\partial \mathcal{V}$ of essentially the product of the square root of the determinant of the $3$-metric on the surface and the divergence of the unit normal to the surface. The procedure of its construction uses (a) the unit normal to $\partial \mathcal{V}$, (b) the induced metric $h_{ab}$ on $\partial \mathcal{V}$ and (c) the covariant derivative on $\partial \mathcal{V}$ compatible with the induced metric $h_{ab}$. These structures are well-defined for spacelike or timelike regions of the boundary surface. The trouble is,  we can have a third kind of region on the boundary surface, viz. a null surface, which is ubiquitous in general relativity. The horizons of black holes, for example, are null surfaces. It is generally accepted that entropy and temperature can be associated with black hole horizons \cite{Bekenstein:1972tm,Bekenstein:1973ur,Bekenstein:1974ax,Hawking:1974sw,Hawking:1976de}. There is also a claim in the literature that entropy and temperature can also be ascribed to any local Rindler horizon \cite{Unruh:1976db,Davies:1974th}, which is also a null surface. These  properties, as far as we know, are not shared by any spacelike or timelike surface. Some physical cases where we might require a counter-term for a null surface are the case of a spacetime with a black hole where we want to consider the black hole horizon as one of the boundaries, the case of the interior of a causal diamond formed by the intersection of the future light cone from a point P and the past light cone from a point Q in the future of P \cite{Gibbons:2007nm}, etc. The trouble with the standard prescription is that it is not directly applicable to a null surface as the normal to a null surface has a zero norm and the natural $3$-metric on a null surface is degenerate. Also, the notion of a covariant derivative on a non-null surface does not naturally extend to a null surface, making the usual procedure inapplicable. 

A possible work-around  is to consider  a timelike surface infinitesimally separated from the null surface (``stretched horizon''), perform the calculations and then take the null limit. As we shall demonstrate, this approach does give a proper, finite expression for the boundary term and the counter-term when the null limit is taken with some reasonable assumptions about the limiting behaviour of the metric components. But, given the importance of this problem, one would like to have a first-principle approach, leading to a prescription for the counter-term to be added on the null surface purely based only on properties of the surface, without recourse to any limiting procedure. Surprisingly, this issue does not seems to have attracted sufficient attention in the literature. As far as we know, there is no approach available in the literature that allows one to start from first principles and find the counter-term for a null surface. (The only related reference we could find was \cite{Barth:Phd} which treats a very specialized metric and is not applicable to a general null surface and does not address several important technical and conceptual issues.)

There is another important reason for attempting this study. 
For the non-null surfaces, we also obtain, from the variational principle itself, the result that only the components of the induced metric (modulo diffeomorphisms) need to be fixed at the boundary \cite{Padmanabhan:2014BT}. 
So, in a timelike/spacelike surface one can
take the gravitational degrees of freedom as being contained in the 6 components of the induced 3-metric $h_{ab}$. Given the freedom in transforming the 3 coordinates on $\partial \mathcal{V}$, this leaves 3 independent degrees of freedom. (It is possible to reduce it further to 2 using the freedom in foliation, giving two degrees of freedom for the gravitational field per point \cite{MTW,Wald}.) When we try to obtain the results for a null surface  as a limit of the results on a non-null surface,  the question arises as to how to characterize these degrees of freedom in the null limit.
The properly defined induced metric for null surfaces is essentially a $2$-metric $q_{ab}$. It is not clear how the six degrees of freedom in the induced metric $h_{ab}$ on the non-null surface could relate to the three degrees of freedom in the $2$-metric $q_{ab}$ on taking the null limit.

We stress that it should be possible to tackle the above issues from first principles, by just studying the form of the variation of the action $\df \mathcal{A}_{_{\partial  \mathcal \tiny V}}$ on the null boundary. (We shall use the notation $\df \mathcal{A}_{_{\partial  \mathcal \tiny V}}$ for the boundary term in the variation of the Einstein-Hilbert action for a boundary $\partial \mathcal{V}$ of the spacetime region $\mathcal{V}$ under consideration.). The algebraic structure of $\df \mathcal{A}_{_{\partial  \mathcal \tiny V}}$ should suggest (a) the counter-term that may be added to make the variation well-defined and (b) the variables which need to be fixed on the boundary. (For the corresponding analysis in the case of a non-null boundary, see \cite{Padmanabhan:2014BT}.) It is important to perform this computation for a null boundary.

We should also point out that a complete analysis of the well-posedness of the action requires consideration of the equations of motion as well, since well-posedness essentially requires that the boundary conditions be consistent with the equations of motion and that each choice of boundary conditions selects out a unique, different solution of the equations of motion \cite{Dyer:2008hb}. The evolution of geometry from one spatial surface to another under the Einstein equations is well-understood and there is a standard, accepted formalism in the Arnowitt-Deser-Misner (ADM) formalism \cite{Arnowitt:1962hi}. On the other hand, the evolution of geometry starting from a null surface has been analyzed by several authors \cite{sachs1962characteristic,newman1962approach,geroch1973space,Penrose:1980yx,d'Inverno:1980zz,friedrich1981regular,Torre:1985rw,hayward1993dual,Goldberg:1995gb,Brady:1995na}, but no one approach has yet come to be accepted as the standard. Details like the specification of the degrees of freedom and the identification of constraints vary from approach to approach. One of these formalisms has to be chosen and then the well-posedness of the action has to be analyzed in its framework. We intend to pursue this matter in a future publication.

We will provide comprehensive discussion of these and related issues from several different perspectives in this paper. Because of the rather extensive nature of the discussion, we first
summarize the key results obtained in this work. 

\begin{itemize}
	
	\item We briefly review the procedure of obtaining a counter-term and identifying the variables which need to be fixed on the boundary \textit{just from examining the algebraic structure} of $\df \mathcal{A}_{_{\partial  \mathcal \tiny V}}$ on non-null boundary. Then, we perform a very similar analysis and obtain $\df \mathcal{A}_{_{\partial  \mathcal \tiny V}}$ on a general null surface. The final result is given by
	\begin{align} \label{NewLabel030}
	\delta \mathcal{A}_{null}=&\intdx~ \left\{-2 \delta\left[\sqg(\Theta + \kappa)\right]+\sqg \left[\Theta _{ab}-(\Theta + \kappa) q_{ab} \right]\delta q^{ab} + P_{c}\df \ell^{c}\right\}~.
	\end{align}
	(There is also an ignorable total derivative term on the $3$-surface, which has been omitted here for simplicity but can be found in the main discussion later on.) The general result as given in \ref{NewLabel030} can be interpreted by the following identifications: $\ell_a=\partial_a \phi$ is the normal to the $\phi=\textrm{constant}$ boundary surface which is null (other surfaces of constant $\phi$ may not be null), $k^a$ is an auxiliary null vector and $q_{ab}=g_{ab}+\ell_a k_b+k_a \ell_b$ is the induced metric. The symmetric tensor $\Theta_{ab}=q^m _a q^n _b \nabla_m \ell_n$ is the second fundamental form, $\Theta=\Theta^{a}_a$ is the expansion scalar and $\kappa$ is the non-affinity coefficient on the null surface, i.e. $\ell^a \nabla_a \ell^b=\kappa \ell^b$. (Using $\ell_a=\partial_a \phi$, as opposed to $\ell_a=A\partial_a \phi$ for some scalar $A$, will render $\kappa=0$ if \textit{all} the $\phi=\textrm{constant}$ surfaces are null but not when  $\phi=\textrm{constant}$ is null only for a specific value of $\phi$; this is why $\kappa\neq0$ even though $\ell_a=\partial_a \phi$; see discussion later.). In a coordinate system adapted to a null surface with a particular choice of $k_a$, we can write this decomposition in the form
	\begin{align} 
	\STn=&\intlt \left\{ -2 \delta\left[\sqrt{q}(\Theta + \kappa)\right]+\sqrt{q} \left[\Theta _{ab}-(\Theta + \kappa) q_{ab} \right]\delta q^{ab} + P_{c}\df \ell^{c}\right\}~\label{STgeneral}~,
	\end{align}
	The integration is over a parameter $\lambda$, varying along the null geodesic congruence such that $k_{a}=\nabla_a \lambda$ locally, and two coordinates $\left(z^1,z^2\right)$ that are constant along the null geodesics, with $d^2z_\perp=dz^1 dz^2$. The symbol $q$ is used for the determinant of the $2$-metric $q_{AB}$. The conjugate momentum to $q^{ab}$ is $\sqrt{q} \left[\Theta _{ab}-(\Theta + \kappa) q_{ab} \right]$ and the conjugate momenta to $\ell ^{a}$ turns out to be
	\begin{align}
	P_{c}=-\sqrt{q}  k_{b}\left[\nabla _{c}\ell^{b}+\nabla ^{b}\ell_{c}- 2\df^b_c \left(\Theta +\kappa \right) \right]~.
	\end{align}
	We have also carried out the analysis for $\ell _{a}$ having the form $A\partial _{a}\phi$ in \ref{app:gen-normal}. In that case also, we obtain a decomposition of the same structure as above.
	
	The structure of $\STn$ is similar to the familiar decomposition that we obtain for the surface term on a non-null surface:
	\begin{align}\label{ADM-ST}
	\STnn=\int_\mathcal{\partial V} d^3x \left[  \delta ( 2\sqrt{|h|}  K) - \sqrt{|h|} (K_{ab} - K h_{ab}) ~\delta h^{ab}\right]~. 
	\end{align}
	(Again, there is a total $3$-derivative on the surface that we have omitted here but  can be found in the main text.) Thus, \ref{NewLabel030} suggests that the counter-term to be added on a null surface is $2\sqrt{-g}(\Theta + \kappa)$ ($2\sqrt{q}[\Theta + \kappa]$ for \ref{STgeneral}) and that the intrinsic quantities to be fixed on the null surface are $q^{ab}$ and $\ell^a$.
\item We verify the result in \ref{STgeneral} by deriving it through a different route. We start from the standard decomposition of the surface term for a non-null boundary and then take a careful null limit with some reasonable limiting behaviour for the metric components. We obtain back \ref{STgeneral}, reaffirming our faith in the result. 

\item In an appendix, we illustrate these general results in  two natural coordinate systems adapted to describe the metric near an arbitrary null surface. We first review the explicit construction of both these coordinate systems, which we call the  Gaussian Null Coordinates (GNC) and Null Surface Foliation (NSF). The GNC metric has the feature that the null surface arises naturally as the limit of a sequence of non-null surfaces. Thus, we can easily apply the GHY prescription on a nearby non-null surface and take the null limit. In the NSF, the fiducial null surface appears as one member of a set of null surfaces. Hence, the limiting procedure cannot be applied directly (and would need the explicit construction of an infinitesimally separated non-null surface). But the NSF has some other advantages. In both cases, we find that we obtain explicit results in agreement with \ref{STgeneral}.

\item The question of whether the variational principle is well-posed with this counter-term requires deeper analysis. Also, the limitations of our framework, since we impose certain constraints on the variations of the metric to keep, for example, the null surface as a null surface, does not allow us to directly connect these boundary conditions with the degrees of freedom in the theory.

\end{itemize}

The conventions used in this paper are as follows: We use the metric signature $(-,+,+,+)$. The fundamental constants $G$, $\hbar$ and $c$ have been set to unity. The Latin indices, $a,b,\ldots$, run over all space-time indices, and are hence summed over four values. Greek indices, $\alpha ,\beta ,\ldots$, are used when we specialize to indices corresponding to a codimension-1 surface, i.e a $3-$surface, and are summed over three values. Upper case Latin symbols, $A,B,\ldots$, are used for indices corresponding to two-dimensional hypersurfaces, leading to sums going over two values. We shall use $\eqH$ to indicate equalities that are valid only on the null surface.

\section{The information contained in the boundary term of the variation of the action}\label{BT-info}

The variation of an action, generically, contains two terms: (a) A bulk term, the vanishing of which will give us the equations of motion and (b) a boundary term, which is usually ignored. But the boundary term contains significant amount of information! By a careful analysis of this term, we can determine \textit{the structure of the counter-term that may be added to the action principle} to make it well-defined and, with that counter-term, what degrees of freedom are to be fixed at the boundary. Some of these degrees of freedom may be eliminated by using gauge freedom and, once the gauge freedom is exhausted, the number of the remaining degrees of freedom represents \textit{the number of true degrees of freedom in the theory}.  Let us illustrate this idea in electromagnetism and gravity.

\subsection{Warm-up: Electromagnetic field}
As a warm up, let us consider the simple case of electromagnetism (see \cite{MTW}, Chapter 21). The action for the free electromagnetic field is given by
\begin{equation}
	\mathcal{A} = -\frac{1}{16\pi} \int_\mathcal{V}  F_{ik}F^{ik}d^{4}x; \qquad F_{ik}=\partial_iA_k-\partial_kA_i
\end{equation}
Let us begin by assuming that $A_i$ are the dynamical variables and vary them in the action. This leads to
\begin{equation}
	\delta \mathcal{A}=-\frac{1}{4\pi}\int_\mathcal{V} \left[\partial _{k}F^{ik}\right]\delta A_{i}d^{4}x 
	+\frac{1}{4\pi}\int_{\partial\mathcal{V}} F^{ik}\delta A_{i}d\sigma _{k}
	=-\frac{1}{4\pi}\int_\mathcal{V} \left[\partial _{k}F^{ik}\right]\delta A_{i}d^{4}x 
	+\frac{1}{4\pi}\int \mathbf{E}\cdot\delta \mathbf{A}d^3\mathbf{x}
\end{equation}
where we have assumed that the boundary contributions arise from $t=$ constant surfaces.
Since bulk and boundary variations are independent, $\delta \mathcal{A}=0$ will require vanishing of these two terms individually. 
Let us first assume (rather naively) that $\delta A_{i}$ is completely arbitrary in the bulk $\mathcal{V}$. Then the vanishing of the bulk term  will lead to the equations of motion $\partial _{k}F^{ik}=0$. Once this is granted, we next want the on-shell boundary term to vanish. This is, of course, possible if we specify $A_i$ on the boundary; but that is not required. The algebraic structure of the boundary term already tells us that we need not fix $A_0$ at all. Further, even as regards $\mathbf{A}$ we only need to fix it modulo the addition of a gradient: $\mathbf{A}\to \mathbf{A}+\nabla F$. This is because, $\mathbf{E}\cdot \nabla F=\nabla\cdot(\mathbf{E} F)$ on-shell (since $\nabla\cdot \mathbf{E}=0$) which allows this term to be converted into a surface integral at spatial infinity and ignored. Thus, we only need to fix the part of $\mathbf{A}$ which is unaffected by the addition of a gradient; viz., the $\mathbf{B}=\nabla \times \mathbf{A}$. The magnetic field, since it is a spatial vector, has three components, of which one component is eliminated by the constraint $\mathbf{\nabla}. \mathbf{B}=0$ arising from the equations of motion. Thus, there are two degrees of freedom to be fixed at a boundary. Further, it can be verified that each choice of these two degrees of freedom selects out a single solution of the equations of motion. Thus, the action is well-posed and does not require the addition of a counter-term.  
In other words: 

\textit{We do not tell the action principle what needs to be fixed at the boundary; instead, the action principle tells us what should be fixed. If the action is well-posed, the number of degrees of freedom that needs to be fixed at a boundary will correspond to the number of true degrees of freedom in the theory.}

In the case of the electromagnetism, it is the magnetic field that is to be fixed at the boundary. We naively thought it is $A_i$ but the algebraic structure of the boundary term in the variation of the action corrects our mistake and tells us that it is the magnetic field.

\subsection{Gravity: Non-null boundary}
 
One can do exactly the same in the case of gravity. We will first assume (again naively)
that $g_{ab}$ are the dynamical variables and vary them in the Einstein-Hilbert action
\begin{equation}
16\pi \mathcal{A}_{\rm EH}= \int_\mathcal{V} d^4x\sqrt{-g}\, R,
\label{A_EH}
\end{equation} 
where $\mathcal{V}$ indicates the spacetime region under consideration, over which the integration is performed. On varying the metric, we get
\begin{align}
16\pi \delta \mathcal{A}_{\rm EH} &= -\int_\mathcal{V} d^4x \sqrt{-g}\, G^{ab} \delta g_{ab} + \int_\mathcal{V} d^4x \, \sqg g^{ab} \delta R_{ab}  \nn \\
&\equiv -\int_\mathcal{V} d^4x \sqrt{-g}\, G^{ab} \delta g_{ab} + \int_\mathcal{V} d^4x \sqrt{-g}\, \nabla_c  w^c \nn \\
&= -\int_\mathcal{V} d^4x \sqrt{-g}\, G^{ab} \delta g_{ab} + \int_\mathcal{\partial V} d^3S \, n_c w^c \nn \\
&\equiv -\int_\mathcal{V} d^4x \sqrt{-g}\, G^{ab} \delta g_{ab} +\int_\mathcal{\partial V} d^3S \, Q[n_c] 
\label{var_A_EH}
\end{align} 
where $d^3S$ is the appropriate integration measure on the boundary, $n_c$ is the normal with constant norm (0 or $\pm 1$) on the surface,
$
w^c\equiv \left(g^{ab} \delta \Gamma^{c}_{ab}-g^{ck} \delta \Gamma^{a}_{ak}\right)
$
and  
we have defined the function $Q[A_c]$ for any one-form $A_c$ as 
\begin{align}\label{Q_def}
Q[A_{c}] &\equiv A_{c} (g^{ab} \delta \Gamma^{c}_{ab}-g^{ck} \delta \Gamma^{a}_{ak})~.
\end{align}
which will prove to be useful throughout the paper.
Straightforward algebra now leads to the following alternate expression for $Q[n_c]$:
\begin{equation}
Q[n_{c}] = \nabla_{a}(\delta \un^{a})- \delta (2 \nabla_a n^{a}) + \nabla_{a} n_{b} ~\delta g^{ab}; \qquad \delta \un^{a} \equiv \delta n^{a} + g^{ab} \delta n_{b}
\label{eq:q_n_exp1}
\end{equation}
This expression can be easily verified by expanding out and simplifying the RHS  (see \ref{BC-novel} for a derivation). Note that 
$
\delta \un^{a} n_{a}=0,
$
which implies that the vector $\delta \un^{a}$ lives on the boundary.
The description so far is completely general and holds for \textit{both null and non-null} boundaries with suitable measure $n_cd^3S $ for integrating vector fields. 

For the non-null case (\cite{Padmanabhan:2014BT}, also see \ref{BC-novel}), we have  the relation $\nabla_{a} V^{a} = D_{a} V^{a} - \epsilon a_{b}  V^{b}$  valid for any $V^a$ which satisfies $V^an_a=0$, where $a_i\equiv n^j\nabla_j n_i$. Further,
the integration measure is now $d^3S=\sqrt{h} d^3x$ with $n_a$  normalized to $\pm 1$ so that
the surface term  becomes the integral of $d^3x\ \sqrt{|h|} Q[n_{c}]$. 
This will lead to the expression:
 \begin{equation}
Q[n_{c}] = D_{a}(\delta \un^a)- \delta (2 \nabla_a n^{a}) + (\nabla_{a} n_{b}-\epsilon n_{a} a_{b}) ~\delta h^{ab}
\label{Qalt}
\end{equation}
\textit{Note that this algebraic manipulation has naturally led us to the combination $(\nabla_{a} n_{b}-\epsilon n_{a} a_{b})$ without us having to introduce any geometrical considerations.} This combination, of course, is the negative of the extrinsic curvature \cite{gravitation} defined by
$
K_{ab}= - h^{c}_{a} \nabla_c n_{b}
$
and has the following properties:
$
(a) K_{ij} = K_{ji}; (b) K_{ij} n^{j}=0;(c) K = g^{ij} K_{ij} = - \nabla_a n^{a}~.
$
The second property tells us that $K_{ij} \delta g^{ij} = K_{ij} \delta h^{ij}$ which has been used to arrive at \ref{Qalt}.
The surface term is an integral of $Q[n_c]$ over $d^3x \sqrt{|h|}$ which allows the first term    
$D_{a}(\delta \un^a)$ in \ref{Qalt} to be converted to a 2-dimensional surface integral and ignored. The rest of the terms will reduce to the following expression for the boundary term of EH action:
\begin{align}
\STnn 
&=\int_\mathcal{\partial V} d^3x \, \left[ \delta ( 2\sqrt{|h|}  K) - \sqrt{|h|} (K_{ab} - K h_{ab}) ~\delta h^{ab}\right]~. \label{S.T-nonnull1}
\end{align}
We need  to be set $\STnn$ to zero with appropriate boundary conditions to obtain the equations of motion in the bulk from the action principle $\delta \mathcal{A}=0$. Unlike in the case of e.g., electromagnetism, this is not straightforward because we will need to set both $\delta[ K\sqrt{h}]=0$ and $\delta h^{ab}=0$ on the boundary.  
This  is not acceptable, since $\delta K$ contains  variations of normal derivatives of the metric as well. Setting both $\delta[ K\sqrt{h}]=0$ and $\delta h^{ab}=0$, in general, will lead to inconsistency with the Einstein's equations. (For a discussion of this and other issues with fixing the metric and its derivatives at the boundaries, see, e.g, \cite{Dyer:2008hb}.)

\textit{Incredibly enough, the algebraic structure of \ref{S.T-nonnull1} itself suggests a way out.} Since the term involving $\delta [K\sqrt{h}]$ is the variation of a boundary integral, we can add this as a  counter-term to the action such that the surface term arising from the new action does not depend on the variations of the normal derivatives of the metric. Once this is done, we only have to set $\delta h^{ab}=0$ on the boundary and this leads to a well-posed action principle.
This leads to the GHY counter-term \cite{York:1972sj,Gibbons:1976ue}, viz. integral of $-2 K\sqrt{h}$ over the boundary, which is the most commonly used one for this purpose. 

Thus, the algebraic structure of the boundary term in the variation of the action itself can tell us \textit{not only} (i) what counter-term can be added to the action to make the variation  well-defined \textit{but also} (ii) what degrees of freedom need to be fixed  on the boundary [viz. $h_{ab}$] after the addition of this counter-term. Even here, eliminating four degrees of freedom using the gauge freedom that is the freedom of diffeomorphisms, one should end up with two remaining degrees \cite{MTW}, matching the number of degrees of freedom of the gravitational field obtained from considering gravitational radiation in the linearized approximation. 

\section{Boundary Term of the Einstein-Hilbert Action: Null Surfaces} 

The crucial difference between non-null and null boundaries is the following: If the surface is non-null, we can define an induced metric $h_{ab}=g_{ab}-\epsilon n_a n_b$ and a covariant derivative operator $D_a$ compatible with it (i.e., $D_ah_{bc}=0$). We can then re-express
$ \nabla_{a}(\delta \un^{a})$ in terms of the covariant derivative $ D_{a}(\delta \un^{a})$ on the boundary which leads to significant simplification. {We cannot do this for a null surface, which makes a vital algebraic difference.}
We, therefore, need to use some other procedure to analyse the algebraic structure of the boundary variation on a null-surface. 

One possible approach to handle a null surface would be to treat the null surface as the limit of a sequence of non-null surfaces, apply the  prescription given above on the non-null surfaces and then take the limit in which this sequence goes over to the null surface. This can be done with the help of a parameter which labels the surfaces such that a particular value of the parameter corresponds to the null surface; we will do this in \ref{ADMtoNull}. However,  we would also like to formulate a first-principle prescription that can be applied to the null surface as a surface in its own right and not as a fringe member of some family. This will be done in \ref{paddy_deriv-null_surf}. As a preamble to these two derivations, we shall first recall several features of null surfaces relevant to our task.

\subsection{Null surfaces}

 A null surface in a four-dimensional spacetime $M$ is a three-dimensional submanifold $N$ with the criterion that the metric $\gamma _{\mu \nu}$ obtained by the restriction of the full metric $g_{ab}$ to the hypersurface $N$ is degenerate, which means that it is possible to find non-zero vectors $v^{\mu}$ on $N$ such that $\gamma _{\mu \nu}v^{\mu}=0$. We shall assume that the full $g_{ab}$ is non-degenerate, i.e every non-zero four-dimensional vector is mapped by the metric to a non-zero one-form. If $v^{a}$ is the four-dimensional push-forward of $v^{\mu}$ and $w^{a}$ is the four-dimensional push-forward of $w^{\mu}$, another vector at the same point on $N$, then $g_{ab}v^{a}w^{b}= \gamma_{\mu \nu} v^{\mu} w^{\nu}=0$. Thus, $v^{a}$ is orthogonal to every vector at that point on the surface including itself. A set of such null vectors on the surface, one vector per point, gives a null vector field $\ell^{a}$ normal to $N$. (There will be only one null curve on the surface passing through each point on $N$. So the only freedom we have in the choice of the null vector is the freedom of scaling by a scalar factor.) Note that $\ell^a$ has the peculiar property that it is both tangent and normal to the surface: tangent since we had started with the vector being on the null surface and hence we will be moving along the surface if we move along the orbits of $\ell^a$; and normal since we have $g_{ab}w^a\ell^b=0$ for any vector $w^a$ on the surface. 

If $N$ corresponds to some $\phi =$constant surface, the corresponding normal one-form can be expressed as $\ell_{a}=A\partial _{a}\phi$, where $A$ is a scalar. Since we have assumed the four-dimensional metric $g_{ab}$ to be non-degenerate, $A \neq 0$. As proved in \ref{inaffinity}, we will then have 
\begin{equation}\label{kappa_eq_1}
\ell^a\nabla_a\ell_b \eqH \kappa \ell_b; \qquad \kappa\equiv  \ell^a \partial_a (ln A) - \frac{k^c}{2} \partial_c (\ell^{a}\ell_{a})
\end{equation} 
Note that while we assume $\ell^2=0$ on the null surface, it can in general be non-zero off the surface. So $\kappa$ has two contributions: one arises because $A\neq 1$ and the other because $\ell^2\neq0$ off the surface. Very often, we will set $A \eqH 1$ on the null surface making $\ell_a \eqH \nabla_a\phi$ on it. The above result shows that $\kappa$ can be still non-zero on the null-surface due to the second term in its expression in \ref{kappa_eq_1}.

Since $\ell^a\nabla_a\ell_b \eqH \kappa \ell_b$ is a geodesic equation we see that the orbits of $\ell^{a}$ are geodesics. We may call $\kappa$ as the non-affinity coefficient because $\kappa =0$ if we choose an affine parametrization. Since $\ell^{a}$ lies on the null surface, the null surface can be thought of as being filled by a congruence of null geodesics. (If e.g., $\ell^{a}$ is the Killing field normal to the horizon of a stationary black hole in an asymptotically flat spacetime, we can identify $\kappa$ with the surface gravity of the horizon \cite{Wald}). Two specific parametrizations of the metric on a general null surface are provided in \ref{app:BT-GNC+NSF}.

\subsection{Structure of the boundary term}

To study the boundary term on a null surface, it is more convenient to use the notion of a \textit{surface gradient} rather than normal to the surface and rewrite the variation of the Einstein-Hilbert action in the following  form:
\begin{align}
16\pi \delta \mathcal{A}_{\rm EH} 
= -\int_\mathcal{V} d^4x \sqrt{-g}\, G^{ab} \delta g_{ab} + \int_\mathcal{\partial V} d^3x \, v_c \left[ \sqg \left(g^{ab} \delta \Gamma^{c}_{ab}-g^{ck} \delta \Gamma^{a}_{ak}\right)\right]~,
\label{var_A_EH1}
\end{align} 
We have again converted the total divergence $\nabla_a w^a$ in \ref{var_A_EH} to a term on the boundary and defined $v_c$ to be the \textit{surface gradient} (not the unit normal) to the surface that bounds the volume. If the surface is given by the equation $\phi=\textrm{constant}$, then \ref{var_A_EH1} is valid in any coordinate system where $\phi$ is one of the coordinates with $v_{c}\equiv \pm \partial_c \phi$, the appropriate sign chosen as per the conventions of Gauss' theorem (see \ref{app:Gauss}). When the volume integral is done as usual with the higher value of $\phi$ at the upper limit of the integration, $\partial_c \phi$ or $-\partial_c \phi$ is used depending on whether $\phi$ or $-\phi$ increases on going from inside the volume to outside through the surface. (Note that $\sqrt{-g}$ appears in \ref{var_A_EH1} since we have used the surface gradient instead of the unit normal. This is appropriate at this stage since $\mathcal{V}$ could be null in which case there is no natural notion of "unit" normal $n_c$ or $\sqrt{|h|}$. If we choose to use the unit normal $n_{a}$, then $n_{a}\sqrt{|h|}$ will replace $v_{a} \sqg$; see \ref{BC-novel}.) 

We have earlier defined the useful function $Q[A_c]$ in \ref{Q_def} for any one-form $A_c$ as 
\begin{align}
Q[A_{c}] &\equiv A_{c} (g^{ab} \delta \Gamma^{c}_{ab}-g^{ck} \delta \Gamma^{a}_{ak})
=\nabla_{a}(\delta \uA^a)- \delta (2 \nabla_a A^{a}) + \left(\nabla_{a} A_{b}\right) ~\delta g^{ab}
\end{align}
where $\delta \uA^{a}\equiv \delta A^{a} + g^{ab} \delta A_{b}$.
Adding a $\sqg$ and manipulating, we can obtain
\begin{equation}
\sqrt{-g} Q [A_c] = \sqrt{-g} \nabla_c [\delta \uA^{c}] - 2 \delta (\sqrt{-g} \nabla_a A^{a}) + \sqrt{-g} (\nabla_a A_b -g_{ab} \nabla_c A^{c}) \delta g^{ab}~, \label{qgen}
\end{equation}
When $A_c=v_{c}=\partial_c \phi$ is the surface gradient, the integral of this expression gives the surface term of the Einstein-Hilbert action on a $\phi=\textrm{constant}$ surface, \textit{irrespective of whether the surface is timelike, spacelike or null}.  The surface term $\sqrt{-g}Q$ depends only on the surface gradient, $\partial_a \phi$, at the $\phi=\phi_{0}$ surface. It does not depend on the normalization we may choose for $v_c$, i.e, it does not depend on $B$ if we choose to write $v_c= B \partial_c \phi$. It also does not depend on the behaviour of $v_c$ away from the surface. But the decomposition in \ref{qgen} needs to be rewritten if we choose to work with $v_c= B \partial_c \phi$. Also, the decomposition cares about the behaviour of $v_c$ away from the surface due to the presence of derivatives of $v_{c}$. 

 Let us now concentrate on the null-surface and denote the surface gradient on it by $v_c\equiv\ell_c=\nabla_c\phi$. Following the route we took for non-null surfaces, we shall first convert the first term in \ref{qgen} into a surface term on the $3-$surface $\phi=\phi_{0}$. For this, we require the notion of surface covariant derivative on the null surface, which, in turn, requires a projector on to the surface. Based on the analogy with non-null surfaces (where $h_{ab}=g_{ab}- \epsilon n_{a}n_{b}$ is the projector), if we try  $h_{ab} = g_{ab}- \epsilon \ell_{a} \ell_{b}$, it does not work because now 
 $
 h_{ab}\ell^{a} = \ell_{b} - \epsilon \ell^{a}\ell_{a} \ell_{b} = \ell_{b} \neq 0, 
 $
 since $\ell_{a}\ell^{a}=0$ on the null surface. Thus, there is no straightforward extension of the projector in the non-null case to the null case, essentially because the metric is degenerate.

The solution to this problem was suggested by Carter in \cite{Hawking:2010mca}. Since we are not able to define a projector by using $\ell_{a}$ alone, we choose an auxiliary vector $k^{a}$ such that $\ell_{a}k^{a}=-1$ at $\phi=\phi_0$. Then, as we prove in \ref{app:q_for_null}, $q_{ab}= g_{ab} + \ell_{a}k_{b}+\ell_{b}k_{a}$ is the object on the null surface analogous to $h_{ab}$ (see also \cite{Poisson}). Since $k^a$ is under our control, we shall assume that $\ell_{a}k^{a}=-1$ and $k_{a}k^{a}=0$ to be valid everywhere. Thus, we have
\begin{equation} \label{l_k_conditions}
\ell_{a}\ell^{a} \eqH 0; \qquad~ \ell_{a}k^{a}=-1;~\qquad k_{a}k^{a}=0~. 
\end{equation}
We shall enforce the first relation only on the boundary null surface. Further, we shall demand that these relations are respected by the variations. Since $\ell_a=\partial_a \phi$, this demand on the first relation requires $\df g^{\phi \phi}=0$. This is a restriction on the variations of the metric and translates to the statement that we are only considering variations of the metric that keep the null surface null. We should ideally consider arbitrary variations of the metric, but the geometrical setup that we are using will lose its validity for the varied metric if the boundary surface ceases to be null. We hope to address this issue in a future publication. The second and third relations do not require any constraint on the metric variations. This is because these relations are the \textit{only} restrictions on $k^a$ and we can \textit{choose} an appropriate $k'^a= k^a+ \df k^a$ in the varied metric. (Compare this with the case in usual double null formulations where $k^a$ is also chosen to be hypersurface-orthogonal \cite{vickers2011double}.) Our analysis in \ref{paddy_deriv-null_surf} will be general while we shall choose a form for $k_a$ in \ref{spec_coord}.

\subsection{Boundary contribution in terms of 2-surface variables}\label{paddy_deriv-null_surf}
In order to find an expression for the boundary term on a null surface in terms of the $2$-surface variables, similar to the expression for non-null surfaces in \ref{S.T-nonnull1}, we shall start with the expression for the surface term in \ref{qgen}, with $v_{a}=\partial_a \phi$ being the surface gradient. We may take our null normal to be $\ell_a = A \partial_a \phi=A v_{a}$, where $A$ is some scalar. Then, recalling the definition of $Q$ in \ref{Q_def}, we can write the boundary term on the null surface in the following form:
\begin{equation}\label{null-ST-1}
\sqg Q[v_{a}]=\frac{\sqg}{A}Q[\ell_{a}]= \frac{1}{A}\left\{ \sqrt{-g} \nabla_c [\delta \ul^{c}] - 2 \delta (\sqrt{-g} \nabla_a \ell^{a}) + \sqrt{-g} (\nabla_a \ell_b -g_{ab} \nabla_c \ell^{c}) \delta g^{ab} \right\}~.
\end{equation}
But since we do not seem to have any natural way of fixing the factor $A$ (unlike the case in the non-null case where the condition $n_{a}n^{a}=\epsilon$ was a natural choice) we shall make the choice $A=1$. (The counter-term for a general $A$ is derived in \ref{app:gen-normal}.) This offers two advantages in our manipulations. First, it eliminates the $1/A$ factor in \ref{null-ST-1}. Secondly, we will have $\df \ell_{a}=\partial_a \df \phi =0$ as the scalar $\phi$ labelling the null surface is not being varied. With the choice $A=1$, the boundary term on the null surface becomes
\begin{equation}\label{null-ST-2}
\sqg Q[\ell_{a}]=  \sqrt{-g} \nabla_c [\delta \ul^{c}] - 2 \delta (\sqrt{-g} \nabla_a \ell^{a}) + \sqrt{-g} (\nabla_a \ell_b -g_{ab} \nabla_c \ell^{c}) \delta g^{ab}~.
\end{equation}
Our first task will be to separate out a surface term from the first term in \ref{null-ST-2}. We shall label this term as $\sqg Q_1$. We have
\begin{equation}
\sqg Q_1 =\sqrt{-g} \nabla_a [\delta \ul^{a}] = \partial_{a}[\sqg \delta \ul^{a}]
\end{equation}
Let us consider this expression in the coordinates $(\phi, y_1,y_2,y_3)$, where $(y_1,y_2,y_3)$ are some coordinates introduced on the $\phi=$constant null surface under consideration. Then, the partial derivatives with respect to $(y_1,y_2,y_3)$ are on the null surface and the $\phi$-derivative is off the surface. Thus, we have
\begin{equation}\label{null-ST-3}
\sqg Q_1 = \partial_{\phi}[\sqg \delta \ul^{\phi}] +\partial_{\alpha}[\sqg \delta \ul^{\alpha}],
\end{equation}
where $\alpha$ runs over $(y_1,y_2,y_3)$. Now, $\delta \ul^{\phi}=\df \ul^{a} \ell_a=\df \left(\ell_a \ell^a\right)$. We have assumed this to be zero on the null surface but not off it. Hence, the first term in \ref{null-ST-3} is not zero and $\sqg Q_1$ is not a total surface derivative on the null surface. 

In order to separate out a total surface derivative on the null surface from $\sqg Q_1$, there are two projectors, $\Pi^a_{\phantom{b}b}$ and $q^{a}_{b}$, that we can use on $\delta \ul^{a}$. They are defined by (see \ref{app:q_for_null})
\begin{align}
\Pi^a_{\phantom{b}b}= \df^{a}_b+k^a \ell_b~, \\
q^a_b =\df^{a}_b+k^a \ell_b +\ell^a k_b~.
\end{align}
Now, $\Pi^a_{\phantom{b}b}$ projects orthogonal to $\ell_a$ but $q^{a}_{b}$ projects orthogonal to both $\ell_a$ and $k_a$. Since our aim is to separate out derivatives along the surface (i.e, orthogonal to $\ell_a$) even if they are not orthogonal to $k_a$, we shall use the projector $\Pi^a_{\phantom{b}b}$. Thus,
\begin{align}
\sqg Q_1 = \partial_{a}[\sqg \delta \ul^{a}] &=\partial_{a}[\sqg \Pi^a_{\phantom{b}b}\delta \ul^{b}]-\partial_{a}[\sqg k^a \ell_b \delta \ul^{b}]  \nn\\
&= \partial_{a}[\sqg \Pi^a_{\phantom{b}b}\delta \ul^{b}]-\sqg \df\left(\ell_a \ell^a\right)\nabla_b k^b -\sqg k^b \nabla_b \left[\df\left(\ell_a \ell^a\right) \right] \nn\\ 
&= \partial_{a}[\sqg \Pi^a_{\phantom{b}b}\delta \ul^{b}]-\sqg k^b \partial_b \left[\df\left(\ell_a \ell^a\right) \right]\label{null-ST-4}, 
\end{align}
where the last step was obtained by using our assumption $\df\left(\ell_a \ell^a\right)=0$ on the null surface. The first term in \ref{null-ST-4} is a surface derivative on the null surface as $\Pi^a_{\phantom{b}b}\ell_a=0$. The second term in \ref{null-ST-4} is a bit of a bother. With $\ell_a=\partial_a 	\phi$, $\df\left(\ell_a \ell^a\right)= \df g^{\phi \phi}$. Hence, this term has variations of the derivatives of the metric. We shall take out the $\df$ to obtain
\begin{equation}\label{null-ST-5}
-k^b \partial_b \left[\df\left(\ell_a \ell^a\right) \right] = -\df \left[k^b \partial_b \left(\ell_a \ell^a\right) \right]+ \df k^b \partial_b \left(\ell_a \ell^a\right) ~.
\end{equation}   
The vector $\df k^a$ lies on the null surface as $\ell_a\df k^a=\df \left(\ell_a k^a\right)=0$ as per our assumption $\df \ell_a=0$. Since $\ell_a \ell^a=0$ on the null surface, the last term in \ref{null-ST-5} is zero and we are left with
\begin{equation}\label{null-ST-6}
-k^b \partial_b \left[\df\left(\ell_a \ell^a\right) \right] = -\df \left[ k^b \partial_b \left(\ell_a \ell^a\right) \right] ~.
\end{equation}   
Substituting in \ref{null-ST-4}, we obtain
\begin{align}
\sqg Q_1 &=\partial_{a}[\sqg \Pi^a_{\phantom{b}b}\delta \ul^{b}]-\sqg \df \left[k^b \partial_b \left(\ell_a \ell^a\right) \right] \nn\\
&=\partial_{a}[\sqg \Pi^a_{\phantom{b}b}\delta \ul^{b}]- \df \left[ \sqg k^b \partial_b \left(\ell_a \ell^a\right) \right] - \frac{\sqg}{2} \left[ k^b \partial_b \left(\ell_a \ell^a\right) \right] g_{ij} \df g^{ij} \label{Q1-simplified}
\end{align}
Thus, we have written $\sqg Q_1$ in a form where all the variations of the derivatives of the metric have been smuggled into surface derivatives and total variations.

Substituting \ref{Q1-simplified} back in \ref{null-ST-2}, we find the following expression occurring in two places:
\begin{equation}
\nabla_a \ell^a+\frac{k^a}{2}  \partial_a \left(\ell_b \ell^b\right)= \df^{a}_{b} \nabla_a \ell^b + k^a \ell_b \nabla_a \ell^b = \Pi^a_{\phantom{b}b}\nabla_a \ell^b~.
\end{equation}
Thus, the boundary term on the null surface reduces to
\begin{equation} \label{null-ST-intermediate}
\sqg Q[\ell_{a}]=\partial_{a}[\sqg \Pi^a_{\phantom{b}b}\delta \ul^{b}] - 2 \delta (\sqrt{-g} \Pi^a_{\phantom{b}b}\nabla_a \ell^b) + \sqrt{-g} (\nabla_a \ell_b -g_{ab} \Pi^c_{\phantom{b}d}\nabla_c \ell^d) \delta g^{ab}~.
\end{equation}
We have succeeded in separating out a total derivative on the surface and a total variation to remove all derivatives of the metric. In order to understand what degrees of freedom need to be fixed on the null surface, we shall now decompose $g^{ab}$ into $q^{ab}$, $\ell^a$ and $k^a$.

First, note that the relations $q^{ab}\ell_a=0$ and $q^{ab}k_a=0$ are respected by the variations. This is because we have assumed \ref{l_k_conditions} to be valid even on variation. Thus, terms of the form $\ell_a \ell_b\df q^{ab} $, $\ell_a k_b\df q^{ab}$ and $k_a k_b\df q^{ab}$ would reduce to zero. For example,
\begin{align}\label{lkdfq}
\ell_a k_b \df q^{ab}  = \df \left(q^{ab} \ell_a\right)k_b - q^{ab} k_b \df \ell_a =0~. 
\end{align}
Using this result, we can simplify $g_{ab}\df g^{ab}$ as follows:
\begin{align}
g_{ab} \df g^{ab}&=g_{ab}\left[\df q^{ab}-\df \left(\ell^{a}  k^{b}\right)-\df \left(\ell^{b} k^{a}\right) \right]= q_{ab}\df q^{ab}+2 \left(\ell_{a}k_{b} + \ell_{b}k_{a}\right) \df \left(\ell^{a}  k^{b}\right) \nn\\ &= q_{ab}\df q^{ab}+2 \ell_{b}k_{a} \df \left(\ell^{a}  k^{b}\right)= q_{ab}\df q^{ab} - 2 k_{a}\df \ell^{a},  \label{gdfg}
\end{align}
where we have also used $\ell_{a} \df k^{a}=0$, arising from our assumptions of $\df \ell_{a}=0$ and $\df( \ell_a k^a)=0$.

Next, we shall simplify $\left(\nabla_{a} \ell_b \right) \df g^{ab}$. This is a symmetric object since $\ell_b=\nabla_b \phi$. We have
\begin{align}
\left(\nabla_{a} \ell_b \right) \df g^{ab} &= \left(\nabla_{a} \ell_b \right) \df q^{ab} - 2 \df\left(\ell^a k^b \right) \nabla_a \ell_b \nn\\
                             &= \left(\nabla_{a} \ell_b \right) \df q^{ab} - 2 \df{\ell^a} k^b \nabla_a \ell_b - 2 \df{k^b }\ell^a \nabla_a \ell_b\nn\\
                            &= \left(\nabla_{a} \ell_b \right) \df q^{ab} - 2 \df{\ell^a} k^b \nabla_a \ell_b - 2\kappa \df{k^b } \ell_b\nn\\
                            &= \left(\nabla_{a} \ell_b \right) \df q^{ab} - 2 \df{\ell^a} k^b \nabla_a \ell_b, \label{dldg-1}
\end{align}
where we have used $\ell^a \nabla_a \ell_b=\kappa \ell_b$ (see \ref{kappa_eq_1}) and $\ell_a\df k^a=- k^a \df \ell_a=0$. The expression $\left(\nabla_{a} \ell_b \right) \df q^{ab}$ can be simplified as follows:
\begin{align}\label{NewLabel10}
\left(\nabla_{a} \ell_b \right) \df q^{ab} = \df^m_a \df^n_b \left(\nabla_{m} \ell_n \right) \df q^{ab} &= \left(q^m_a -\ell^m k_a-k^m \ell_a\right) \left(q^n_b-\ell^n k_b-k^n \ell_b\right) \left(\nabla_{m} \ell_n \right) \df q^{ab}   \nn\\
&= \left(q^m_a -\ell^m k_a \right) \left(q^n_b-\ell^n k_b \right) \left(\nabla_m \ell_n \right) \df q^{ab} \nn \\
&=\left(q^m_a q^n_b \nabla_m \ell_n - q^m_a \ell^n k_b \nabla_m \ell_n - q^n_b \ell^m k_a \nabla_m \ell_n \right) \df q^{ab} \nn \\
&=\left(\Theta_{ab}-2\kappa q^m_a  k_b  \ell_m \right) \df q^{ab} \nn \\
&=\Theta_{ab} \df q^{ab}
\end{align}
where we have used $\ell_a\df q^{ab}=\df \left(q^{ab} \ell_a \right)=0$ to get to the second line, $k_a k_b \df q^{ab}=0$ (see \ref{lkdfq})to get to the third line, the definitions of $\Theta_{ab}$ and $\kappa$ (see \ref{app:SFF+exp} and \ref{kappa_eq_1} in this paper) and the symmetry of $\nabla_a \ell_b$ to get to the fourth line and $\ell_a q^{ab}=0$ to get the final result.

Substituting back in \ref{dldg-1}, we get
\begin{equation} \label{dldg}
\left(\nabla_{a} \ell_b \right) \df g^{ab} = \Theta_{ab} \df q^{ab} - 2  \left(k^b \nabla_a \ell_b\right) \df{\ell^a}~.
\end{equation}
We shall also write $\Pi^a_{\phantom{b}b}\nabla_a \ell^b$ in the following form:
\begin{equation} \label{P-tk}
\Pi^a_{\phantom{b}b}\nabla_a \ell^b = \left(q^{a}_b -\ell^a k_b \right)\nabla_a \ell^b = \Theta + \kappa,
\end{equation}
making use of the definitions of $\Pi^a_{\phantom{b}b}$ and $\kappa$ and introducing the expansion scalar $\Theta$ (see \ref{app:SFF+exp}).

We have thus derived all the results that we need. Substituting \ref{gdfg}, \ref{dldg} and \ref{P-tk} in \ref{null-ST-intermediate}, we get
\begin{align}
\sqg Q[\ell_{a}] =&\partial_{a}\left( \sqg \Pi^{a}_{\phantom{a}b}\df \ul^{b}\right)-2 \delta\left[\sqg(\Theta + \kappa)\right]+\sqg \left[\Theta _{ab}-(\Theta + \kappa) q_{ab} \right]\delta q^{ab}  \nn \\
& +2 \sqg \left[ \left(\Theta+ \kappa\right) k_c+ \ell^{b}\nabla _{c}k_{b} \right]\df \ell^{c} \label{q_with_A_1}
\end{align}
Thus, the surface term in the action for a null surface takes the following form:
\begin{align} \label{NewLabel03}
\delta \mathcal{A}_{null}=& \intdx~ \sqg Q[\ell_{a}] \nn\\=&\intdx~ \left\{\sqg\nabla_{a}\left( \Pi^{a}_{\phantom{a}b}\df \ul^{b}\right)-2 \delta\left[\sqg(\Theta + \kappa)\right]+\sqg \left[\Theta _{ab}-(\Theta + \kappa) q_{ab} \right]\delta q^{ab} \right. \nn \\
&\left. +2 \sqg \left[ \left(\Theta+ \kappa\right) k_c+ \ell^{b}\nabla _{c}k_{b} \right]\df \ell^{c}\right\}~.
\end{align}
In the first line, we have a structure like we have in the non-null case in \ref{S.T-nonnull-1} (in \ref{BC-novel}), reproduced below for convenience:
\begin{align}
\STnn =\int_\mathcal{\partial V} d^3x~ \left[\sqrt{|h|} D_{a}(\delta \un^a)+ \delta ( 2\sqrt{|h|}  K) - \sqrt{|h|} (K_{ab} - K h_{ab}) ~\delta h^{ab}\right]~. \label{S.T-nonnull-2}
\end{align}
The first term in \ref{NewLabel03} is the total $3$-derivative term (since $\Pi^{\phi}_{\phantom{\phi}b}=\Pi^{a}_{\phantom{a}b}\ell_a=0$), the second term can be cancelled by a counter-term and the third term is the term that may be killed by fixing $q^{ab}$ on the surface. We have a crucial extra term with $\df \ell^{c}$ which we shall deal with when we discuss degrees of freedom. 

The result in \ref{NewLabel03} is invariant under coordinate transformations on the three surface. That is, it is invariant under coordinate transformation from $\left(\phi,x^1,x^2,x^3\right)$ to $\left(\phi,x'^1,x'^2,x'^3\right)$. The integration volume $\sqg d^3x$ is invariant once $\phi$ is fixed since $\sqg d^3x d \phi$ is invariant under general coordinate transformations, while the integrand is a scalar once $\ell_a=\nabla_a \phi$ and a particular $k^a$ is chosen.

The result in \ref{NewLabel03}, with its $\sqrt{-g}$ and $d^3x$, may seem unusual for readers with some familiarity to the literature on null surfaces. To remedy this, we shall write this result in a specific coordinate system $\left(\phi,\lambda,z^1,z^2\right)$ adapted to the null surface. This will bring the integration element to the form given, e.g., in \cite{Poisson} (also see \ref{null_coord}). In the process, we shall also choose a form for $k_a$. The double null coordinate system commonly found in the literature (see, e.g., \cite{vickers2011double}) will be a special case. This coordinate system, as well as the meaning of our constraints on variations in this coordinate system, is described in the next section. 

\subsection{Writing the Results in A Special Coordinate System}\label{spec_coord}
In this section, we shall detail the construction of a coordinate system adapted to a null surface and also choose a form for $k_a$. This coordinate system has the double null coordinate system \cite{vickers2011double} as a special case. In fact, the form of the metric on the null surface will be identical to that in double null coordinates. Please note that we require the form of the metric only on the fiducial null surface and that is what we shall write down. The expression for the metric in this section is valid only at $\phi=\phi_0$ and should not be taken to represent the form of the metric in a region around the null surface.

The object $q_{ab}$ has been constructed to represent the induced metric on a $2$-dimensional space orthogonal to $\ell_a$ and $k_a$. Let us choose $k^a$ such that $k_a\eqH \nabla_a \lambda=\partial_a \lambda$ for some $\lambda$ at the boundary null surface $\phi=\phi_0$. This is always possible. (The GNC coordinates, whose construction is given in \ref{GNC}, is an example. In these coordinates, $\ell_a=\partial_a r$ and $k_a=-\partial_a u$.) Like $\phi$, $\lambda$ is not varied when we vary the metric and hence $\df k_a= \partial_a (\df \lambda)=0$. This is an additional restriction which was not present in the derivation leading up to \ref{NewLabel03}. Note that the relation $k_a= \partial_a \lambda$ is assumed to hold only at $\phi=\phi_0$. If it is assumed to hold everywhere, then we are specializing to a double null foliation. But that is not required for our purpose. (Again, the GNC metric is a good example. $k_{a}=(2r \alpha,-1,r \beta^A)$ in the coordinate system $(u,r,x^A)$ and is equal to $-\partial_a u$ on the null surface $r=0$.) In this set-up, $q_{ab}$ represents the induced metric on the $2$-surfaces at the intersection of our fiducial $\phi=\phi_0$ null surface and the many $\lambda=$constant surfaces (whose normal coincides with $k_a$ at $\phi=\phi_0$). If we choose coordinates $(z_1,z_2)$ on a particular $2$-surface and carry these coordinates along integral curves of $\ell^a$ on the null surface, then the inverse metric \textit{on the null surface} in the coordinate system $\left(\phi,\lambda,z^1,z^2\right)$, using $\ell_a\ell^a=0$, $k_ak^a=0$, $\ell_ak^a=-1$, valid on the null surface, and the fact that both $k^a$ and $\ell^a$ are orthogonal to the basis vectors on the $\phi=\phi_0$, $\lambda=$constant $2$-surfaces (which means that the metric is block-diagonal), can be written down as   
\begin{equation}
g^{ab}=
\left(\begin{array}{llll}
0          & -1     & 0 & 0\\
-1          &0 & 0 & 0\\
0 &0 & q^{11} & q^{12}\\
0 &0 & q^{12} & q^{22}
\end{array}\right)~,
\end{equation} 
giving the metric, \textit{on the null surface $\phi=\phi_0$,} in the form
\begin{equation}
g_{ab}=
\left(\begin{array}{llll}
0          & -1     & 0 & 0\\
-1         &0 & 0 & 0\\
0 &0 & q_{11} & q_{12}\\
0 &0 & q_{12} & q_{22}
\end{array}\right)~,
\end{equation}
where $q_{11}$, $q_{12}$, etc. have the appropriate expressions in terms of $q^{11}$, $q^{12}$, etc. For this metric, the determinant satisfies $-g=q$, where $q$ is the determinant of the $2$-metric $q_{AB}$. The relation between the determinant of a $2$-metric to the determinant of the $4$-metric in a general case has been derived in \ref{g_to_2} in \ref{app:g_to_2} as
\begin{equation} \label{g_to_2-2}
g=\frac{q}{g^{\phi \phi}g^{\lambda \lambda} -(g^{\lambda \phi})^2}~,
\end{equation}
where we have inserted $\lambda$ instead of the $1$ that stood for the coordinate $x^1$ in the appendix. Then, since we are keeping $g^{ab}\ell_a\ell_b=g^{\phi \phi}=0$, $g^{ab}k_ak_b=g^{\lambda \lambda}=0$ and $g^{ab}\ell_a k_b=g^{\phi \lambda}=-1$ fixed even under variation, \ref{g_to_2-2} gives 
\begin{equation}
\sqrt{-g}= \sqrt{q},~~~ \delta \left(\sqrt{-g}\right)= \delta \left(\sqrt{q}\right).
\end{equation} 
Note that our constraints fix $g^{\phi \phi}$, $g^{\lambda \lambda}$ and $g^{\phi \lambda}$ while the other components are free to vary. In particular, these variations may disturb the relations $\ell^a=-\partial x^a/\partial \lambda$ and $k^a=-\partial x^a/\partial \phi$ valid for the on-shell metric.

In these coordinates, we can write the boundary term in \ref{NewLabel03} as 
\begin{align} \label{NewLabel04-2}
\delta \mathcal{A}_{null}=& \intlt~ \sqrt{q}~ Q[\ell_{a}] \nn\\=& \intlt~ \left\{\sqrt{q} \nabla_{a}\left( \Pi^{a}_{\phantom{a}b}\df \ul^{b}\right)-2 \delta\left[\sqrt{q}(\Theta + \kappa)\right]+\sqrt{q} \left[\Theta _{ab}-(\Theta + \kappa) q_{ab} \right]\delta q^{ab} \right.\nn \\
&\left.+2 \sqrt{q} ~\left[ \left(\Theta+ \kappa\right) k_c+ \ell^{b}\nabla _{c}k_{b} \right]\df \ell^{c}\right\}~.
\end{align}
Here, $d^2 z_{\perp}=dz^1 dz^2$. 
\subsection{Analyzing The Results}
The final decomposition of the boundary term in the variation of the Einstein-Hilbert action is given in \ref{NewLabel03} and it is written in specific coordinates and a particular choice of $k_a$ in \ref{NewLabel04-2}. We see that \textit{the action principle has again suggested what can be done!} On a null surface, we can add the integral of $2 \sqrt{-g}(\Theta + \kappa)$ ($2 \sqrt{q}[\Theta + \kappa]$ with the coordinates and choice of $k_a$ of \ref{spec_coord}) as the counter-term. Further, we need to set $\delta q^{ab}=0$ and $\delta \ell^a=0$ on the boundary, which requires fixing the set $(q^{ab},\ell^a)$ (six independent components, since $\ell^2=0$) on the boundary. This is analogous to fixing $h^{ab}$ (six independent components) on the non-null surface. To know for certain whether this furnishes a well-posed variational problem, we need to analyse the structure of the equations of motion on the null surface. As we have mentioned in the introduction, there is no standard way of writing the Einstein equations on null surfaces. Hence, we postpone this analysis for a future work. We shall briefly discuss about the degrees of freedom in \ref{dof}.

The expression \ref{NewLabel04-2} also tells us that the momentum conjugate to $q^{ab}$ is $\sqrt{q} \left[\Theta _{ab}-(\Theta + \kappa) q_{ab} \right]$ and the momentum conjugate to $\ell^{a}$ is $2 \sqrt{q} \left[ \left(\Theta+ \kappa\right) k_c+ \ell^{b}\nabla _{c}k_{b} \right]$. We can write the momentum conjugate to $\ell^{a}$ in another form as
\begin{align}
P_{c}&=-2 \sqrt{q}  k_{b}\left[\nabla _{c}\ell^{b}- \df^b_c \left(\Theta+ \kappa \right) \right]~.
\end{align}
In the cases where $\kappa=0$, as would happen if $\ell^2=0$ everywhere, the counter-term would be the integral of just $2 \sqrt{q} \Theta$. In this case, we can make use of $\Theta = (1/\sqrt{q})(d\sqrt{q}/d\lambda)$ (see \ref{theta-dqdl} in \ref{app:SFF+exp}) to write the counter-term as
\begin{equation}\label{CT-diffA}
2 \intlt~\sqrt{q}\Theta  =2\intlt~ \frac{\partial\sqrt{q}}{\partial \lambda}=2 \int_\mathcal{\partial \partial V} d^2z_{\perp}\sqrt{q}\vbar^{\lambda_2}_{\lambda_1}=2 \left[S(\lambda_2)-S(\lambda_1)\right]~,
\end{equation}
where ${\lambda_1}$ and ${\lambda_2}$ are the limits of integration of $\lambda$. Thus, the counter-term in this case, apart from an overall factor,  is the difference in the area of the $2$-surface orthogonal to $\boldsymbol{\ell}$ and $\boldsymbol{k}$ on the null surface, denoted by $S(\lambda)$, at the limits of $\lambda$ integration. A similar interpretation can be given to the counter-term for the \textit{non-null} case, the integral of $-2 \sqh K$ over the boundary surface, if we assume that the Lie bracket of the normal vector $n^a$ with any of the coordinate basis vectors on the surface is zero (see \ref{nonnull_CT_interp}). This, in turn, requires $n^a$ to be a tangent vector to a \textit{geodesic} congruence, which may not be true in general. With this assumption, $-\sqh K$ is the rate of change along $n^a$ of the volume of a $3$-surface element orthogonal to $n^a$. 
\newline
We have provided a decomposition of the boundary term on the null surface working purely on the null surface. This decomposition can also be obtained by treating the null surface as a limit of a sequence of non-null surfaces. In the next section, we shall demonstrate that a particular method of taking the null limit of \ref{S.T-nonnull-3} gives back the decomposition in \ref{NewLabel03}.


\subsection{Null surface as a limit of sequence of non-null surfaces} \label{ADMtoNull}

The previous sections provided a decomposition of the boundary term on the null surface working purely on the null surface. In this section, we shall show that this decomposition can also be obtained by treating the null surface as a limit of a sequence of non-null surfaces.

Let the relevant null surface  be part of a family of surfaces, not all of which are necessarily null, characterized by the constant values of a scalar function $\phi (x)$. The null surface is specified by   $\phi(x)=\phi_{0}$. (That is,  the surface $\phi= \phi_{0}$ is null but the other $\phi=$constant surfaces may be spacelike, timelike or even null.) We shall label the null surface as $\mathcal{N}$ and also introduce the scalar function $r=\phi-\phi_0$. Thus, we can represent the null surface by the symbol $\mathcal{N}$ or by one of the two equations, $\phi=\phi_0$ or $r=0$, as per convenience. The normal to the surfaces has the form $\ell_{a}=A\nabla _{a}\phi$, for some scalar function $A(x)$. If the surfaces were spacelike or timelike, we could have fixed $A(x)$ by demanding $\ell^{a}\ell_{a}=\pm 1$. But now, we have $\ell_{a}\ell^{a}= A^2 g^{\phi \phi}$ with $g^{\phi \phi}=0$ at $\phi=\phi_0$. Thus, imposing this constraint would make $A$ diverge at $\phi=\phi_{0}$. Hence, we cannot impose any such normalization condition without introducing infinities. The expression for the normalized normal for a non-null surface is given in \ref{eq:n} in \ref{BC-novel}: $n_{b}= N \nabla _{b}\phi$. On the other hand, we have used the expression $\ell_{a}= \partial_a \phi$ in obtaining \ref{NewLabel03}. So, we have the identification $n_{a}= N \ell_{a}.$ Now, $N$ is defined in terms of $g^{\phi \phi}$ in \ref{eq:gphiphi} in \ref{BC-novel}: $g^{\phi \phi} = \epsilon /N^2~$. So, the null limit can be imposed by demanding that $N\rightarrow \infty$. Near the null surface, we shall take $g^{\phi \phi}$ to have the behaviour $g^{\phi \phi} = B r^n +O (r^{n+1})$ for a finite non-zero $B$ and some positive integer $n$. 

We shall now examine  the decomposition of the surface term of the Einstein-Hilbert action for a non-null surface, provided in \ref{S.T-nonnull-1} in \ref{BC-novel}, and obtain the null surface limit of each of the terms. For convenience, we reproduce \ref{S.T-nonnull-1} below:
\begin{align}
\STnn =\int_\mathcal{\partial V} d^3x~ \left[\sqrt{|h|} D_{a}(\delta \un^a)+ \delta ( 2\sqrt{|h|}  K) - \sqrt{|h|} (K_{ab} - K h_{ab}) ~\delta h^{ab}\right]~. \label{S.T-nonnull-3}
\end{align}
In obtaining the limit, we shall assume that the metric components are finite and that the metric determinant is finite (and non-zero) when we take the null limit. (Later on, we will demonstrate these conditions and the final result in specific coordinate systems.)

Let us start with the surface boundary term in \ref{S.T-nonnull-3}. This term has been analyzed in \ref{APPdervST2} to show that it can be written in the form:
\begin{align}
\sqrt{|h|}D_{a}\left(\delta \un^{a}\right)&=\partial_{\alpha}\left[\sqrt{-g}\delta \ell^{\alpha}\right]+2\partial _{\alpha}\left[\sqg \ell ^{\alpha}\delta \ln N \right]~. \label{ST21}
\end{align}
The first term in \ref{ST2} is finite in the $N\rightarrow \infty$ limit. In the second term in \ref{ST2},  we use the expansion for $g^{\phi \phi}$ near the null surface and see that the $r$ factor cancels (as we have taken the variation to not affect the coordinates) and the term is finite in the null limit of $N\rightarrow \infty$ or $r\rightarrow 0$. Thus, we have decomposed $\sqrt{|h|}D_{a}\left(\delta \un^{a}\right)$ in such a way that each term in the decomposition is clearly finite in the null limit.

On the other hand, the surface boundary term for a null surface in \ref{NewLabel03} has been shown in \ref{APPdervST2} to reduce to
\begin{equation}\label{ST22}
\partial_{a}\left[\sqrt{-g}\Pi ^{a}_{\phantom{a}b}\delta \ul^{b}\right]= \partial_{\alpha}\left[\sqrt{-g}\delta \ell^{\alpha}\right]
\end{equation}
Thus, from \ref{ST21} and \ref{ST22}, we obtain the result that the null surface limit of the surface boundary term in \ref{S.T-nonnull-3} is
\begin{equation}\label{ST2}
 \sqrt{|h|}D_{a}\left(\delta \un^{a}\right) \overset{r\rightarrow0}= \partial_{a}\left[\sqrt{-g}\Pi ^{a}_{\phantom{a}b}\delta \ul^{b}\right]+2\partial _{\alpha}\left[\sqg \ell ^{\alpha}\delta \ln N \right]~.
\end{equation}

Next, let us look at the  counter-term in \ref{S.T-nonnull-3}. From \ref{APPdervK2}, a similar analysis shows that  the null limit of the variation of the counter-term turns out to be 
\begin{align}
\delta \left(2\sqrt{|h|}K\right)
\overset{r\rightarrow0}= -2\delta \left[\sqrt{-g}\left(\Theta +\kappa \right) \right]
-2\delta \left[ \sqg~ \ell^{\alpha} \partial_{\alpha} \ln N \right] ~.\label{K2}
\end{align}
Consider now the last set of terms in \ref{S.T-nonnull-3}. 
Among them, the $h_{ij}\delta h^{ij}$ term is manipulated in \ref{APPdervKhh} leading to
\begin{align}
\sqrt{|h|}Kh_{ij}\delta h^{ij}
\overset{r\rightarrow0}=&-\sqrt{-g}\left(\Theta +\kappa \right)q_{ab}\delta q^{ab} +2\sqrt{-g} \left(\Theta +\kappa \right)k_{b}\delta \ell ^{b}
-2\sqrt{-g}\left(\Theta +\kappa \right){\delta \ln N}
\nonumber
\\
&-\sqg \left(\ell^{\alpha} \partial_{\alpha} \ln N \right)
g_{ij}\delta g^{ij}
-2\sqrt{-g}\left(\ell^{\alpha}\partial _{\alpha}\ln N\right) \delta \ln N \label{Khh}
\end{align}
Again, one can see that each term is finite in the null limit under our assumptions.

The last term to be considered in \ref{S.T-nonnull-3} is the $K_{ij}\delta h^{ij}$ term. The pertinent expression is derived in \ref{APPdervKij2} as 
\begin{eqnarray}
-\sqrt{|h|}K_{ij}\delta h^{ij}\overset{r\rightarrow0}=\sqrt{-g}\Big[\Theta _{ij}\delta q^{ij}+2\delta \ell ^{i}\ell ^{j}\nabla _{i}k_{j}
+2 \left(\partial _{i}\ln N\right) \df \ell^{i} + 2 \ell^{i} \left(\partial_{i} \ln N \right)\df \ln N \Big]~. \label{Kij2}
\end{eqnarray}
Adding \ref{ST2}, \ref{K2}, \ref{Khh} and \ref{Kij2}, we obtain the 
following result when we take the null limit of the integrand on the right hand side of \ref{S.T-nonnull-3}:
\begin{align}
 &\sqrt{|h|} D_{a}(\delta \un^a)+ \delta ( 2\sqrt{|h|}  K) - \sqrt{|h|} (K_{ab} - K h_{ab}) ~\delta h^{ab}\nn\\ \overset{r\rightarrow0}=~~&\partial_{\alpha}\left[\sqrt{-g}\Pi ^{\alpha}_{\phantom{a}b}\delta \ul^{b}\right]-2\delta \left[\sqrt{-g}\left(\Theta +\kappa \right) \right] +\sqrt{-g}\left[\Theta _{ab}-\left(\Theta +\kappa \right)q_{ab}\right]\delta q^{ab} \nn \\
 &+2 \sqrt{-g}\delta \ell ^{i} \left[\left(\Theta+\kappa \right)k_{i}+  \ell ^{j}\nabla _{i}k_{j}\right] \nn \\
 &+2\partial _{\alpha}\left[\sqg \ell ^{\alpha}\delta \ln N \right]-2\delta \left[ \sqg~ \ell^{\alpha} \partial_{\alpha} \ln N \right] -2\sqrt{-g}\left(\Theta +\kappa \right){\delta \ln N}
\nonumber
\\
&-\sqg \left( \ell^{\alpha} \partial_{\alpha} \ln N \right)
g_{ij}\delta g^{ij}
-2\sqrt{-g}\left(\ell^{\alpha}\partial _{\alpha}\ln N \right)\delta \ln N  \nn \\
&+\sqrt{-g}\Big[2 \left(\partial _{i}\ln N\right) \df \ell^{i} + 2 \ell^{i} \left(\partial_{i} \ln N \right)\df \ln N \Big] \label{result1}
\end{align}
We find that the first two lines reproduce the result in \ref{NewLabel03}. Thus, we should be able to show that the rest of the terms cancel out.
This is proved in \ref{APPdervADM-2-Null} and we obtain 
the following expression for the surface term:
\begin{align}
\STnn \overset{r\rightarrow0}=&~~ \STn \nn \\ 
\overset{\phantom{r\rightarrow0}}=& ~~\intdx~ \left\{\sqrt{-g}\nabla_{a}\left[\Pi ^{a}_{\phantom{a}b}\delta \ul^{b}\right]-2\delta \left[\sqrt{-g}\left(\Theta +\kappa \right) \right] +\sqrt{-g}\left[\Theta _{ab}-\left(\Theta +\kappa \right)q_{ab}\right]\delta q^{ab} \right.\nn \\
 &\left.+2 \sqrt{-g}\delta \ell ^{i} \left[\left(\Theta+\kappa \right)k_{i}+  \ell ^{j}\nabla _{i}k_{j}\right]\right\}
  \label{ADM-2-Null}~. 
\end{align}
where we have replaced the surface index $\alpha$ in the first term in \ref{result1} with the four-dimensional index $a$ as $\Pi ^{\phi}_{\phantom{a}b}=0$. Thus, we have re-derived the result in \ref{NewLabel03}, with proper canonical momenta conjugate to $\ell ^{a}$,
reaffirming our faith in the correctness of this result.

In \ref{app:BT-GNC+NSF}, we have illustrated the result in \ref{NewLabel03} or \ref{ADM-2-Null} for two specific parametrizations of a general null surface.


\section{Discussion of the Number of Degrees of Freedom}\label{dof}

We shall now try to enumerate and identify the  metric degrees of freedom that we have to fix on the null boundary in our approach. The number of degrees of freedom that we have to fix on the boundary modulo gauge invariance (i.e, after making use of gauge invariance to eliminate certain degrees of freedom) corresponds to the number of physical degrees of freedom in the theory for a well-posed action (see \ref{BT-info} and \cite{Dyer:2008hb}). The addition of a boundary term to the action can change the number of degrees of freedom to be fixed on the surface. But, for a well-posed action, the number of degrees of freedom to be fixed on the boundary should match with the number of degrees of freedom obtained from the analysis of the initial value problem of the equations of motion. For an initial value problem, the number of degrees of freedom is half of the number of initial data to be specified. For a variational problem, the number of degrees of freedom is the number of boundary data to be fixed at the boundary. It is well-known from the analysis of the initial value problem of general relativity, with an initial spacelike surface, that the theory has two physical degrees of freedom per spacetime point (see, e.g., \cite{MTW,Wald}). The same conclusion can be reached by analysing the variational problem for a non-null boundary. In the next paragraph, we shall demonstrate this for a spacelike boundary. The analysis for a timelike boundary is similar.

Consider the case when the boundary is a spacelike surface in a $3+1$ framework. Initially, the metric has ten degrees of freedom but we have the freedom of making four coordinate choices. We shall effect a $3+1$ split by slicing the spacetime into $t=\textrm{constant}$ surfaces with the fiducial boundary being a $t=\textrm{constant}$ surface. The $t$-coordinate of each spacetime point is fixed. Next, we introduce the normal to the $t$-constant surfaces. There are four components to the normal $n_{a}$. But since it is the normal to $t=\textrm{constant}$ surfaces, $n_{a}=-N \partial_a t$ and there is only one functional degree of freedom in $n_a$ in the choice of the function $N$. Imposing the normalization condition $n_a n^a=-1$ fixes $N$ in terms of the metric by the relation $g^{tt}=-1/N^2$. This relation is preserved on variation since we take the varied normal, $n_{a}+\df n_{a}$, also to be normalized. The variation of the normal is $\df n_a =- (\df N) \partial_a t$, since coordinates are kept constant during the variation, and essentially contains the variation $\df g^{tt}$. On the other hand, the upper components are $n^a= -N g^{ab} \partial_a t = -Ng^{at}= \left(1/N , -Ng^{\alpha t}\right)$, where we have substituted for $g^{tt}$ in terms of $N$. From this relation, we observe that the time evolution vector $t^a= (1,0,0,0)$ can be written as $t^a= N n^a + N^{\alpha}$, for the purely spatial shift vector $N^{\alpha}$ with components $N^2 g^{\alpha t}$ (for more details, see Chapter 12 in \cite{gravitation}). Thus, $n^{a}=\left(1/N, -N^{\alpha}/N\right)$ and $\df n^a$ basically contains $\df g^{tt}$ and the three $\df g^{t \alpha}$. Introducing the notation $h_{\alpha \beta}$ for the spatial components of the metric $g_{\alpha \beta}$, the degrees of freedom in the 3+1 decomposition can then be said to be in $(N, N^{\alpha},h_{\alpha \beta})$ or equivalently in $(n^a, h_{\alpha \beta})$. That is, the ten degrees of freedom in the metric have been inherited by the normal $n^a$ (four degrees of freedom) and the induced metric $h_{\alpha \beta}$ (six degrees of freedom). Out of these ten, variations of only six, the degrees of freedom in $ h_{\alpha \beta}=g_{\alpha \beta}$, appear in the boundary term in the variation of the Einstein-Hilbert action, after we remove the counter-term part and the surface term on the $3-$surface, and hence only these need to be fixed on the $t=\textrm{constant}$ boundary surface. But we have three more coordinate choices to make. The choice of three coordinates on the initial $t=\textrm{constant}$ surface can be used to remove three of the six degrees of freedom in $h_{\alpha \beta}$ on the surface. This leaves 3 degrees of freedom. It is generally argued that one of these three degrees of freedom just gives the location of the $3$-surface in the $4$-dimensional spacetime, as determined by the slicing of the spacetime into $t=$constant surfaces, and can be tuned by appropriate choice of the $t$ coordinate \cite{MTW,Wald}. Thus, only two degrees out of the ten degrees of freedom in the metric are real physical degrees of freedom and only these need to be fixed on the boundary if we use our freedom of choice of coordinates appropriately.

Coming to the case of null surfaces, the initial value problem on a null surface has been analysed by several authors \cite{sachs1962characteristic,newman1962approach,geroch1973space,Penrose:1980yx,d'Inverno:1980zz,friedrich1981regular,Torre:1985rw,hayward1993dual,Goldberg:1995gb,Brady:1995na}. But unlike the case for the initial value problem on a spacelike surface, there appears to be no one standard formulation for null surfaces. The constraint structure and the initial data to be specified varies with the formalism. Thus, the analysis of the question of whether the action is well-posed with the counter-term that we have proposed is not as straightforward as in the case of non-null surfaces. We intend to pursue this question in a future publication.

For the time being, we can ask which degrees of freedom in the metric have to be fixed to set our null boundary term in \ref{NewLabel03} to zero. But even this question cannot be satisfactorily answered in our framework. The degrees of freedom to be fixed on the boundary appear to be in the three components of $q^{ab}$ (coefficient of $\df q^{ab}$ has only three components since it is symmetric and orthogonal to $\ell^a$ and $k^a$) and the three components of $\ell^a$ ($\df \ell^{\phi}=\ell_a \df \ell^{a}=\df \left(\ell_a\ell^{a}\right)=0$ as per our assumptions). This gives six components and one might think that four components might be eliminated by diffeomorphisms to match with the two degrees of freedom that one obtains in a non-null case. But notice that we have imposed the constraints $\ell^a \ell_a=0$, $k^a k_a=0$ and $\ell_a k^a=-1$ throughout, even under variations. The effect of the last two constraints on the variations of the metric can be specified only once $k^a$ is specified. But, with $\ell_a = \partial_a \phi$, $\df\left(\ell_a\ell^{a}\right)=0$ imposes the constraint $\df g^{\phi \phi}=0$. In \ref{spec_coord}, we take $k_a=\partial_a \lambda$ on the null boundary to obtain the other constraints as $\df g^{\phi \lambda}=0$ and $\df g^{\lambda \lambda}=0$. Hence, in our framework, variations of certain degrees of freedom of the metric are constrained. To do a rigorous analysis, either the framework has to be extended to eliminate these constraints or these constraints have to be explicitly taken into account (by adding Lagrange multipliers, for example).

\section{Conclusions} \label{conclusions}

Our aim in this work was to find out what boundary term can be added to the Einstein-Hilbert action for a null boundary to make the action well-posed. We wanted to do this from first principles, adhering to the philosophy: \textit{The action can itself suggest the counter-term to be added to make it well-defined and also what boundary conditions go along with this counter-term.} In the case of the Einstein-Hilbert action with a null boundary, we expect the action itself to suggest what counter-term needs to be added to make the action well-defined.

We first undertook a general analysis of the surface term for a null surface $\phi=\phi_{0}$, or $r=\phi-\phi_0=0$, with $\ell_{a}=\partial_{a} \phi$ and $k^{a}$ being the auxiliary null vector satisfying $k^a \ell_a=-1$. We took $\phi$ as one of the coordinates and considered variations that vary the metric but do not affect the coordinates. (This assumption also means $\df \ell_a =0$.) The variations were taken to respect the following constraints: \begin{inparaenum}[i)] \item $\boldsymbol{\ell}.\boldsymbol{\ell} \eqrtoz 0$, \item $\boldsymbol{k}.\boldsymbol{k}=0$ and \item $\boldsymbol{\ell}.\boldsymbol{k}=-1~,$\end{inparaenum}\phantom{a}with the first constraint being imposed only on the null surface. In this case, we found that the surface term on the null surface can be expressed in the following form (see \ref{NewLabel03}):
\begin{align} \label{NewLabel031}
\delta \mathcal{A}_{null}=&\intdx~ \left\{\sqg\nabla_{a}\left( \Pi^{a}_{\phantom{a}b}\df \ul^{b}\right)-2 \delta\left[\sqg(\Theta + \kappa)\right]+\sqg \left[\Theta _{ab}-(\Theta + \kappa) q_{ab} \right]\delta q^{ab} +P_c\df \ell^{c}\right\}~.
\end{align}
where, taking the normal $\ell_a=\partial_a \phi$ for the null boundary being a $\phi=\textrm{constant}$ surface and $k^a$ to be the auxiliary null vector, we have $\Pi^a_{\phantom{a}b}=\df^a_b +k^a \ell_b$, $\df \ul^a=\df \ell^a +g^{ab}\df l_b$ and $q_{ab}=g_{ab}+\ell_a k_b+k_a \ell_b$, the induced metric on the null surface. $\Theta_{ab}=q^m _a q^n _b \nabla_m \ell_n$ is the second fundamental form, $\Theta=\Theta^{a}_a$ is the expansion scalar, $\kappa$ is the non-affinity coefficient on the null surface and $P_c=2 \sqg \left[ \left(\Theta+ \kappa\right) k_c+ \ell^{b}\nabla _{c}k_{b} \right]$. (The definitions of $\Theta_{ab}$, $\Theta$ and $\kappa$ can be found in \ref{app:SFF+exp} and \ref{inaffinity}.) In a coordinate system adapted to the null surface and with a specific choice of $k_a$, we can write the above decomposition as (see \ref{NewLabel04-2})
\begin{align} 
\delta \mathcal{A}_{null}=& \intlt~ \left\{ \sqrt{q}\nabla_{a}\left( \Pi^{a}_{\phantom{a}b}\df \ul^{b}\right)-2 \delta\left[\sqrt{q}(\Theta + \kappa)\right]+\sqrt{q} \left[\Theta _{ab}-(\Theta + \kappa) q_{ab} \right]\delta q^{ab} +~P_{c}\df \ell^{c} \right\}~\label{final_result},
\end{align}
The integration is over a parameter $\lambda$, varying along the null geodesics such that $k_a = \nabla_a \lambda$ locally, and two coordinates $\left(z^1,z^2\right)$ that are constant along the null geodesics. $q$ is the determinant of the $2$-metric $q_{AB}$. The conjugate momentum to $q^{ab}$ turns out to be $\sqrt{q} \left[\Theta _{ab}-(\Theta + \kappa) q_{ab} \right]$ while the conjugate momentum to $\ell^a$ is 
$P_c=2 \sqrt{q} \left[ \left(\Theta+ \kappa\right) k_c+ \ell^{b}\nabla _{c}k_{b} \right]$. In \ref{app:gen-normal}, we have repeated the whole analysis for a null normal of the general form $\ell _{a}=A\partial _{a}\phi$ to obtain a decomposition with the same structure as in \ref{NewLabel031} or \ref{final_result}. We have also verified the result in \ref{NewLabel031} using two parametrizations of the metric near an arbitrary null surface, the GNC and NSF metrics, in \ref{app:BT-GNC+NSF}.

We note that the structure is very similar to the structure in the non-null boundary (see \ref{S.T-nonnull-3}). The first term in \ref{NewLabel031} or \ref{final_result} is the total $3$-derivative term (due to the properties of $\Pi^{a}_{\phantom{a}b}$), the second term can be cancelled by a counter-term, $2\sqrt{-g}(\Theta + \kappa)$ or $2\sqrt{q}(\Theta + \kappa)$ respectively, added to the action and the remaining terms can be killed by fixing $q^{ab}$ and $\ell^a$ on the surface.  

The question of whether this prescription leads to an action that is well-posed requires answering the question whether the suggested boundary conditions and the equations of motion are consistent and whether there is a unique solution for any choice of the boundary conditions. This task is complicated by the plethora of frameworks available for dealing with Einstein equations on a null surface. Even if the action turns out to be well-posed, the analysis of degrees of freedom of the theory on the null surface requires careful analysis of the constraints that we have imposed. There are six degrees of freedom in $q^{ab}$ and $\ell^a$, which matches the number of degrees of freedom found in $h_{ab}$ for a non-null boundary. But, unlike the standard analysis for non-null boundary, our framework constrains certain variations of the metric. For example, the constraint that the null surface remains a null surface means $\df g^{\phi \phi}=0$. There is also the matter of whether we can find a better method to normalize the normal on the null surface than the $A=1$ fiat that we imposed in this paper. Detailed investigations in these directions have been kept for a future work.

We summarise in \ref{comp_null-non} the similarities and differences between non-null surfaces and null surfaces with respect to the intrinsic geometry and treatment of the boundary term.

\begin{table}[h]
	\begin{center}
		\caption{\bf Comparison Between Non-null and Null Surfaces. \label{comp_null-non}}
		\centering
		
		\begin{tabular}{|c|c|c|}
			
			\hline
			\hline
			{\bf Properties} & {\bf Non-null} & {\bf Null} \\[0.7ex]
			
			\hline
			\hline
			
			Normal
			&
			$n_{a}$; $n_{a}n^{a}=\epsilon=\pm 1$
			&
			$\ell _{a}$; $\ell _{a}\ell ^{a}=0$\\
			
			``Dimension"
			&
			3
			&
			2\\
			
			Induced Metric
			&
			$h_{\alpha \beta}=g_{ab}e^{a}_{\alpha}e^{b}_{\beta}$
			&
			$q_{AB}=g_{ab}e^{a}_{A}e^{b}_{B}$\\

			Auxiliary Vector
			&
			$- \epsilon n^{a}$
			&
			$k^{a}$; $\ell _{a}k^{a}=-1$\\
			
			Integration Measure
			&
			$d^{3}x\sqrt{|h|}$
			&
			$d^{2}xd\lambda \sqrt{q}$\\			
			
			Second Fundamental Form
			&
			$K_{ab}=-h^{m}_{a}\nabla _{m}n_{b}$
			&
			$\Theta _{ab}=q^{m}_{a}q^{n}_{b}\nabla _{m}n_{n}$\\
			
			Counter-term
			&
			$-\int d^3x\sqrt{|h|} ~2K$
			&
			$\int d\lambda d^2z_{\perp}\sqrt{q}~ 2\left(\Theta +\kappa \right)$\\

			Boundary Term
			&
			$\sqh D_{a}\delta \un^{a}$
			&
			$\sqrt{q}\nabla _{a}\left(\Pi ^{a}_{\phantom{a}b}\delta \ul^{b}\right)$\\
			
			Degrees of Freedom
			&
			$h_{\alpha \beta}$
			&
			$q_{AB}$, $\ell ^{c}$\\			
			
			Conjugate Momentum to Induced Metric
			&
			$-\sqh \left[K^{ab}-h^{ab}K\right]$
			&
			$\sqrt{q} \left[\Theta ^{ab}-q^{ab}\left(\Theta+\kappa\right)\right]$\\
			
			Conjugate Momentum to Normal
			&
			0
			&
			$-\sqrt{q}~  k_{b}\left[\nabla _{c}\ell^{b}+\nabla ^{b}\ell_{c}- 2\df^b_c \left(\nabla_a \ell^a \right) \right]$\\

			\hline
			\hline
			
		\end{tabular}
	\end{center}
\end{table}
The defining characteristic that distinguished a non-null surface and a null surface is the surface gradient, whose norm is non-zero for a non-null surface and zero for a null surface. The norm of the surface gradient for a non-null surface can be normalized to $\pm 1$ by multiplying with a suitable scalar. The induced metric on a non-null surface is a $3$-metric $h_{\alpha \beta}$ while the induced metric on a null surface is a $2$-metric $q_{ab}$. Hence, the ``dimension" of the null surface has been listed as 2. For a null surface, we define an auxiliary vector $k^a$ to be such that $\ell_a k^a =-1$ and $k^a k_a=0$. For a non-null surface, the term auxiliary vector is not generally used but we have put $-\epsilon n^a$ in the respective place in the table as this vector satisfies $\left(-\epsilon n^a\right) n_a =-1$. The rest of the entries in the table are quantities which we have already discussed in the analysis of our results on the decomposition of the boundary term on a null surface.


\section*{Acknowledgements}

The research of TP is partially supported by J.C.Bose research grant of DST, India. 
KP and SC are supported by the Shyama Prasad Mukherjee Fellowship from the Council of Scientific and 
Industrial Research (CSIR), India. KP would like to thank Kinjalk Lochan for discussions.

\appendix

\labelformat{section}{Appendix #1}
\labelformat{subsection}{Appendix #1}
\labelformat{subsubsection}{Appendix #1}

\section*{Appendices}

\section{Some Requisite Pedagogical and Background Material}

\subsection{Gauss' Theorem} \label{app:Gauss}

In order to explain the conventions regarding the use of the Gauss' theorem, we refer to the proof of the theorem given in Chapter 3 of \cite{Poisson}. The Gauss' theorem is stated in the following form:
\begin{equation}
\int_{\mathcal{V}}d^4 x\,\sqg\,\nabla_{a} A^{a}   = \int_{\partial V}d \Sigma_{a} A^{a}~,
\end{equation}
with $d \Sigma_{a}$ being the directed surface element on the boundary of the integration volume. Note that \cite{Poisson} uses Greek indices to run over all spacetime indices and Latin indices to run over indices on a surface, opposite of the convention we have adopted. Then, the proof is illustrated using a set of $x^{0}=$constant surfaces. The following expression is obtained on integrating over a volume between $x^{0}=0$ and $x^{0}=1$ surfaces:
\begin{equation}
\int_{\mathcal{V}}d^4 x\,\sqg\,\nabla_{a} A^{a} = \oint d^3 x\,\sqg\, A^{0} \vbar^{x^{0}=1}_{x^{0}=0}~,
\end{equation}
This may be rewritten as
\begin{equation}
\int_{\mathcal{V}}d^4 x\,\sqg\,\nabla_{a} A^{a} = \oint d^3 x\,\sqg\, A^{0} \vbar_{x^{0}=1}-\oint d^3 x\,\sqg\, A^{0} \vbar_{x^{0}=0}=\oint d^3 x\,\sqg\, A^{a}\ell_{a} ~,
\end{equation}
with $\ell_{a}=\partial_a x^{0}$ at $x^{0}=1$ and $\ell_{a}=-\partial_a x^{0}$ at $x^{0}=0$. This depends on the fact that the integration is carried out from $x^{0}=0$ to $x^{0}=1$. If the integration was done the other way, the signs would have flipped.

So we may formulate the following rule. Let us assume that $x^{0}$ is going to be integrated from lower value to higher value. In that case, $\partial_a x^{0}$ or $-\partial_a x^{0}$ is to be used at an $x^{0}=$constant surface according to whether $x^{0}$ or $-x^{0}$ is increasing as we move from inside the integration volume to outside through the surface.  

\subsection{Decomposition of $\sqg$ in Terms of the Determinant of Metric on a $2$-surface} \label{app:g_to_2}

One relevant question is whether there is a decomposition of $\sqg$ in terms of $\sqrt{q}$, $q$ being the determinant of the $2$-metric $q_{AB}$ on the null surface, akin to the decomposition
$\sqg= N \sqh$ in the timelike and spacelike case. We shall prove in this appendix a preliminary result on the decomposition of the determinant of a $4\times 4$ metric in terms of the determinant of a $2\times 2$ submatrix, which will be later applied to a null surface. 

We start by writing down a general result relating the determinant of a $2\times2$ submatrix to the determinant of the whole matrix. We shall prove the result working with the metric written in the coordinates $\left(\phi,x^1,x^2,x^3\right)$ with the components on the $\phi=$constant surface being denoted by $h_{\alpha \beta}$ (refer \ref{BC-novel}):
\begin{equation}
g_{ab}=
\left(\begin{array}{llll}
g_{\phi \phi} & g_{\phi 1} & g_{\phi 2} & g_{\phi 3}\\
g_{\phi 1} &h_{11} & h_{12} & h_{13}\\
g_{\phi 2} &h_{21} & h_{22} & h_{23}\\
g_{\phi 3} &h_{31} & h_{32} & h_{33}
\end{array}\right)~.
\end{equation} 	
In this case, we can use the definition of an inverse matrix element applied to $g^{\phi \phi}$ to write
\begin{equation}\label{g_h_gphiphi}
g= \frac{h}{g^{\phi \phi}},
\end{equation}
where $h$ is the determinant of $h_{\alpha \beta}$, the $3 \times 3$-matrix obtained by deleting the $\phi$-column and $\phi$-row from $g_{ab}$. Now, we can play the same game again with $h_{\alpha \beta}$. The determinant of the $2\times2$ matrix $q_{AB}$, $A,B=2,3$, defined by $q_{AB}=h_{AB}$ satisfies an analogue of \ref{g_h_gphiphi}:
\begin{equation}
h= \frac{q}{h^{11}}~,
\end{equation}
where $h^{11}$ is the $11$-th component of the matrix $h^{\alpha \beta}$, the inverse of the matrix $h_{\alpha \beta}$. Substituting for $h$ in \ref{g_h_gphiphi}, we obtain
\begin{equation}
g=\frac{q}{g^{\phi \phi}h^{11}}~.
\end{equation}
Now, the denominator above can be expanded as follows:
\begin{equation}
g^{\phi \phi}h^{11} = g^{\phi \phi}g^{11} -(g^{1 \phi})^2~,
\end{equation}
which is easiest to obtain by using the formula $h^{ab}=g^{ab}-\epsilon n^a n^b$. Thus, we obtain a relation relating the determinant of a $2\times 2$ submatrix with the determinant of the full $4\times4$ matrix:
\begin{equation} \label{g_to_2}
g=\frac{q}{g^{\phi \phi}g^{11} -(g^{1 \phi})^2}~.
\end{equation}

\subsection{Null Surfaces}

In this section, we shall discuss various pedagogical and otherwise useful material on null surfaces that we will be needing in the main text.
\subsubsection{The Non-affinity Coefficient $\boldsymbol{\kappa}$}\label{inaffinity}

In this appendix, we shall prove that $\ell^{a}\nabla_{a}\ell_{b}\propto \ell_{b}$, for $\ell_{a}$ being the null normal, $\ell_{a}=A \partial_a \phi$, to a null surface $\phi = \phi_{0}$. We have
\begin{align}
\ell^{a} \nabla_a \ell_b &= \ell^{a} \nabla_a (A \partial_b \phi) 
= \ell^{a}\frac{\ell_{b}}{A}\partial_a A + \ell^{a} A \nabla_a \nabla_b \phi  \nn\\
&= \ell^{a}\frac{\ell_{b}}{A}\partial_a A + \ell^{a} A \nabla_b \nabla_a \phi  
= \ell^{a}\frac{\ell_{b}}{A}\partial_a A + \ell^{a} A \nabla_b (\frac{\ell_a}{A})  \nn\\
&= \ell^{a}\frac{\ell_{b}}{A}\partial_a A + \ell^{a} \nabla_b \ell_a  
=[\ell^a \partial_a (ln A)] \ell_b + \frac{1}{2} \partial_b (\ell^{a}\ell_{a})
\end{align}
Consider some coordinate system with $\phi$ as one of the coordinates, say $\left(\phi,x_1,x_2,x_3\right)$. In such a coordinate system, 
\begin{equation} \label{l_as_dphi}
\ell_{a}=A\partial_a \phi=\left(A,0,0,0\right)~.
\end{equation}
Now, $\partial_b (\ell^{a}\ell_{a})$ will only have the $\phi-$component at the null surface. This is because $\ell^{a}\ell_{a}=0$ all along the null surface and hence only $\partial_\phi (\ell^{a}\ell_{a})\neq 0$ at the null surface. Thus, we obtain $\partial_b (\ell^{a}\ell_{a})\propto \ell_{b}$ and
\begin{equation}
\ell^{a} \nabla_a \ell_b = \kappa \ell_{b},
\end{equation}
where $\kappa$ is a scalar. It may be termed the non-affinity coefficient as it will be zero for an affine parametrization of the null geodesics \cite{Gourgoulhon:2005ng}. An explicit expression for $\kappa$ is derived in \ref{app:kappatheta}.
\subsubsection{Induced Metric on a Null Surface}\label{app:q_for_null}

In this appendix, we shall discuss how to find the induced metric on a null surface. If we choose an auxiliary vector $k^{a}$ such that $k^{a}\ell_{a}=-1$ on the null surface $\phi=\phi_{0}$ (note that we have not specified $k^a k_a$ yet), then we have the following results on the null surface. First, consider the object 
\begin{equation}\label{Pi}
\Pi^{a}_{\phantom{a}b} = \delta^{a}_{b}+ k^{a}\ell_{b}~.
\end{equation}
It is easy to verify that $\Pi^{a}_{\phantom{a}b}\ell_{a}=0$, so any vector $L^{a}$ can be acted on by $\Pi^{b}_{\phantom{a}a}$ to give $M^{b} = \Pi^{b}_{\phantom{a}a} L^{a}$ such that $M^{a}\ell_{a}=0$. To verify that an operator $P$ is a projector, we need to verify $P^2 =P$, which is satisfied in this case as $\Pi^{c}_{\phantom{a}b}\Pi^{b}_{\phantom{a}a}=\Pi^{c}_{\phantom{a}a}$.

While $\Pi^{a}_{\phantom{a}b}$ may be good as a projector, $h_{ab}$ had the additional status of being the induced metric on the surface. Now, $\Pi_{ab}$ does have the property that if we look at the components on the surface (represented by Greek letters, meant to run over $(x^1,x^2,x^3)$ in a coordinate system $(\phi, x^1, x^2, x^3)$), they do satisfy $\Pi_{ab} = g_{\alpha \beta}$ as $\ell_{\alpha}=0$. But if we want a symmetric object, we may turn to
\begin{equation} \label{qab}
q_{ab} = g_{ab}+\ell_{a}k_{b}+k_{a}\ell_{b} = \Pi_{ab} + \ell_{a} k_{b}~.
\end{equation}
This also has the property $q^{a}_{b} \ell_{a}=0$ and, in addition, $q^{a}_{b} \ell^{b}=0$ at $\phi=\phi_0$ (while $\Pi^{a}_{b} \ell^{b}=\ell^{a}$). To see if this is a projector, consider
\begin{equation}
q^{a}_{b}q^{b}_{c} = q^{a}_{c}+\ell^{a}k_{b}q^{b}_{c}~.
\end{equation}
Thus, in order for $q^{a}_{b}$ to be a projector, we need the additional condition $\ell^{a}k_{b}q^{b}_{c}=0$, or $k_{b}q^{b}_{c}=0$. This means
\begin{equation}
k_{b}q^{b}_{c}=k_{c}-k_{c}+ k_{b}k^{b} \ell_{c}=0~, 
\end{equation}
which requires $k^{a}k_{a}=0$, i.e, we need to choose $k^{a}$ to be a null vector. 

Thus, for a null surface with normal $\ell_{a}$, we choose an auxiliary vector $k^{a}$ such that \begin{inparaenum}[(i)] 
	\item $\ell_{a}k^{a}=-1$ and
	\item $k_{a}k^{a}=0$
\end{inparaenum}. Then, the symmetric object $q_{ab}$ is such that $q^{a}_{b}$ acts as a projector to the space orthogonal to $\ell^{a}$. In fact, it also acts a projector to the space orthogonal to $k^{a}$. Thus, it projects to a $2$-dimensional subspace of the tangent space on the $3$-surface. We may say that it is a projector to the $2$-surface orthogonal to both $\ell^{a}$ and $k^{a}$. The $3$-dimensional space orthogonal to $\ell^a$ at $\phi=\phi_0$ corresponds to the tangent space of the null surface under consideration while a further projection orthogonal to $k^a$ takes us down to a $2$-surface on the null surface. To formalize this statement, let us define a set of 4 basis vectors $(\ell^{a},k^{a},e^{a}_{A})$, with $A=1,2$, at each point on the null surface. $e^{a}_{A}$ are a pair of linearly independent spacelike vectors which satisfy $\ell_{a}e^{a}_{A}=0$ and $k_{a}e^{a}_{A}=0$. We shall call such a basis as \textit{a} canonical null basis, and introduce the notation
\begin{equation}\label{can_null_bas}
\textbf{v}_{(a)}=\left(\boldsymbol{\ell},\boldsymbol{k},\textbf{e}_{A}\right),
\end{equation}
with, e.g, $\textbf{v}_{(3)}=\textbf{e}_{1}$. Now, we have
\begin{equation}
q^{a}_{b}\ell^{b}=0; ~ q^{a}_{b}k^{b}=0;~ q^{a}_{b}e^{b}_{A}= e^{a}_{A}~. 
\end{equation}
Thus, $q^{a}_{b}$ projects to the subspace spanned by $e^{a}_{A}$. The metric induced on the surface formed by the $e^{a}_{A}$ is  
\begin{equation} \label{q_AB}
g_{ab}e^{a}_{A}e^{b}_{B}=\left( q_{ab} - \ell_{a} k_{b}-\ell_{b}k_{a} \right)e^{a}_{A}e^{b}_{B} =q_{ab}e^{a}_{A}e^{b}_{B} \equiv q_{AB}~.
\end{equation}
Now, $q_{AB}$ contains the whole information about $q_{ab}$ since $q_{ab}\ell^{a}$ and $q_{ab}k^{b}$ are already constrained to be zero.
Thus, $q_{AB}$, and hence $q_{ab}$, represents the induced metric on the two-dimensional surface spanned by $e_{A}$ and orthogonal to $\boldsymbol{\ell}$ and $\textbf{k}$.

To get the induced metric on the $3$-surface, note that the vector space on the $3$-surface is spanned by $(\boldsymbol{\ell},\textbf{e}_{A})$ as $\ell^{a}\ell_{a}=0$, $e^{a}_{A}\ell_{a}=0$ but $k^{a}\ell_{a}=-1$. Thus, the induced metric on the $3$-surface consists of the components $g_{ab}e^{a}_{A}e^{b}_{B}=q_{AB}$, $g_{ab}\ell^{a}\ell^{b}=0$ and $g_{ab}\ell^{a}e^{b}_{B}=0$. Hence, the $3$-metric is also effectively $q_{AB}$. 

Let us now introduce the dual basis \cite{Wald} to the canonical null basis. We need a set of four linearly independent one-forms, $\textbf{v}_*^{(a)}$, such that $\textbf{v}^i_{(a)}\textbf{v}_{*i}^{(b)}=\df^{b}_a$. Denoting the inverse of the $2$-metric $q_{AB}$ by $q^{AB}$, we introduce a set of two one-forms, $e^{A}_a$, such that 
\begin{equation}\label{e^A_a}
e^{A}_a=q^{AB}g_{ab}e_{B}^b=q^{AB}(e_{B})_a~.
\end{equation}
Multiplying both sides by $q_{AC}$ and $g^{ac}$, we get the inverse relation
\begin{equation}\label{e^c_C}
e^c_C=q_{AC}g^{ac}e^{A}_a~.
\end{equation}
Then, it can be easily checked that
\begin{equation}\label{dual_can_null}
\textbf{v}_*^{(a)}=\left(-\underline{\textbf{k}},-\underline{\boldsymbol{\ell}},~\underline{\textbf{e}}^{A}\right),
\end{equation} 
with $\underline{\textbf{k}}$ representing the one-form with components $k_a$ etc., provides the required dual basis. In particular, we have
\begin{align}
e_{A}^a e^{B}_a = e_{A}^a q^{BC}(e_{C})_a = q^{BC}g_{ab} e_{C}^b e_{A}^a=q^{BC}q_{CA}=\df^{B}_{A}~.
\end{align} 
The canonical null basis and the dual basis allows us to write down the following decomposition of the Kronecker tensor:
\begin{equation}  \label{delta_decomp}
\df^{a}_b= -\ell^a k_b - k^a\ell_b+e^a_{A}e^A_b~,
\end{equation}
which can be easily checked by contracting the lower index with each member of the canonical null basis and the upper index with each member of the dual one-form basis. Raising the lower index, we obtain
\begin{align}
g^{ab}&=-\ell^a k^b - k^a\ell^b+e^a_{A}(e^A)^b \nn\\
&=-\ell^a k^b - k^a\ell^b+ q^{AB}e^a_{A}e^b_B,
\end{align}		
which implies
\begin{equation}
q^{ab}=q^{AB}e^a_{A}e^b_B~.
\end{equation}
Contracting both sides with $e^{C}_{a}$ and $e^{D}_{b}$, we can invert the above relation to obtain
\begin{equation}
q^{CD}=q^{ab}e^{C}_a e^{D}_b~.
\end{equation}
If we lower one index in \ref{delta_decomp}, we can use \ref{e^c_C} to obtain
\begin{equation}
q_{ab}=q_{AB}e^{A}_a e^{B}_b~,
\end{equation}
the inverse of which, $q_{AB}=q_{ab}e^{a}_{A}e^b_B$, just gives us back our original definition of $q_{AB}$ in \ref{q_AB}. So, finally, let us list out all the relations between the 4-dimensional tensors, $q_{ab}$ and $q^{ab}$, and the 2-dimensional tensors, $q_{AB}$ and $q^{AB}$:
\begin{align}\label{q2<->q4}
q_{AB}&=q_{ab}e^{a}_{A}e^b_B~,\qquad
q^{AB}=q^{ab}e^{A}_a e^{B}_b~,\nn \\
q_{ab}&=q_{AB}e^{A}_a e^{B}_b~,\qquad
q^{ab}=q^{AB}e^a_{A}e^b_B~.
\end{align}
\subsubsection{Erecting a Coordinate System on the Null Surface} \label{null_coord}
Suppose we have a set of coordinates $x^{a}=\left(x^{1},x^{2},x^{3},x^{4}\right)$ charting the four-dimensional spacetime with the null surface under consideration. Any set of three continuous, infinitely differentiable functions, say $y^\alpha=(y^1,y^2,y^3)$, of the spacetime coordinates $x^a$ constitutes a system of coordinates on the null surface provided the set of values of these functions at each and every point on the null surface is unique. Then, the coordinate basis is the set of three vectors
\begin{equation}
e^{a}_{\alpha}=\frac{\partial x^a}{\partial y^\alpha}~.
\end{equation}
If $g_{ab}$ be the components of the metric of the ambient spacetime in the coordinates $x^a$, the induced metric on the null surface is given by \cite{Poisson}
\begin{equation}
h_{\alpha \beta }= g_{ab}e^a_\alpha e^b_\beta~. 
\end{equation}
This is a $3$-metric and the determinant of this $3$-metric, $h$, will be zero as the surface is a null surface. The easiest way to see this and to work on the null surface is to erect a coordinate system naturally suited to the null nature of the surface \cite{Poisson}. The null surface is filled by a congruence of null geodesics, the integral curves of the normal vector $\ell^a$. We choose a parameter $\lambda$ varying smoothly on the null generators such that the displacements along the generators are of the form $dx^a=\ell^a d \lambda$. (In this paper, the null surface is taken as a $\phi=\textrm{constant}$ surface and the normal is taken as $\ell_a=A \partial_a \phi$ for some $A$. $\ell^a$ is then fixed and $\lambda$ has to be chosen appropriately. If our aim is just to erect a coordinate system on a given null surface, we may first choose $\lambda$ to be some parameter which varies along the null geodesics and then choose $\ell^a=dx^a/d\lambda$.) If we further ensure that $\lambda$ varies smoothly for displacements across geodesics, we may choose it as one of the coordinates on the null surface. The other two coordinates are to be chosen as two smooth functions $z^A=\left(z^1,z^2\right)$ that are constant on each null geodesic. They act as a unique label for each null geodesic. In this coordinate system, varying $\lambda$ would correspond to a displacement along a particular null geodesic while varying the set $z^A$ would correspond to displacements across the generators along points of equal $\lambda$. The basis vectors in the coordinate system $\left(\lambda, z^A \right)$ are
\begin{equation}
e^a_\lambda= \frac{\partial x^a}{\partial \lambda};~~~~ e^a_A= \frac{\partial x^a}{\partial z^A}, ~ A=1,2~.
\end{equation} 
Note that the identification $\ell^a=\partial x^a/\partial \lambda$ is possible because the coordinates $z^A$ have been chosen to be constant along the null geodesics. Previously, we had chosen an auxiliary vector $k^a$ and then demanded that $e^{a}_A$ satisfy $e^a_A k_a=0$. But while we are working purely on the null surface, there is no notion of an auxiliary null vector, as no vector on the null surface will satisfy $k^{a}\ell_a=-1$. In fact, having defined the coordinate basis vectors on the null surface, if we now we want to move into the ambient 4-dimensional spacetime and define $q_{ab}$, etc., we can specify $k^a$ uniquely by the four conditions $k^a k_a=0$, $k^a \ell_a=-1$ and $g_{ab}k^a e^b_A=0$. (This is how we defined $k^a$ in the case of GNC metric in \ref{GNC}.)

The components of the induced metric are $h_{\lambda \lambda}=g_{ab}\ell^a \ell^b=0$ (since $\ell^a$ is null), $h_{\lambda A}=g_{ab}\ell^a e^{b}_A=0$ (since $e^{b}_A$ lies on the surface and $\ell^a$ is the normal) and $q_{AB}\equiv g_{ab}e^a_{A}e^b_{B}$. In the coordinate order $\left(\lambda, z^1,z^2\right)$, the matrix form is
\begin{equation}
h_{ab}=
\left(\begin{array}{lll}
0 & 0 & 0\\
0 & q_{11} & q_{12}\\
0 & q_{12} & q_{22}
\end{array}\right)~.
\end{equation}
The determinant of $h_{ab}$, $h$, is obviously zero and will remain zero even if we transform to a different coordinate system on the null surface as the determinant changes only by a Jacobian factor. The line element is $2$-dimensional:
\begin{equation}
ds^2= q_{AB}dz^A dz^B~.
\end{equation}

Keeping this coordinate system as a reference, we can explore other coordinate systems on the null surface. The metric in any coordinate system $y^{\alpha}=\left(y^1,y^2,y^3\right)$ on the surface is given by
\begin{equation}
h'_{\alpha \beta}= q_{AB} \frac{\partial x^A}{\partial y^{\alpha}}\frac{\partial x^B}{\partial y^{\beta}}
\end{equation}
In general, none of the components need to be zero, although the determinant will vanish. But consider the special case $\left(y^1(\lambda,z^1,z^2),y^2(z^1,z^2),y^3(z^1,z^2)\right)$. This coordinate system has coordinates $y^2$ and $y^3$ constant on the null geodesics. $y^1$ may now be considered as the parameter varying along the null geodesics and we will have $\ell^a=M(y^1,y^2,y^3) \partial x^a/\partial y^1$ (with $M= \partial y^1/\partial \lambda$). In this case, the metric will again take the form
\begin{equation}
h'_{ab}=
\left(\begin{array}{lll}
0 & 0 & 0\\
0 & q'_{11} & q'_{12}\\
0 & q'_{12} & q'_{22}
\end{array}\right)~.
\end{equation} 	
\subsubsection{Directed Surface Element for the Null Surface}
For any set of coordinates $y^\alpha=(y^1,y^2,y^3)$ on the null surface with associated basis vectors $e^{a}_{\alpha}=\partial x^a / \partial y^\alpha$, the invariant directed surface element for the null surface is given by \cite{Poisson}
\begin{eqnarray}
d\Sigma_{a}= \epsilon_{abcd}e^{b}_{1}e^{c}_{2}e^{d}_{3} dy^{1}dy^{2}dy^{3}~.
\end{eqnarray}
Now, $\epsilon_{abcd}e^{b}_{1}e^{c}_{2}e^{d}_{3}$ is proportional to the normal $\ell_a$ (for a null surface, there is no unique normalized normal, but our treatment here will work for any choice of the normal) as its contraction with any vector on the null surface, expressible as a linear combination of $e^{a}_{\alpha}$, is zero. Thus, we may write 
\begin{equation} \label{surf_cov}
\epsilon_{abcd}e^{b}_{1}e^{c}_{2}e^{d}_{3}=f \ell_{a}
\end{equation} 
for some scalar function $f$. Contracting with $k^{a}$, we obtain
\begin{equation}\label{f}
f=-\epsilon_{abcd}k^a e^{b}_{1}e^{c}_{2}e^{d}_{3}~.
\end{equation}
Since $f$ is a scalar, it can be evaluated in any coordinate system. So we choose the $4$-dimensional coordinates $x^a$ to be $\left(\phi,y^1,y^2,y^3\right)$ with $\phi$ constant on the null surface. The form of the normal will then be $\ell_a=A \partial_a \phi$ for some scalar function $A$. In this coordinate system, borrowing the notation $\eqscs$ from \cite{Poisson} to indicate equalities valid in a specified coordinate system,
\begin{equation}
f\eqscs-\epsilon_{\phi 1 2 3}k^\phi \eqscs \frac{\epsilon_{\phi 1 2 3}}{A}~,
\end{equation}
where we have used the condition $k^a \ell_a\eqscs k^\phi \ell_\phi=-1$ to find $k^\phi$. If we choose the $\epsilon_{abcd}$ tensor so that $\epsilon_{\phi 1 2 3}$ is positive, then we shall have 
\begin{equation}
f \eqscs \frac{\sqg}{A} ,
\end{equation} 
with the choice $\ell_a=\partial_a \phi$ used in our treatment of the boundary term leading to
\begin{equation}\label{feqsqg}
f \eqscs \sqg ~.
\end{equation}
Thus, for the null surface represented as $\phi=\textrm{constant}$ and the normal being specified as $\ell_a=A \partial_a \phi$, the directed null surface element, in a coordinate system where $\phi$ is one of the coordinates, becomes
\begin{eqnarray}\label{dse}
d \Sigma_a \eqscs \frac{\sqg}{A} \ell_{a} dy^{1}dy^{2}dy^{3} = \sqg \partial_a \phi dy^{1}dy^{2}dy^{3}~.
\end{eqnarray} 
Note that, while this result is not generally covariant, it is valid in any coordinate system with $\phi$ as one of the coordinates. In the case of $\ell_a= \partial_a \phi$, it reduces to simply
\begin{eqnarray}\label{dse_spec}
d \Sigma_a \eqscs \sqg  \ell_{a} dy^{1}dy^{2}dy^{3}~.
\end{eqnarray}
In fact, nowhere in the derivation of \ref{dse} and \ref{dse_spec} did we have to make use of the assumption of the surface being null. In the case of non-null surfaces, we have the extra luxury of specifying the normalization of $\ell_a$ uniquely to give $n_a=\ell_a= N \partial_a \phi$ (see \ref{eq:n}). The auxiliary vector $k^a$ should then be taken as $k^a=-\epsilon n^a$ as the only condition we demand of $k^a$ in the above derivation is $k^a\ell_a=-1.$. Then, $A=N$ and $\sqg/A=\sqrt{|h|}$ and we arrive back at the familiar result for surface element for a non-null case.

We have mentioned that \ref{dse} and \ref{dse_spec} are valid only when $\phi$ is taken as a coordinate system. One way to formulate a fully covariant expression for the surface element is as follows. Note that all the treatment till \ref{f} is fully covariant. The non-covariance sneaked in at the evaluation of $f$. Now, given $\ell_a=\partial_a\phi$ and the triad of vectors $e^{a}_{\alpha}$, we can formulate a tetrad of vectors $(-k^a,e^{a}_{\alpha})$, with $k^a=\left(-1,0,0,0\right)$ in $\left(\phi, y^1,y^2,y^3\right)$ coordinate system. Thus, $k^a$ satisfies $k^a \ell_a=-1$ but is not necessarily null. Using this tetrad and the metric, we can construct ten independent scalars $\bar{g}_{\phi \phi}=g_{ab} (-k^a) (-k^b)$, $\bar{g}_{\phi 1}=g_{ab} (-k^a) e^b_{1}$, etc. In the coordinate system $\left(\phi, y^1,y^2,y^3\right)$, these scalars are just the components of $g_{ab}$. But now imagine taking the matrix corresponding to the metric $g_{ab}$ and replacing each component with the corresponding scalar to get a matrix of scalars. The determinant of this matrix is also a scalar. Let us denote it by $\bar{g}$. Hence, from \ref{feqsqg}, we have
\begin{equation}
f\eqscs \sqg \Rightarrow f \eqscs \sqrt{-\bar{g}}\Rightarrow f = \sqrt{-\bar{g}},
\end{equation} 
which is an equality between scalars and hence valid in all coordinate systems if valid in one. Note that this also means
\begin{equation}
\sqrt{-\bar{g}}=-\epsilon_{abcd}k^a e^{b}_{1}e^{c}_{2}e^{d}_{3}~,
\end{equation}
from \ref{f}.

In the special coordinate system $\left(\lambda, z^1,z^2\right)$ introduced in \ref{null_coord} with $\ell_a=\partial_a \phi$, the directed surface element reduces to \cite{Poisson}
\begin{eqnarray}
d \Sigma_a= \ell_a \sqrt{q} d\lambda d z^1 d z^1~,
\end{eqnarray}
where $q$ is the determinant of the $2$-metric $q_{AB}$. Note the difference in minus sign from the analogous expression in \cite{Poisson}. This minus sign difference arose because \cite{Poisson} defines the normal as $\ell_a=-\partial_a \phi$ so that $\ell^a$ is future-directed. 


\subsubsection{Second Fundamental Form $\Theta_{ab}$ and Expansion Scalar $\Theta$} \label{app:SFF+exp}

In this section, we shall introduce the second fundamental form $\Theta_{ab}$ and the expansion scalar $\Theta$ for a null surface. The terminology follows \cite{Gourgoulhon:2005ng}. But \cite{Gourgoulhon:2005ng} works with a foliation of null surfaces while we have only specified the $\phi=\phi_0$ surface to be a null surface among the $\phi=\textrm{constant}$ surfaces. Hence, we cannot blindly carry forward the results in \cite{Gourgoulhon:2005ng}. But we shall see that the results that we require are in fact unaltered.

Following \cite{Gourgoulhon:2005ng}, we use the projector in \ref{Pi} to introduce the extension of the second fundamental form for the null surface at any point, with respect to the chosen normal $\boldsymbol{\ell}$ (since, unlike the non-null case, the normalization of the normal is not fixed and hence the choice of $\boldsymbol{\ell}$ is not unique), to the tangent space of the four-dimensional spacetime manifold at that point:
\begin{equation}
\Theta_{ab} \equiv \Pi^{c}_{\phantom{a}a}\Pi^{d}_{\phantom{a}b} \nabla_{c}\ell_d~.
\end{equation}
Now, consider the following object on the null surface:
\begin{align}
q^{c}_{a}q^{d}_{b} \nabla_{c}\ell_d &= \left(\Pi^{c}_{\phantom{a}a}+\ell^c k_a\right)\left(\Pi^{d}_{\phantom{a}b}+ \ell^d k_b\right) \nabla_{c}\ell_d \nn\\
&= \Theta_{ab}+ \Pi^{c}_{\phantom{a}a}\ell^d k_b\nabla_{c}\ell_d+\ell^c k_a \Pi^{d}_{\phantom{a}b} \nabla_{c}\ell_d+\ell^c k_a \ell^d k_b \nabla_{c}\ell_d \nn \\
&= \Theta_{ab}+ \Pi^{c}_{\phantom{a}a}\ell^d k_b\nabla_{c}\ell_d+\kappa k_a \Pi^{d}_{\phantom{a}b} \ell_d+ \kappa k_a \ell^d k_b \ell_d \nn \\
&= \Theta_{ab}+ \frac{1}{2}\Pi^{c}_{\phantom{a}a}k_b\nabla_{c}\left(\ell^d \ell_d \right), \label{Thetaqqint}
\end{align}
where we have made use of the results $\Pi^{a}_{\phantom{a}b}\ell_a=0$ and $\ell^a \ell_a=0$ on the null surface. The second term in \ref{Thetaqqint} contains an expression of the form $\nabla_b (\ell^{a} \ell_{a})$. Let us simplify this expression to a form that will be more useful to us. We write
\begin{align}
\nabla_b (\ell^{a} \ell_{a}) &= g_{bc} \nabla^c (\ell^a \ell_a) \nn\\
&= (q_{bc}-\ell_b k_c - k_b \ell_c)\nabla^c (\ell^a \ell_a), \label{kappa2}
\end{align}
where we have used \ref{qab} in the second line. In \ref{kappa2}, the third term has the combination
\begin{equation}
\ell^{c} \nabla_c (\ell^a \ell_a)=\ell^{c} \partial_c (\ell^a \ell_a)=0, 
\end{equation}
since $\ell^c \partial_c$ is a derivative along the null surface and the value of $\ell^a \ell_a$ is zero throughout the null surface.
Next, consider the first term in \ref{kappa2}, which has
\begin{align}
q_{bc} \nabla^c (\ell^a \ell_a) =& q^{c}_{b}\partial_c (\ell^a \ell_a) \nn\\
=& q^{\phi}_{b}\partial_\phi (\ell^a \ell_a),
\end{align}
in the coordinate system in which $\phi$ is one of the coordinates, since the only direction along which $\ell^a \ell_a$ varies on the null surface is off the surface i.e the direction along which $\phi$ varies. 

Now, using \ref{l_as_dphi}, 
\begin{equation}
q^{\phi}_b \propto q^{a}_{b}l_{b}=0~. 
\end{equation}
Thus, $q^{\phi}_{b}$ is identically zero on the null surface and the first term in \ref{kappa2} also vanishes. (To prove that a tensor quantity is identically zero, it is sufficient to prove that it is zero in one coordinate system.) So, we arrive at the result that
\begin{equation} \label{grad_lsq}
\nabla_b (\ell^{a} \ell_{a}) = - k^c \nabla_c (\ell^a \ell_a)\ell_b~. 
\end{equation}
Using this result in \ref{Thetaqqint}, we obtain
\begin{equation}\label{Thetaqq}
\Theta_{ab}=q^{c}_{a}q^{d}_{b} \nabla_{c}\ell_d~.
\end{equation}
Instead of labouring forth with "the four-dimensional extension of the second fundamental form", we shall henceforth take the liberty of referring to $\Theta_{ab}$ as just the second fundamental form of the null surface. To be precise, we should also add that it is the second fundamental form with respect to the normal $\ell_a$ under consideration but we shall consider this to be understood henceforth.

The $\Theta_{ab}$ is a symmetric object. To see this, we use \ref{l_as_dphi} and decompose $\nabla_a \ell_b$ as
\begin{align}\label{dl_for_theta}
\nabla_a \ell_{b}= A \nabla_a \nabla_b \phi + \ell_b \partial_a \ln A~. 
\end{align}	
The first term in \ref{dl_for_theta} is symmetric while the second term does not contribute to $\Theta_{ab}$. Hence, proved.

We shall next prove a relation between $\Theta_{ab}$ and the Lie derivative of $q_{ab}$ along $\boldsymbol{\ell}$. The Lie derivative formula is
\begin{equation}
\pounds_{\ell} q_{ab} = \ell^d \nabla_d q_{ab} + q_{ad} \nabla_b \ell^d + + q_{db} \nabla_a \ell^d~.
\end{equation}
Substituting for $q_{ab}$ from \ref{qab}, we have
\begin{equation}
\pounds_{\ell} q_{ab} =\ell^d \left(\ell_a \nabla_d k_b+ k_b \nabla_d \ell_a + \ell_b \nabla_d k_a + k_a \nabla_d \ell_b\right)+ \nabla_b \ell_a + (\ell_a k_d +k_a \ell_d) \nabla_b \ell^d+ \nabla_a \ell_b + (\ell_b k_d +k_b \ell_d) \nabla_a \ell^d~.
\end{equation}
Contracting with $q^{a}_{m}q^{b}_n$, and using $q^{a}_b \ell_a=0$ and $q^{a}_b k_a=0$, we obtain
\begin{align}
q^{a}_{m}q^{b}_n\pounds_{\ell} q_{ab} &=q^{a}_{m}q^{b}_n \left( \nabla_a \ell_b+\nabla_b \ell_a\right) = \Theta_{mn}+\Theta_{nm}=2 \Theta_{mn} 
\end{align} 
which implies
\begin{equation}
\Theta_{mn} = \frac{1}{2}q^{a}_{m}q^{b}_n\pounds_{\ell} q_{ab}~.\label{thetainlie}
\end{equation}

Finally, let us look at the trace of $\Theta_{ab}$:
\begin{equation}
\Theta = g^{ab}\Theta_{ab}=q^{ab}\Theta_{ab}~.
\end{equation}	
We shall refer to $\Theta$ as the expansion scalar on the null surface. More specifically, it is the expansion along $\ell^a$ of the $2$-surface on the null surface orthogonal to $k^{a}$. The reason for this terminology becomes clear if we take the trace of \ref{thetainlie}. We obtain
\begin{align}
\Theta &= \frac{1}{2} q^{ab}\pounds_{\ell}q_{ab}~.
\end{align}
This equation may be further manipulated to a form easier to interpret. We can make use of the canonical null basis introduced in \ref{app:q_for_null}. We have
\begin{align}
q_{ab}&=q_{AB}e^{A}_a e^{B}_b~,\nn \\
q^{ab}&=q^{AB}e^a_{A}e^b_B~.
\end{align}
where summation over $A,B=1,2$ is implied. Then,
\begin{align}
\Theta &= \frac{1}{2} q^{AB}e^a_{A}e^b_B\pounds_{\ell}\left(q_{CD}e^{C}_a e^{D}_b\right)
= \frac{1}{2} q^{AB}\pounds_{\ell}q_{AB} + q^{AB} q_{BC} e^a_{A}\pounds_{\ell} e^{C}_a~\nn \\ 
&= \frac{1}{2} q^{AB}\pounds_{\ell}q_{AB} + e^a_{A}\pounds_{\ell} e^{A}_a
= \frac{1}{2} q^{AB}\pounds_{\ell}q_{AB} - e^{A}_a\pounds_{\ell}e^a_{A} ~.
\end{align} 
where we have used $e^{a}_{A}e^{B}_{a}= \df^{B}_{A}$, $q^{AB} q_{BC}= \df^{A}_{C}$ and $\pounds_{\ell} \left(e^a_{A}e^{A}_a\right)=\pounds_{\ell} 2=0$ along the way. Now, let $e^{a}_A$ be the coordinate vectors $e^a_{z_1}$ and $e^a_{z_2}$ in the special coordinate system $\left(\lambda, z^1,z^2\right)$ introduced in \ref{null_coord}. Since $e^{a}_{A}$ and $\ell^a$ are then members of a coordinate basis, their Lie bracket should be zero, i.e,
\begin{equation}
[\boldsymbol{\ell}, \mathbf{e_A}] = 0  \Rightarrow \pounds_{\ell}e^a_{A}=0~. 
\end{equation}
Enforcing this condition, we obtain
\begin{align}\label{theta-Llq}
\Theta &=\frac{1}{2} q^{AB}\pounds_{\ell}q_{AB}= \frac{\pounds_{\ell} \sqrt{q}}{\sqrt{q}}~,
\end{align} 
where $q$ is the determinant of the $2$-metric $q_{AB}$. Thus, we see that $\Theta$ represents the fractional change in an area element of the $\lambda=\textrm{constant}$ $2$-surface on the null surface as it is Lie dragged along $\ell^a$.
We can replace the Lie derivative in the above expression with an ordinary derivative since $q_{AB}=q_{ab}e^{a}_{A}e^b_B$ is a scalar under $4$-dimensional coordinate transformations with the $e^{a}_{A}$ kept as the same physical vectors. Thus, $\pounds_{\ell}q_{AB}= \ell^a \partial_a q_{AB}$ which becomes $\pounds_{\ell}q_{AB}=d q_{AB}/d \lambda$ for the parameter $\lambda$ introduced in \ref{null_coord}. Hence,
\begin{equation}\label{theta-dqdl}
\Theta = \frac{1}{\sqrt{q}}\frac{d \sqrt{q}}{d \lambda}~.
\end{equation}

\subsubsection{ $\kappa$ and $\Theta$ in terms of $(\boldsymbol{\ell},\textbf{k})$} \label{app:kappatheta}

In this appendix, we shall derive expressions for the non-affinity coefficient $\kappa$ and $\Theta$ in terms of $\ell_{a}$ and auxiliary vector $k_{a}$, for $\ell_{a}$ being the null normal, $\ell_{a}=A \partial_a \phi$, to a surface $\phi = \phi_{0}$.  

Let us first define the auxiliary vector $k^{a}$ (need not be null for manipulations in this section) such that $k^{a}\ell_{a} = -1$. Then,
\begin{align}
\ell^{a} \nabla_a \ell_b &= \ell^{a} \nabla_a (A \partial_b \phi) 
= \ell^{a}\frac{\ell_{b}}{A}\partial_a A + \ell^{a} A \nabla_a \nabla_b \phi  \nn\\
&= \ell^{a}\frac{\ell_{b}}{A}\partial_a A + \ell^{a} A \nabla_b \nabla_a \phi  
= \ell^{a}\frac{\ell_{b}}{A}\partial_a A + \ell^{a} A \nabla_b (\frac{\ell_a}{A})  \nn\\
&= \ell^{a}\frac{\ell_{b}}{A}\partial_a A + \ell^{a} \nabla_b \ell_a 
=[\ell^a \partial_a (ln A)] \ell_b + \frac{1}{2} \partial_b (\ell^{a}\ell_{a}) \label{kappa1}
\end{align}
Substituting in \ref{kappa1}, we obtain
\begin{eqnarray}
\ell^{a} \nabla_a \ell_b&= [\ell^a \partial_a (ln A)] \ell_b - \frac{k^c}{2} \partial_c (\ell^{a}\ell_{a}) \ell_b \nn\\
&= [\ell^a \partial_a (ln A) - \frac{k^c}{2} \partial_c (\ell^{a}\ell_{a})] \ell_b
\end{eqnarray}
Hence, we see that $\ell^{a} \nabla_a \ell_b = \kappa \ell_b$ with the non-affinity coefficient $\kappa$ given by the formula
\begin{equation}
\kappa = \ell^a \partial_a (ln A) - \frac{k^c}{2} \partial_c (\ell^{a}\ell_{a}) \label{kappa}
\end{equation}
Using this expression, we can also evaluate
\begin{align}
\nabla_a \ell^a &= q^{ab} \nabla_a \ell_b - \ell^a k^b \nabla_a \ell_b - \ell^b k^a \nabla_a \ell_b \\
&= \Theta + \kappa - \frac{k^{a}}{2} \partial_a (\ell^{b}\ell_{b})~. \label{exp_nabla_l}
\end{align}
Let us define
\begin{equation} \label{kappatilde}
\tilde{\kappa}\equiv- \frac{k^{a}}{2} \partial_a (\ell^{b}\ell_{b})=- k^{a}\ell^{b} \nabla_a \ell_{b}~,
\end{equation}
a measure of how much $\ell_{a}\ell^{a}$ varies as we move away from the null surface. If $\nabla_{a}\ell_{b}$ is symmetric, like in the case when we have $\ell_{b}=\nabla_{b} \phi$, we shall get $\tilde{\kappa}=\kappa$. If $\nabla_{a}\ell_{b}$ is antisymmetric, i.e if $\ell_{b}$ is a Killing vector, then we get $\tilde{\kappa}=-\kappa$. 
\ref{exp_nabla_l} now takes the form
\begin{equation}\label{exp_nabla_l_1}
\nabla_{a}\ell^{a}= \Theta + \kappa + \tilde{\kappa}~.
\end{equation}

\subsubsection{Decomposition of $\sqg$ in Terms of $\sqrt{q}$ for a Null Surface} \label{app:decomp_sqg_null}

We shall use the result \ref{g_to_2} which is applicable even when we take the null limit. In the limit the $\phi=\textrm{constant}$ surface under consideration is null, $g^{\phi \phi}=0$ and $h=0$. Taking the limit $g^{\phi \phi }\rightarrow 0$ on \ref{g_to_2}, we get
\begin{equation}\label{q_to_g}
g=\frac{-q}{(g^{1 \phi})^2}~.
\end{equation}

For the null surface, we need $q$ to be the determinant of the $2$-metric $q_{AB}=g_{ab}e^a_A e^b_B$ (see \ref{q_AB}). To apply result \ref{q_to_g}, we shall specialize to a coordinate system such that $e^a_A$ are coordinate basis vectors and $\ell_a=\partial_a \phi$. On the null surface, we shall introduce the two coordinates $z^A=\left(z^1,z^2\right)$, constant on the null geodesics and a third coordinate $\mu$ which is a parameter varying along the null geodesics such that the vectors $e^{a}_A$ lie on a $\mu=\textrm{constant}$ surface and $e^a_A=\partial x^a/\partial z^A$. We shall also have $\ell^a= \left(1/M\right)\partial x^a/\partial \mu$ for some scalar $M$. In the case where we choose the special coordinate system $\left(\lambda,z^1,z^2\right)$ introduced in \ref{null_coord}, we will have $M=1$ and $\ell^a= \partial x^a/\partial \lambda$. In the coordinates $\left(\phi,\mu,z^1,z^2\right)$, $g_{AB}=q_{AB}$, $g_{\mu A}=g_{ab}\left(\ell^a/M\right)e^a_A=0$ and $g_{\mu \mu} \propto g_{ab}\ell^a \ell^b=0$. Also, consider $g_{ab} \ell^a k^b= \left(g_{\lambda b}/M\right)k^b$. Now, $g_{ab} \ell^a k^b=\ell_a k^{a}=k^{\phi}$. Hence, only $k^{\phi}$ contributes to $g_{ab} \ell^a k^b$. Since $g_{ab} \ell^a k^b=-1$, we have $k^{\phi}=-1$ and we obtain $g_{\lambda b}=M$. Thus, in the coordinates $\left(\phi,\mu,z^1,z^2\right)$, the metric takes the form
\begin{equation}
g_{ab}=
\left(\begin{array}{llll}
g_{\phi \phi} & M& g_{\phi 1} & g_{\phi 2}\\
M & 0& 0 & 0\\
g_{\phi 1} &0 & q_{11} & q_{12}\\
g_{\phi 2} &0 & q_{12} & q_{22}
\end{array}\right)~.
\end{equation} 	
Using $\ref{q_to_g}$ (or by direct calculation of the determinant from the above matrix), we can write down 
\begin{equation}
g=-M^2 q~,
\end{equation}
and
\begin{equation}\label{g_eq_Mq}
\sqrt{-g}=|M|\sqrt{q}~.
\end{equation}
If we specialize to $\mu=\lambda$, we would have $g=q$ and
\begin{equation}\label{g_eq_q}
\sqrt{-g}=\sqrt{q}~.
\end{equation}

\section{Boundary Conditions for Spacelike and Timelike Surfaces: An Alternate Approach} \label{BC-novel}
In this appendix, we shall detail an alternate approach to arrive at the standard prescription for fixing the boundary conditions on timelike and spacelike surfaces \cite{Padmanabhan:2014BT}. As we have seen in \ref{var_A_EH}, the surface term of the Einstein-Hilbert action is the integral of the quantity
\begin{equation}\label{surf_term}
\sqrt{-g} Q [v_c] = \sqrt{-g} v_{c} (g^{ab} \delta \Gamma^{c}_{ab}-g^{ck} \delta \Gamma^{a}_{ak})~ 
\end{equation}
over the boundary of the spacetime region under consideration. Here, $v_{c}$ is the surface gradient. To be concrete, let us take $\phi=\phi_{0}$ to be the surface, with $\phi_{0}$ a constant. Then $v_{c}= \pm \partial_c \phi$, with the sign decided according to the conventions of the Gauss' theorem (see \ref{app:Gauss}). Let us assume that our surface is such that $\phi$ increases on going from inside the integration volume to outside through the surface. Hence, we shall use $v_{c}=\partial_c \phi$ from now on. 
It is clear that \ref{surf_term} contains the variations of the metric and its derivatives. We can separate out the terms with the variations of the metric and its derivatives in \ref{surf_term} and write
\begin{equation}\label{ST-metric+deriv}
\sqrt{-g} Q [v_c] =  v_c \left[2 \sqg P^{cdea} \Gamma^{b}_{de} \delta g_{ab} - 4 \sqg P^{cadb} \delta(\partial_d g_{ab})     \right]~,
\end{equation}
where $P_{a}^{~bcd}=1/2\left(g^{bd}\delta ^{c}_{a}-g^{bc}\delta ^{d}_{a}\right)$. Since $v_c=\partial_c \phi$, the coefficient of $\delta(\partial_d g_{ab})$ is $- 4 \sqg P^{\phi adb}$. We can easily check that $P^{\phi a \phi b} \neq 0$ in general. Hence, \ref{surf_term} contains variations of the normal derivatives of the metric.

Now, let us consider the case of timelike and spacelike surfaces as the boundary and see how we can fix boundary conditions without having to fix both the metric and its normal derivatives at the boundary. Generalizing \ref{surf_term}, we shall define $Q[A_{c}]$ for any vector $A_{c}$ to be
\begin{equation}
Q [A_c] \equiv A_{c} (g^{ab} \delta \Gamma^{c}_{ab}-g^{ck} \delta \Gamma^{a}_{ak})~. \label{q}
\end{equation}
For timelike or spacelike boundaries, we can normalize the surface gradient to obtain the unit normal $n_{a}$ such that 
\begin{equation}
n_a=\frac{v_{a}}{\sqrt{|g^{\phi \phi}|}};~~ n_a n^{a}=\epsilon, 
\end{equation}
with $\epsilon=1$ for a timelike surface and $\epsilon=-1$ for a spacelike surface. We shall also demand that the normalization be preserved under the variation of the metric, i.e,
\begin{equation} \label{eq:var_normaliz}
\delta (n_{a}n^{a}) = n_{a} \delta n^{a} + \delta n_{a} n^{a} = 0~.
\end{equation}
We choose a coordinate system with $\phi$ taken as one of the coordinates. Let the other coordinates, the coordinates charting the $\phi= \textrm{constant}$ surface, be labelled $(x^1, x^2, x^3)$. As mentioned in the introduction, the Greek indices (other than $\phi$) run over $(x^1, x^2, x^3)$ while the Latin indices run over $(\phi,x^1, x^2, x^3)$. Let us use $N$ for the normalization factor in $n_{a}$, i.e
\begin{equation} \label{eq:n}
n_{a} = N \partial_a \phi= Nv_{a}~, 
\end{equation}
where $N$ is to be assumed positive so that $n_{a}$ and $\partial_a \phi$ point in the same direction. The normalization $n_{a}n^{a}=\epsilon$ then relates $N$ to the metric as
\begin{equation}\label{eq:gphiphi}
g^{\phi \phi} = \epsilon /N^2~. 
\end{equation}
Another convention that is used often is to demand that $\phi$ increases in the direction of the normal vector $n^{a}$, i.e., $n^{a} \partial_a \phi >0$. In this case, we would write
\begin{equation} \label{eq:ntilde}
\tilde{n}_{a} = \epsilon N v_{a},
\end{equation}
with $N$ positive, where we have put the tilde just to distinguish the normal in this convention from the $n_{a}$ in \ref{eq:n}.
Now, from the standard procedure of calculating the inverse of a metric we know that
\begin{equation} \label{eq:gphiphidef}
g^{\phi \phi} = \frac{\textrm{Cofactor} [g_{\phi \phi}]}{g}=\frac{\textrm{Det} [h_{\alpha \beta}]}{g}\equiv\frac{h}{g}~,
\end{equation}
 where the $3 \times 3$ matrix corresponding to the metric on the $\phi=\phi_{0}$ surface, obtained by deleting the $\phi$-row and $\phi$-column from the matrix representing the metric, was denoted by $h_{\alpha \beta}$ and the symbol $h$ was defined to be the determinant of this matrix.
We can rewrite \ref{eq:gphiphidef}, using \ref{eq:gphiphi}, as
\begin{equation}\label{g_decomp_Nh}
g= \epsilon N^2 h =- N^2 |h|~. 
\end{equation}  
Thus, we arrive at the following decomposition for $\sqg$:
\begin{equation}
\sqg = N \sqrt{|h|}~. 
\end{equation}
Using this expression and \ref{eq:n}, we can rewrite the expression for the surface term in \ref{surf_term}, in the case of timelike or spacelike surfaces, as
\begin{align} \label{ST_hn}
\sqrt{-g} Q [v_c] &= \sqrt{|h|} n_{c} (g^{ab} \delta \Gamma^{c}_{ab}-g^{ck} \delta \Gamma^{a}_{ak})~ \nn\\
&= \sqrt{|h|} Q [n_c]~.
\end{align}
As we can see from \ref{ST-metric+deriv}, this expression contains the variations of the metric as well as its normal derivatives. 
Our aim is to discover a counter-term such that, when added to the Einstein-Hilbert action, the surface term obtained in the variation will contain only the variations of the metric. That is, we would like to express the surface term in the form
\begin{equation}
\delta \mathcal{A}_{_{\partial  \mathcal \tiny V}} = \delta [X] + Y_{ab}\delta g^{ab}~.
\end{equation}
Then, it is $-X$ that we have to add to the action as the counter-term. 

Proceeding towards this goal, let us first manipulate $Q[A_{c}]$ for an arbitrary vector $A^{c}$. We have the following relations:
\begin{equation}
\delta(\nabla_{a} A_{b}) = \nabla_{a} \delta A_{b} - \delta \Gamma^{c}_{ab} A_{c} ~; \quad \quad \delta(\nabla_{a}A^{a}) = \nabla_{a} \delta A^{a} + A^{c} \delta \Gamma^{a}_{ac}~.
\end{equation}
Using these relations, we can rewrite the expression for $Q[A_{c}]$ in \ref{q} as follows:
\begin{align}
Q[A_{c}] &= \left\{\left[ \nabla_{a}, \delta \right], g^{ab}\right\} A_{b}\nn \\
&= \nabla_{a}(\delta \uA^{a})- \delta (2 \nabla_a A^{a}) + \nabla_{a} A_{b} ~\delta g^{ab}, \label{eq:q_exp}
\end{align}
where
\begin{equation}\label{eq:u}
\delta \uA^{a} = \delta A^{a} + g^{ab} \delta A_{b}~. 
\end{equation}
In the first line of \ref{eq:q_exp}, we have used the notation $[A,B]$ for the commutator, $AB-BA$, and $\{A,B\}$ for the anticommutator, $AB+BA$.  

Let us now specialize to the case of $n_{a}$. Then, we have
\begin{equation}
Q[n_{c}] = \nabla_{a}(\delta \un^{a})- \delta (2 \nabla_a n^{a}) + \nabla_{a} n_{b} ~\delta g^{ab}, \label{eq:q_n_exp}
\end{equation}
where $\delta \un^{a} = \delta n^{a} + g^{ab} \delta n_{b}$. Using \ref{eq:var_normaliz}, we can see that
\begin{equation} \label{eq:u_n_orth}
\delta \un^{a} n_{a}=0,
\end{equation}
which means that the vector $\delta \un^{a}$ lies on the surface $\phi = \phi_{0}$. This property can be used to decompose the first term of \ref{eq:q_n_exp} in the following manner:
\begin{align}
\nabla_{a} (\delta \un^{a}) &= \delta^{a}_{b} \nabla_{a} (\delta \un^{b}) \nn\\
&= (h^{a}_{b} + \epsilon n^{a}n_{b})  \nabla_{a} (\delta \un^{b}) \nn\\
&= h^{a}_{b} \nabla_{a} (\delta \un^{b}) + \epsilon n^{a}n_{b} \nabla_{a} (\delta \un^{b})\nn\\
&= D_{a}(\delta \un^{a}) - \epsilon (n^{a}\nabla_{a}n_{b})\delta \un^{b} \nn\\
&= D_{a} (\delta \un^{a}) - \epsilon a_{b}  \delta \un^{b}~. \label{eq:du_decomp}
\end{align}
On the way to obtaining the above expression, we have introduced the induced metric 
\begin{equation}\label{hab}
h_{ab}= g_{ab} - \epsilon n_{a} n_{b} ~,
\end{equation}
and a derivative operator $D_{a}$ such that $D_{a} V^{b} = h^{c}_{a}h^{b}_{d}\nabla_{c} V^{d}$. For vectors $V^{a}$ on the surface ( i.e satisfying $V^{a}n_{a}=0$), $D_{a}$ is the natural covariant derivative on the $3-$surface (see \cite{Poisson} and Chapter 12 in \cite{gravitation}) compatible with the induced surface metric.

To proceed further, let us look at the nature of the variations we are considering. The $\delta$s are field variations on the metric and leave the coordinates unchanged, which is why they could be taken inside integrals over spacetime and derivatives with respect to the coordinates. The scalar $\phi$ is taken to be fixed during the variation, which means that the foliating surfaces $\phi=\textrm{constant}$ are kept fixed during the variation. In our case, this is not an extra assumption as we have already taken $\phi$ to be one of our coordinates. Now, from \ref{eq:n}, 
\begin{equation} \label{eq:var_n}
n_{a}= N \partial_a \phi \Rightarrow \delta n_{a} = \delta N  \partial_{a} \phi = \delta(\ln N) n_{a}~. 
\end{equation}
This shows that $\delta n_{a}$ is in the direction of $n_{a}$. We can write this in terms of variations of the metric using the constraint $n^{a}n_{a}=\epsilon$, equivalent to \ref{eq:gphiphi}. We see that 
\begin{equation}
\df \ln N=\frac{1}{2}\df \ln N^2=-\frac{1}{2}\df \ln g^{\phi \phi}= -\frac{\epsilon N^2}{2} \df g^{\phi \phi}= -\frac{\epsilon}{2} \df g^{a b}n_{a}n_{b}~.
\end{equation}
Thus, we obtain the following explicit expression for the variation of $n_a$ in terms of the metric:
\begin{equation}
\delta n_{a} = -\frac{\epsilon}{2}\delta g^{ij} n_{i}n_{j}  n_{a} ~.
\end{equation}
Now, the second term in \ref{eq:du_decomp} contains
\begin{equation}
a_{b}  \delta \un^a = a_{b} \delta n^{b} + a^{b} \delta n_{b} = a_{b} n_{a} \delta g^{ab} + a_{b} \delta n_{a} g^{ab}  =   a_{b} n_{a} \delta g^{ab},
\end{equation}
where we have first used the definition of $\delta \un^{b}$ from \ref{eq:u}, and then applied the result $a^{b}\delta n_{b} =0$, which we know from \ref{eq:var_n}, twice. Thus, \ref{eq:du_decomp} can be rewritten as
\begin{equation}
\nabla_{a} (\delta \un^a) =D_{a} (\delta \un^a) - \epsilon a_{b} n_{a} \delta g^{ab}~.
\end{equation}
So, \ref{eq:q_n_exp} takes on the form
\begin{equation}
Q[n_{c}] = D_{a}(\delta \un^a)- \delta (2 \nabla_a n^{a}) + (\nabla_{a} n_{b}-\epsilon n_{a} a_{b}) ~\delta g^{ab}
\end{equation}
Our manipulations have naturally produced the quantity $\nabla_{a} n_{b}-\epsilon n_{a} a_{b}$. This quantity is nothing but the negative of the extrinsic curvature \cite{gravitation} defined by
\begin{equation}
K_{ab}= - h^{c}_{a} \nabla_c n_{b} = - \nabla_a n_{b} + \epsilon n_{a} a_{b} ~,
\end{equation}
where we have used the definition of the induced metric given in \ref{hab}. $K_{ab}$ has the following properties:
\begin{equation}
K_{ij} = K_{ji};\quad \quad K_{ij} n^{j}=0;\quad \quad K = g^{ij} K_{ij} = - \nabla_a n^{a}
\end{equation}
The second equation tells us that $K_{ij} \delta g^{ij} = K_{ij} \delta h^{ij}$. Using this result and the results in the last three equations, we can rewrite the surface term of the Einstein-Hilbert action, in the following useful form:
\begin{align}
\STnn &=\int_\mathcal{\partial V} d^3x ~ \sqrt{|h|} Q[n_{c}]\nn\\ &=\int_\mathcal{\partial V} d^3x~ \left[\sqrt{|h|} D_{a}(\delta \un^a)+ \delta (2 \sqrt{|h|}  K) - 2K \delta(\sqrt{|h|})- \sqrt{|h|} K_{ab} ~\delta h^{ab}\right] \nn\\
&=\int_\mathcal{\partial V} d^3x~ \left[\sqrt{|h|} D_{a}(\delta \un^a)+ \delta ( 2\sqrt{|h|}  K) - \sqrt{|h|} (K_{ab} - K h_{ab}) ~\delta h^{ab}\right]~. \label{S.T-nonnull-1}
\end{align}
On integrating over the $3-$surface, the first term in \ref{S.T-nonnull-1} is a $3-$divergence which can be converted to a boundary term. If the surface bounding the integration volume is a closed surface, then this term goes to zero. The second term in \ref{S.T-nonnull-1} can be killed off by adding the counter-term $- 2 \sqh K$ to the Einstein-Hilbert Lagrangian. The last term in \ref{S.T-nonnull-1} can be put to zero by fixing the induced metric $h^{ab}$ on the boundary. 

If we choose to work in the convention where the normal is chosen to be the one in \ref{eq:ntilde}, then we would have the surface term in the form
\begin{align}
\STnn &=\int_\mathcal{\partial V} d^3x~ \epsilon \sqrt{|h|} Q[\tilde{n}_{c}] \nn\\&=\int_\mathcal{\partial V} d^3x~ \left[\epsilon\sqrt{|h|} D_{a}(\delta \unt^a)+ \delta ( 2\epsilon\sqrt{|h|}  K) - \epsilon\sqrt{|h|} (K_{ab} - K h_{ab}) ~\delta h^{ab}\right]~. \label{S.T-nonnull-alt-1}
\end{align}

\section{Boundary Term on The Null Surface for Specific Parametrizations} \label{app:BT-GNC+NSF}

In this appendix, we shall verify the validity of the result in \ref{NewLabel03} by working out two specific cases. We shall introduce two parametrizations for a general null surface and calculate the boundary term of the Einstein-Hilbert action on the null surface. In both cases, we find results in agreement with \ref{NewLabel03}.

\subsection{Gaussian Null Coordinates (GNC)}\label{GNC}

The first parametrization that we shall discuss is what is commonly known as Gaussian null coordinates (GNC), in analogy with the Gaussian normal coordinates \cite{MTW}. Gaussian normal coordinates are constructed by extending the coordinates on a non-null hypersurface to a spacetime neigbourhood using geodesics normal to the surface. This prescription does not work for a null surface as the ``normal'' geodesics lie on the null surface itself. Hence, the construction of Gaussian null coordinates is carried out by making use of certain, uniquely defined, auxiliary null geodesics. As far as we know, this coordinate chart in the neighbourhood of a null surface was introduced by Moncrief and Isenberg in \cite{Moncrief:1983} and hence may also be termed as Moncreif-Isenberg coordinates. The construction of GNC is also described in \cite{Friedrich:1998wq}, \cite{Racz:2007pv} and \cite{Morales}. 

\subsubsection{Construction of the Coordinates}

We shall now describe the construction of GNC. Our description will be less technical and more intuitive. We shall mostly follow the notation of \cite{Morales}. Note that \cite{Morales} has the GNC metric in Eq. 3.2.7 but then claims that it can be further constrained and writes it as in Eq. 3.2.8, i.e makes the replacement $\alpha \rightarrow r \alpha$. We do not subscribe to this claim and will use Eq. 3.2.7. 

Consider a smooth null surface $N$ in a spacetime manifold $M$. Let $g_{ab}$ represent the metric on $M$. We shall take the spacetime to be four-dimensional but the following construction can be easily extended to an $n$-dimensional Lorentzian spacetime as long as the extra dimensions are spacelike. Take a spacelike $2$-surface, $\zeta$, on $N$ and introduce coordinates $(x^1, x^2)$ on that surface. We have shown that the null surface $N$ is generated by null geodesics. The null geodesics cannot be along the spacelike surface at any point, since any vector that lies on the spacelike surface has to be spacelike. Thus, one necessarily moves away from the spacelike surface by travelling along any of these null geodesic. Therefore, the null geodesics can be used to define coordinates on the null surface as follows. Let $u$ be a parameter, not necessarily affine, along the null geodesics with $u=0$ on $\zeta$ and increasing towards future. Let each null geodesic be labelled by the coordinates $(x^1, x^2)$ of the point at which it intersects $\zeta$. Then, any point on $N$ in a neighbourhood of $\zeta$, sufficiently small so that the geodesics do not cross, can be assigned the coordinates $(u,x^1,x^2)$, where $(x^1,x^2)$ corresponds to the label given to the null geodesic passing through that point and $u$ corresponds to the value of our chosen parameter at that point. Henceforth, when we talk about, say, coordinates on the null surface, it is to be understood that we are talking about such a neighbourhood. We shall call the future-directed vector field tangent to the null geodesics, $\partial/ \partial u$, as $\boldsymbol{\ell}$ and the basis vectors corresponding to the coordinates $(x^1,x^2)$ as $\textbf{X}_{A}=\partial/\partial x^{A}$ with $A=1,2$. So far, we have not introduced the fourth coordinate, and hence the above definitions, with the partial derivatives, will continue to hold only if the fourth coordinate is chosen to be constant on the null surfaces. But such a choice will indeed be made, as we see next. 

Having thus constructed a coordinate chart on the null surface, it is time for us to move out into the surrounding spacetime. We shall do this with the help of a new set of null geodesics. At each point on the null surface, there is a unique vector $k^{a}$ (\textit{note the interchange of notation of $\boldsymbol{\ell}$ and $\boldsymbol{k}$ compared to \cite{Morales}}) satisfying the following four constraints:
\begin{inparaenum}[(i)]
	\item $k_{a}k^{a}=0$, i.e it is a null vector;
	\item $\ell_{a}k^{a}=-1$ and
	\item $X^{a}_{A}k_{a}=0$ for $A=1,2$.
\end{inparaenum}
From $\ell_{a}k^{a}=-1$, it is clear that this vector does not lie on the surface at any point but ``sticks out''. Thus, we can make use of this vector to ``go off'' the null surface. From each point on the null surface, send out null geodesics in the direction of $k^{a}$ labelled by the coordinates, $(u,x^1,x^2)$, of that point. The fourth coordinate may be chosen as an affine parameter along the null geodesic. Let $r$ be that affine parameter such that $r=0$ represents the null surface and $\boldsymbol{k}=-\partial/\partial r$. (The sole purpose of the minus sign is to reproduce the exact form of the metric in \cite{Morales} by off-setting the fact that we demand $\ell_{a}k^{a}$ to be $-1$ while \cite{Morales} equates the same to $+1$.) These constraints uniquely determine the affine parameter. Any $r'= A(u,x^1,x^2)r+B(u,x^1,x^2)$ is also an affine parameter, but the condition $r'=0$ on the null surface will lead to $B=0$ and the condition $\ell_{a}k^{a}=-1$ for $\boldsymbol{k}=-\partial/\partial r'$ fixes the value of $A$ to be unity. Having chosen this affine parameter, any point in a sufficiently small neighbourhood of the null surface (so that the null geodesics do not cross) can be assigned the coordinates $(u,r,x^1,x^2)$, where the set $(u,x^1,x^2)$ corresponds to the label of the null geodesic that passes through that point while $r$ represents the affine parameter value at that point.   

Let us now turn our attention to the form of the metric in the neighbourhood of the null surface. We shall start by writing down the form of the metric on the null surface. Since $\boldsymbol{\ell}$ is null everywhere on the null surface, we can write down our first metric component: $g_{uu}=g_{ab}\ell^{a}\ell^{b}=0,$ everywhere on the null surface. Next, since the null surface is represented by $r=\textrm{constant}$, $\textbf{X}_A=\partial/ \partial x^{A}$ at the null surface has to lie on the null surface. This means that $X^{a}_{A}\ell_{a}=0$, which implies $g_{ab}X^{a}_{A} \ell^{b}=g_{uA}=0$ everywhere on the null surface. Finally, denoting the rest of the components relevant for vectors on the null surface, $g_{AB}=g_{ab}X^{a}_{A}X^{b}_{B}$, by $q_{AB}$, we have the following list of components of the metric on the null surface:
\begin{equation}\label{GNC_on_null}
g_{uu}=0; ~ g_{uA}=0; ~ g_{AB}=q_{AB}; 
\end{equation}

Let us next determine the metric components in the neighbourhood of the null surface that we are considering. First, since $\boldsymbol{k}= -\partial/\partial r$ is null everywhere, we have $g_{ab}k^{a}k^{b}=g_{rr}=0$ everywhere. To find the rest of the components, let us write down the geodesic equation governing the null geodesics that are integral curves of $\boldsymbol{k}$. Since $r$ is an affine parameter, we have
\begin{equation}
k^{a} \nabla_a k^{b} =0 \Longrightarrow \Gamma^{b}_{rr}=0 \longrightarrow \Gamma_{arr}=0,
\end{equation}
where we contracted with $g_{ab}$ in the last step. Putting $a=r$ tells us that $g_{rr}$ remains zero along the geodesics. But if we put $a=\mu$ where $x_{\mu}$ is any one of the three surface coordinates $(u,x^1,x^2)$, we have
\begin{equation}\label{eq3}
\Gamma_{\mu rr}=\frac{1}{2}\left( -\partial_{\mu}g_{rr}+ 2\partial_{r}g_{r\mu}\right) = \partial_{r}g_{r\mu}=0~. 
\end{equation}
Thus, $g_{r \mu}$ is also constant along the geodesics. (This is a common feature for affinely parametrized geodesics. For null geodesics, if $\partial/ \partial \phi$ is the tangent vector with $\phi$ being the affine parameter, then we shall have $\partial_{\phi}g_{\phi a}=0$. For non-null geodesics, we can get the same constraint provided $\phi$ is taken as the proper time or proper length or a constant multiple thereof, so that we can put $\partial_{\mu}g_{\phi \phi}=0$ as in the last step above.) To fix the values of $g_{r \mu}$, note that we have the constraints $k^{a}\ell_{a}=-1$ and $k_{a}X^{a}_{A}=0$ at $r=0$, which translate into $g_{ru}=1$ and $g_{rA}=0$. From \ref{eq3}, we conclude that $g_{r u}=1$ and $g_{rA}=0$ everywhere in the region under consideration. Finally, defining $q_{AB}=g_{ab}X^{a}_{A}X^{b}_{B}$, with the definition $X_{A}=\partial/\partial x^{A}$ extended everywhere, and respecting \ref{GNC_on_null}, we have all the components of the metric as follows:
\begin{equation}
g_{rr}=g_{rA}=0;~g_{ru}= 1;~g_{uu}=-2 r \alpha;~g_{uA}=-r \beta_{A};~g_{AB}=q_{AB}~,
\end{equation}
where $\alpha$ and $\beta^{A}$ are finite at $r=0$ so that $g_{uu}$ and $g_{ru}$ vanish at $r=0$. The  $2\times2$ matrix $q_{AB}$ is positive definite (since $\textbf{X}_{A}$ are spacelike vectors). The choice of the sign for $g_{uA}$ is not of importance as the sign can be flipped by changing the coordinates on the spatial surface, $x^{A} \rightarrow -x^{A}$. On the other hand, the sign of $g_{uu}$ does have a physical significance. With this particular choice, the vector $\ell^{a}$ becomes timelike for $r>0$ and spacelike for $r<0$ when $\alpha>0$. Also, 
$$g^{rr}=2 r \alpha + r^2 \beta^2,$$ where $\beta^2= q_{AB} \beta^{A} \beta^{B}$, which is a positive quantity since $q_{AB}$ is positive definite. Thus, the normal to $r$-constant surfaces, $\partial_a r \equiv r_{a}$, satisfies the condition $$r^{a}r_{a}=2 r \alpha + r^2 \beta^2.$$ The second term is strictly positive. But the first term dominates as $r \rightarrow 0$. With $\alpha>0$, $r$-constant surfaces near $r=0$ are time-like in the $r>0$ region and space-like in the $r<0$ region. In fact, all $r$-constant surfaces in the $r>0$ region become time-like by this choice.

So, finally, here is the line element in GNC:
\begin{equation}\label{GNCline}
ds^{2}=-2r\alpha du^{2}+2drdu-2r\beta _{A}dudx^{A}+q _{AB}dx^{A}dx^{B}
\end{equation}
Note that there are six independent functions $\left(\alpha, \beta_{A},q_{AB}\right)$ in this metric. We have used up the freedom of choosing the four coordinates to eliminate four out of the ten degrees of freedom in the metric. [The inverse metric and the Christoffel symbols corresponding to this metric are given in \ref{GNCMetric}. The Ricci tensor components are provided in \cite{Moncrief:1983}.] Introducing a time coordinate by $t=u+r$ in place of $u$, we can rewrite this metric in the standard ADM form \cite{Arnowitt:1962hi} as,
\begin{align}
ds^{2}&=-N^{2}dt^{2}-2\left(1+r\alpha \right)\left(dr-N^{r}dt\right)\left(dr-N^{r}dt\right)
\nonumber
\\
&+2r\beta _{A}\left(dr-N^{r}dt\right)\left(dx^{A}-N^{A}dt\right)
+q_{AB}\left(dx^{A}-N^{A}dt\right)\left(dx^{B}-N^{B}dt\right)
\\
N^{r}&=\frac{1+2r\alpha +r^{2}\beta ^{2}}{-2-2r\alpha +r^{2}\beta ^{2}}
\\
N^{A}&=-r\beta ^{A}\left[\frac{3+4r\alpha}{-2-2r\alpha +r^{2}\beta ^{2}} \right]
\\
N^{2}&=2r\alpha +2r\beta _{A}N^{r}N^{A}-2(1+r\alpha)\left(N^{r}\right)^{2}+q_{AB}N^{A}N^{B}
\end{align}
This expression shows that the relationship between the GNC metric components and the standard ADM variables $(N,N_\alpha, h_{\alpha\beta})$ is not simple. In particular the degrees of freedom in $
h_{\alpha\beta}$ comes in entangled in terms of the other degrees of freedom in GNC variables.
\subsubsection{Surface Term on a Null Surface as a limit in GNC} \label{sec:ST-GNC-null_limit}
	
	Gaussian null coordinates (GNC) provide a situation where $r=\textrm{constant}$ surfaces are 
	time-like for $r>0$ and null for $r=0$. In order to show the validity of our previous derived result for null surfaces, \ref{NewLabel03}, we will obtain the boundary term on the null surface as the limit of the boundary term on a time-like surface, providing a specific example of the abstract manipulations carried out in \ref{ADMtoNull}. Consider the following expression 
	for the term to be integrated on the boundary of a time-like or 
	space-like surface as given in \ref{S.T-nonnull-3}:
	\begin{equation}\label{GNCBegin}
	Q\sqrt{|h|} = \sqrt{|h|} D_{a}(\delta \un^{a})+ \delta (2 K \sqrt{|h|})-\sqrt{|h|}(K_{ij}-K h_{ij})\delta h^{ij},
	\end{equation}
	where $Q$ as given in \ref{q} in \ref{BC-novel} has to be evaluated for the \textit{unit} normal. 
	We shall use the object $P^{ab}=-\sqrt{|h|}(K^{ab}-K h^{ab})$, which, apart from a constant factor, represents the canonical momentum conjugate to $h_{\alpha \beta}$ in ADM formalism \cite{gravitation}.
	Then, the above equation will take the form
	\begin{equation}
	Q \sqrt{|h|} = \sqrt{|h|} D_{a}(\delta \un^{a})+ \delta (2 K \sqrt{|h|})+P _{ij}\delta h^{ij}
	\end{equation}
	Further, we have the following relations
	\begin{equation}
	P ^{ab}h_{ab} = 2 \sqrt{|h|} K; ~~  P _{ab} \delta h^{ab}=-P ^{ab} \delta h_{ab}
	\end{equation}
	Thus, we can also write the boundary term $Q\sqrt{|h|}$ using the above expressions 
	for $P^{ab}h_{ab}$ and $P_{ab} \delta h^{ab}$ leading to
	\begin{equation}
	Q \sqrt{|h|} = \sqrt{|h|} D_{a}(\delta \un^{a})+ h_{ij}\delta P ^{ij}
	\end{equation}
	The idea is to evaluate this term for an $r=\textrm{constant}$ surface with $r>0$, which corresponds 
	to a timelike surface, and then take the $r \rightarrow 0$ limit. 

We shall first assume that the variations preserve the GNC form, i.e, we will only vary the functions present in the GNC metric, viz. $(\alpha, \beta_A, q^{AB})$. Further since the metric only has the combination $r\alpha$ and $r\beta_A$ occurring in it,  the variations $ r\delta \alpha,r\delta\beta_A $  will vanish in the $r\rightarrow 0$ limit. So we are essentially restricting ourselves to variations with $\delta \ell^a=0, \delta q^{AB}\neq0$ at this stage. (We will describe a more general situation later on.)

When the evaluation of $h_{ab}\delta P^{ab}$ is carried out and the limit 
	$r\rightarrow0$ is taken carefully, we get the result
	\begin{equation}
	-h_{ab}\delta (\sqrt{|h|}(K^{ab}-Kh^{ab})) = \partial_u(\sqrt{q}\frac{\delta \alpha}{\alpha})
	+\sqrt{q}[-2 \delta \alpha - 
	\frac{1}{2} \delta q^{AB} \partial_u  q_{AB}-q^{AB}\partial_u(\delta q_{AB})], \label{eq62}
	\end{equation}
	where $q$ is the determinant of the $2$-metric $q_{AB}$. The first term in \ref{eq62} precisely cancels the $\sqrt{|h|} D_{a} (\delta \un^{a})$ term so that we readily obtain
	\begin{equation}
	Q \sqrt{|h|} = \sqrt{q}[-2 \delta \alpha - \frac{1}{2} \delta q^{AB} \partial_u  q_{AB}-q^{AB}\partial_u(\delta q_{AB})] \label{qrGNC-old}
	\end{equation}
	This result may be rewritten in the following form:
	\begin{equation}
	Q \sqrt{|h|} = -2 \sqrt{q} \delta \alpha - \frac{1}{2} \sqrt{q} q^{AB}
	\partial_u \delta q_{AB}-\sqrt{q} \partial_u [\delta (ln \sqrt{q})], 
	\label{qrGNC}
	\end{equation}
	We can also derive the same result starting from \ref{GNCBegin}. Under the variations that we are considering, each of the terms in \ref{GNCBegin} have the following limits as $r\rightarrow0$:
	\begin{eqnarray}
	\sqrt{|h|}D_{a}\left(\delta \un^{a}\right)&=&-\partial _{u}\left(\sqrt{q}\frac{\delta \alpha}{\alpha}\right) \\
	\delta \left(2K\sqrt{|h|}\right)&=&\delta \left(2\partial _{u}\sqrt{q}+2\alpha \sqrt{q}
	-\sqrt{q}\frac{\partial _{u}\alpha}{\alpha}\right) \\
	\sqrt{|h|}(K_{ij}-K h_{ij})\delta h^{ij}&=&\frac{\delta \alpha}{\alpha}\partial _{u}\sqrt{q}
	+\frac{1}{2}\sqrt{q}\delta q ^{CD}\partial _{u}q _{CD}
	\nonumber
	\\
	&&+\frac{2}{\sqrt{q}}\delta \sqrt{q}\partial _{u}\sqrt{q}
	+2\delta \sqrt{q}\left(\alpha -\frac{1}{2}\frac{\partial _{u}\alpha}{\alpha}\right)
	\end{eqnarray}
	The difference of the last two terms can be simplified and written as
	\begin{eqnarray}
	\delta (2 K \sqrt{|h|})-\sqrt{|h|}(K_{ij}-K h_{ij})\delta h^{ij}
	&=&-2\delta \alpha \sqrt{q}+\partial _{u}\left(\sqrt{q}\frac{\delta \alpha}{\alpha}\right)
	\nonumber
	\\
	&&-\frac{1}{2}\sqrt{q}q^{AB}\partial _{u}\delta q _{AB}-\sqrt{q}\partial _{u}\left(\delta \ln \sqrt{q}\right)
	\end{eqnarray}
	Combining with the total derivative term, we again arrive at \ref{qrGNC} as the null limit for \ref{GNCBegin}:
	\begin{eqnarray}
	Q\sqrt{|h|} &=& \sqrt{|h|} D_{a}(\delta \un^{a})+ \delta (2 K \sqrt{|h|})-\sqrt{|h|}(K_{ij}-K h_{ij})\delta h^{ij},
	\nonumber
	\\
	&=&-2\sqrt{q}\alpha
	-\frac{1}{2}\sqrt{q}q ^{AB}\partial _{u}\delta q _{AB}-\sqrt{q}\partial _{u}\left(\delta \ln \sqrt{q}\right)
	\end{eqnarray}
	For the Gaussian null coordinates, 
	we choose the normal $v_a$ to be $\ell_{a}=\partial_a r$, the surface gradient to the $r=0$ surface. 
	Then, the normal vector has the following elements in $(u,r,x^1,x^2)$ coordinates:
	\begin{eqnarray}\label{NewLabel01}
	\ell_{a} = (0, 1, 0, 0) ;\qquad \ell^{a} = g^{ar}
	\end{eqnarray}
	(Note that this is not the $\boldsymbol{\ell}$ we used in \ref{GNC} in constructing the GNC metric.) The auxiliary vector (see \ref{app:q_for_null}), defined by the two conditions $\ell_{a}k^{a}=\ell^{a}k_{a}=-1$ and $k_{a}k^{a}=0$ on the null surface, can be chosen to be $k_{a} = -g_{ar}$ and $k^{a}=(0, -1, 0, 0)$. Then, we shall have the induced metric on the null surface defined as, 
	$q_{ab}=g_{ab}+\ell_{a}k_{b}+\ell_{b}k_{a}$. The explicit form of the induced metric 
	on the $r=0$ surface is
	\begin{equation}
	q_{ab} = q _{AB}\delta ^{A}_{a}\delta ^{B}_{b},\qquad q^{ab}=q ^{AB}\delta ^{a}_{A}\delta ^{b}_{B}  ,              
	\end{equation}
	while $q^{a}_{b}=diag(0,0,1,1)$. Next, let us look at the second fundamental form for the null surface (see \ref{app:SFF+exp}):
	\begin{equation}
	\Theta _{ab}=q^{m}_{a}q^{n}_{b}\nabla _{m}\ell_{n} 
	\end{equation}
	Calculating $\Theta_{ab}$ for our case, we find that the non-zero components are
	\begin{equation}
	\Theta_{AB}= \frac{1}{2} \partial_u q_{AB},
	\end{equation}
	so that the contraction of the above tensor $\Theta _{ab}$ leads to the the trace
	\begin{equation}
	\Theta = \frac{1}{2} q^{AB} \partial_{u} q_{AB}~.
	\end{equation}
	Thus, we get the following expression for $(\Theta_{ab}- \Theta q_{ab}) \delta q^{ab}$ in GNC: 
	\begin{equation}
	(\Theta_{ab}- \Theta q_{ab}) \delta q^{ab} = \frac{1}{2} \partial_u q_{AB} \delta q^{AB} + \frac{2}{q} \partial_{u}(\sqrt{q})\delta(\sqrt{q})
	\end{equation}

	Comparing this result with \ref{qrGNC}, we find that, for GNC, the surface term is expressible 
	in terms of the induced quantities $\Theta _{ab}$ and $\Theta$ in the following fashion:
	\begin{eqnarray}
	\delta \mathcal{A}_{null} &=& \int du d^{2}x\left[-2 \sqrt{q} \delta \alpha -\delta (2 \sqrt{q} \Theta) +(\Theta_{ab} - \Theta q_{ab})\delta q^{ab} \sqrt{q}\right] \nn\\
	&=&\int du d^{2}x\left[ - \delta [2 \sqrt{q}(\Theta + \alpha)] + [\Theta_{AB}- (\Theta + \alpha)q_{AB}]\delta q^{AB} \sqrt{q}\right] \label{ST-GNC}
	\end{eqnarray}

	Therefore, for the variations we have considered,  we only need to fix $q _{AB}$ on the surface, by construction. Even though we varied the six components in the GNC metric, $\alpha$ and $\beta ^{A}$ appear in the combination $r\alpha$ and $r\beta ^{A}$. As the null surface remains $r=0$ even after variation, when we take the limit $r\rightarrow 0$ contributions from $\df \alpha$ and $\df \beta ^{A}$ vanish. Since the components of $\ell^a$ is given by \ref{NewLabel01}, this corresponds to variations with $\delta\ell^a=0$ and our general result in \ref{NewLabel03} shows that we only need to fix $q _{AB}$ on the surface. So everything is consistent. In order to get the full structure of \ref{NewLabel03}, we need to consider unconstrained variations of the GNC line element, which we shall take up next.

\subsubsection{Surface Term in GNC for Unconstrained Variations}
	
	We shall now consider the null surface described in the GNC coordinates but allow the variations to be arbitrary but finite. That is, the variations are not restricted to be of GNC form. We shall start from the expression for the surface term in terms of the variation of the Christoffel symbols given in \ref{surf_term}, \ref{BC-novel}:
	\begin{equation}\label{surf_term1}
	\sqrt{-g} Q [v_c] = \sqrt{-g} v_{c} (g^{ab} \delta \Gamma^{c}_{ab}-g^{ck} \delta \Gamma^{a}_{ak})~. 
	\end{equation}
	Substituting the base metric as the GNC metric and $v_{c}=\partial_c r$, we obtain
	\begin{equation}
	\sqg Q = \sqrt{q}\left(\delta \Gamma^{r}_{ru}+ q^{AB} \delta \Gamma^{r}_{AB}-\delta \Gamma^{u}_{uu}-\delta \Gamma^{A}_{Au} \right)
	\end{equation}
	Evaluating the $\Gamma$s for arbitrary variations from the GNC metric, we obtain the following expansion:
	\begin{align} \label{surf_term2}
	\sqg Q =& \sqrt{q} \left[\partial_r (\delta g_{uu}) - \frac{1}{2}\partial_u q_{AB} \delta q^{AB} - q^{AB}\partial_{u}(\delta q_{AB}) \right.\nn\\ &- \frac{\delta g^{rr}}{2}q^{AB}\partial_{r}q_{AB}-\partial_{u}(\delta g_{ur})\nn \\ &+\delta g^{ru} \left(- 2\alpha - \frac{q^{AB}}{2}\partial_{u}q_{AB}\right)\nn \\ &+ \delta g^{rC}\left(-\beta_{C}-\frac{1}{2}q^{AB}\partial_C q_{AB}+q^{AB}\partial_{A}q_{BC} \right)\nn \\ &+ \left.q^{AB}\partial_{A}(\delta g_{uB}) \right] 
	\end{align}
	It will be fruitful to convert the variations of the inverse metric into variations of the metric. For example,
	\begin{align}
	\delta g^{rr} = -g^{ra}g^{rb}\delta g_{ab} 
	\eqH -g^{ru}g^{ru}\delta g_{uu} 
	= -\delta g_{uu}~.
	\end{align}
	Similarly, we can obtain
	\begin{align}
	\delta g^{ru}\eqH& - \delta g_{ru} \\
	\delta g^{rC}\eqH& -q^{AC} \delta g_{uA}
	\end{align}
	Using these expressions in \ref{surf_term2} and manipulating the terms, we can obtain
	\begin{align} \label{surf_term3}
	\sqg Q =& \partial_{A}\left( \sqrt{q}q^{AB} \delta g_{uB}\right) - \partial_u (\sqrt{q} \delta g_{ur}) \nn \\
	&+ \delta (\sqrt{q}~ \partial_r g_{uu}) + 2 \delta(g_{ur} \partial_u \sqrt{q})+ \delta[2\sqrt{q} ~\alpha (g_{ru}-1)]\nn \\
	&+ \delta g_{uu}\partial_{r}(\sqrt{q})+ \sqrt{q}~\beta^{B} \delta g_{uB}+ \alpha \sqrt{q}~ q^{AB}\delta q_{AB}- \frac{\sqrt{q}}{2}\partial_{u}q_{AB} \delta q^{AB} - \sqrt{q}~q^{AB}\partial_{u}(\delta q_{AB})~.
	\end{align}
	
	We will now show that this result matches with the one in \ref{NewLabel03}. To do this, let us start with the following expression:
\begin{align}
\Theta +\kappa &=\nabla _{a}\ell ^{a}+\frac{1}{2}k^{a}\partial _{a}\ell ^{2}
\nonumber
\\
&=\nabla _{a}\ell ^{a}+\frac{1}{2}k^{a}\partial _{a}g^{rr}
\end{align}
Varying it, we arrive at:
\begin{align}
2\delta \left(\Theta +\kappa \right)&=2\delta \left(\nabla _{a}\ell ^{a}\right)
+\delta k^{a}\partial _{a}g^{rr}+k^{a}\partial _{a}\delta g^{rr}
\nonumber
\\
&=2\partial _{u}\delta g^{ur}+\partial _{r}\delta g^{rr}+2\partial _{A}\delta g^{rA}
+2\delta g^{ur}\partial _{u}\ln \sqrt{-g}+2\delta g^{rr}\partial _{r}\ln \sqrt{-g}
\nonumber
\\
&+2\delta g^{rA}\partial _{A}\ln \sqrt{-g}+2\partial _{u}\delta \ln \sqrt{-g}
-2\alpha \delta g^{ur}
\nonumber
\\
&=-2\partial _{u}\delta g_{ur}-\partial _{r}\delta g_{uu}-2\alpha \delta g_{ur}-2\beta ^{A}\delta g_{uA}
-2\partial _{A}\left(q ^{AB}\delta g_{uB}\right)-\delta g_{ur}\left(q ^{AB}\partial _{u}q _{AB}\right)
\nonumber
\\
&-\delta g_{uu}\left(q ^{AB}\partial _{r}q _{AB}\right)-q ^{AB}\delta g_{uB}\left(q ^{CD}\partial _{A}q_{CD}\right)+2\partial _{u}\delta \ln \sqrt{-g}
\end{align}
Using this result, the variation of the counter-term becomes
\begin{align}
2\delta \left[\sqrt{-g}\left(\Theta +\kappa \right) \right]&=
2\sqrt{-g}\delta \left(\Theta +\kappa \right)+2\sqrt{-g}\left(\Theta +\kappa \right)\delta g_{ur}
-\sqrt{-g}\left(\Theta +\kappa \right)q _{AB}\delta q ^{AB}
\nonumber
\\
&=\sqrt{-g}\Big[-2\partial _{u}\delta g_{ur}-\partial _{r}\delta g_{uu}
-2\partial _{A}\left(q ^{AB}\delta g_{uB}\right)
-\delta g_{ur}\left(2\alpha +2\Theta +2\kappa +q ^{AB}\partial _{u}q _{AB}\right)
\nonumber
\\
&-\delta g_{uu}q ^{AB}\partial _{r}q _{AB}-\delta g_{uB}\left(2\beta ^{B}+q ^{AB}q ^{CD}\partial _{A}q_{CD}\right)+2\partial _{u}\delta \ln \sqrt{-g}-\left(\Theta +\kappa \right)q _{AB}\delta q ^{AB}
\Big]
\end{align}
Adding to \ref{surf_term3}, we get:
\begin{align}
\sqrt{-g}Q+2\delta \left[\sqrt{-g}\left(\Theta +\kappa \right) \right]&=
-3\sqrt{-g}\partial _{u}\delta g_{ur}-\partial _{A}\left(q ^{AB}\delta g_{uB}\right)
-\delta g_{ur}\left(\frac{1}{2}q ^{AB}\partial _{u}q _{AB}+2\Theta +2\kappa \right)
\nonumber
\\
&-\delta g_{uu}\left(\frac{1}{2}q ^{AB}\partial _{r}q _{AB}\right)
-\beta ^{B}\delta g_{uB}-\frac{1}{2}q ^{AB}\delta g_{uB}q ^{CD}\partial _{A}q _{CD}
\nonumber
\\
&+2\partial _{u}\delta \ln \sqrt{-g}-\left(\Theta +\kappa \right)q _{AB}\delta q ^{AB}
-\Theta _{AB}\delta q ^{AB}-q ^{AB}\partial _{u}\delta q _{AB}
\nonumber
\\
&=\left[\Theta _{AB}-\left(\Theta +\kappa \right)q _{AB}\right]\delta q ^{AB}
-2\sqrt{-g}\partial _{u}\delta g_{ur}-\sqrt{-g}\partial _{A}\left(q ^{AB}\delta g_{uB}\right)
\nonumber
\\
&-\sqrt{-g}\Big[\delta g_{ur}\left(3\Theta +2\kappa \right)+\delta g_{uu}\left(\frac{1}{2}q ^{AB}\partial _{r}q _{AB}\right)+\delta g_{uB}\left(\beta ^{B}+\frac{1}{2}q ^{AB}q ^{CD}\partial _{A}q _{CD} \right)
\nonumber
\\
&-2\partial _{u}\delta \ln \sqrt{-g}-\partial _{u}\left(q^{AB}\delta q _{AB}\right)\Big]
\nonumber
\\
&=\left[\Theta _{AB}-\left(\Theta +\kappa \right)q _{AB}\right]\delta q ^{AB}
-\partial _{u}\left(\sqrt{-g}\delta g_{ur}\right)-\partial _{A}\left(\sqrt{-g}q ^{AB}\delta g_{uB}\right)
\nonumber
\\
&-2\sqrt{-g}\left(\Theta +\kappa \right)\delta g_{ur}-\sqrt{-g}\delta g_{uu}\left(\frac{1}{2}q ^{AB}\partial _{r}q _{AB}\right)-\sqrt{-g}\beta ^{B}\delta g_{uB}
\end{align}
In this expression the total divergence term, the counter-term involving $\Theta +\kappa$ and one involving the variation with $\left[\Theta ^{ab}-(\Theta +\kappa)q^{ab} \right]$ are easy to interpret and we have seen them earlier. The  extra term which appears, in addition to these, has the  following form:
\begin{align}
Q_{extra}&=+2\sqrt{-g}\left(\Theta +\kappa \right)\delta g^{ur}+\sqrt{-g}\delta g^{rr}\left(\frac{1}{2}q ^{AB}\partial _{r}q _{AB}\right)+\sqrt{-g}\beta _{B}\delta g^{rB}
\end{align} 
These terms have the variation of  $\delta \ell ^{a}=\delta g^{ar}$. One can also show that the coefficient of $\delta \ell ^{a}$ matches with the one in our general expression derived earlier in \ref{NewLabel03}. Hence the above result can be taken as a verification of our general result. If we had restricted our variations to GNC parameters only, then we would have $\delta g^{rB}= r\delta \beta ^{B}=0$, $\delta g^{rr}=2r\delta \alpha =0$ and $\delta g^{ur}=0$. Hence we would retrieve our earlier result that only $q ^{AB}$ needs to be fixed. 
	
\subsection{Null Surface Foliation (NSF)}\label{NS}

In this section, we introduce another parametrization for the spacetime metric near a null surface. The fiducial null surface is now taken to be a member of a set of null surfaces, unlike the case for GNC where \textit{only} the $r=0$ surface in the family of $r=$constant surfaces was taken to be null. We shall call the resulting form of the metric the null surface foliation (NSF) metric. Here we shall give the main outline of its construction. The details can be found in \cite{Jezierski:2000,Padmanabhan:2010rp}.   

\subsubsection{Construction} \label{NSconstr}
Consider a four-dimensional spacetime manifold specified by $\left(M,g_{ab}\right)$. Let our fiducial null surface be one member of a family of null hypersurfaces in this spacetime. Using the fact that null surfaces are spanned by null geodesics, it is possible to introduce a natural coordinate system adapted to our family of null hypersurfaces in the following manner. We first select one of the co-ordinates, say $x^{3}$, such that $x^{3}=\textrm{constant}$ represents the set of null surfaces with a choice of $x^{3}=0$ on $S$, our fiducial null surface. Then, we choose a spacelike Cauchy surface, $\Sigma _{t}$, and denote the intersection of $S$ with $\Sigma _{t}$, a two dimensional surface, by $S_{t}$. Let us define the coordinates on this surface as $x^{A}\equiv (x^{1},x^{2})$, with the two corresponding basis vectors lying on $S_t$ denoted by $\boldsymbol{e}_A=\partial/\partial x_A$ (with the foreknowledge that the fourth coordinate will be chosen to be constant on $S_t$). At every point $P$ on $S _{t}$, there are exactly two future-pointing null directions orthogonal to it, among which one direction will lie on $S$. Let $\boldsymbol{\ell}$ denote a null vector field tangent to this direction at every point on $S_t$. Thus, $\boldsymbol{\ell}$ will satisfy the relation $\boldsymbol{\ell}.\boldsymbol{e}_A=0$. The same exercise can be repeated replacing $S$ with every other $x_3=\textrm{constant}$ surface so that we will have the coordinates, $x_1$ and $x_2$ (as well as $x_3$, which has already been defined all over spacetime), and the vector field $\boldsymbol{\ell}$ defined all over $\Sigma_t$. Within this arrangement, we can construct a coordinate system near $\Sigma _{t}$ in the following way: (a) Take the coordinates $x^{1},x^{2},x^{3}$ as constant all along the null geodesics passing through every point $P(x^{1},x^{2},x^{3})$ on $\Sigma _{t}$ and moving in the direction of $\ell^{a}$ and (b) take $t$ to be an affine parameter distance along these geodesics with, say, $t=0$ on $\Sigma_t$. Let us order our four coordinates as $(t,x^1,x^2,x^3)$. In this coordinate system, the null vector is given by $\ell^a = (1,0,0,0)$. Thus, the null condition $\ell^2=0$ implies $g_{tt}=0$. On the other hand, the geodesic condition with affine parametrization, $\ell^b\nabla_b\ell^a = 0$, gives $g^{ab}\partial_t g_{tb} = 0$; i.e. $\partial_tg_{tb}=0$. In other words, $g_{tb}$ remains constant along the null geodesics. Since $\mathbf{\boldsymbol{\ell}\cdot e_{A}}=0$ on the initial surface, we must have $g_{tA} = 0$ everywhere. Thus, the form of the line element near the null surface may be written as
\begin{equation}\label{NSline}
ds^{2}=-N^{2}dt^{2}+\left(\frac{M dx^{3}}{N}+\epsilon Ndt \right)^{2}+ q_{AB}\left(dx^{A}+m^{A}dx^{3}\right)\left(dx^{B}+m^{B}dx^{3}\right)~,
\end{equation}
with $\epsilon = \pm 1$. The $t=\textrm{constant}$ surfaces are taken as a spacelike surfaces. The $x^3 =$constant surfaces are null with the metric given by $ds^2 = q_{AB}dx^Adx^B$. The inverse metric and Christoffel symbols of this metric are given in \ref{NSapp}.

This metric may be rewritten in the following condensed form:
\begin{equation} \label{NSline-short}
ds^2 = \widetilde{M}^2 (dx^{3})^2 + 2 M \epsilon dt dx^3 + 2 q_{AB} m^A dx^B dx^3 + q_{AB} dx^A dx^B,
\end{equation}
with $\widetilde{M}^2= M^2 / N^2 + q_{AB} m^A m^B$.
This metric has seven parameters, one more than the GNC metric in \ref{GNCline}. Since it is possible to eliminate four metric coefficients out of the ten independent coefficients required to parametrize a symmetric four-dimensional metric by using the four coordinate transformations, there must be a coordinate transformation that we could make to reduce the number of parameters to six. This coordinate transformation has been carried out in \ref{KillM} leading to
\begin{eqnarray} \label{NSline-short-transf}
ds^2 &=& (\widetilde{M}^2- 2 \epsilon \bar{t}~ \frac{\partial \ln M}{\partial x^3}) (dx^{3})^2 + 2 \epsilon d \overline{t} dx^3 + 2 (q_{AB} m^A- \epsilon \ \overline{t}~ \frac{\partial \ln M}{\partial x^B}) dx^B dx^3 + q_{AB} dx^A dx^B \nn\\
& = &\overline{M}^2 (dx^{3})^2 + 2  \epsilon d\overline{t} dx^3 + 2 q_{AB} \overline{m}^A dx^B dx^3 + q_{AB} dx^A dx^B
\end{eqnarray}
Another way to get rid of $M$ is to go over the construction of the coordinate system once again and note that the condition $\partial_tg_{t3}=0$ out of the four conditions $\partial_tg_{ta}=0$ has not been used. Now, once the coordinate system $(x_1,x_2,x_3)$ has been defined on $\Sigma_t$ and we have decided that $\Sigma_t$ is a $t=\textrm{constant}$ surface for the fourth coordinate $t$, we have the basis vector $\boldsymbol{e}_3 =\partial/\partial x_3$ corresponding to the $x_3$ coordinate at every point on $\Sigma_t$. We can then demand $\boldsymbol{\ell}.\boldsymbol{e}_3=1$ everywhere on $\Sigma_t$. The four conditions, $\boldsymbol{\ell}.\boldsymbol{\ell}=0$, $\boldsymbol{\ell}.\boldsymbol{e}_3=1$ and $\boldsymbol{\ell}.\boldsymbol{e}_A=0$, fixes $\boldsymbol{\ell}$ uniquely. Then, we shall have $M=g_{t3}=\boldsymbol{\ell}.\boldsymbol{e}_3=1$ on $\Sigma_t$ and, by virtue of $\partial_tg_{t3}=0$, everywhere else. Thus, from now on, we will assume that the parameter $M=1$ and do the rest of the calculations. The final form of the metric we shall use is
\begin{equation}\label{NSline1}
ds^{2}=-N^{2}dt^{2}+\left(\frac{ dx^{3}}{N}+\epsilon Ndt \right)^{2}+ q_{AB}\left(dx^{A}+m^{A}dx^{3}\right)\left(dx^{B}+m^{B}dx^{3}\right)~,
\end{equation}
which on expansion becomes
\begin{equation} \label{NSline-short2}
ds^2 = \widetilde{M}^2 (dx^{3})^2 + 2 M \epsilon dt dx^3 + 2 q_{AB} m^A dx^B dx^3 + q_{AB} dx^A dx^B,
\end{equation}
with $\widetilde{M}^2= 1/ N^2 + q_{AB} m^A m^B$.


\subsubsection{Surface Term in the NSF Metric under constrained variation} 

In this case, with the coordinates $(t,x^1,x^2,x^3)$, we shall take the normal to the null surface as
\begin{equation}
\ell_{a} = \partial_{a}x_{3}=(0, 0, 0, 1); ~~ \ell^{a} = (\epsilon, 0, 0, 0).
\end{equation}
We shall choose the auxiliary vector
\begin{equation}
k^{a}=(\frac{\bar{M}^2}{2 \epsilon}, 0, 0, -1);~~k_{a}=(- \epsilon , -m_1, -m_2, -\frac{\bar{M}^2}{2} ),
\end{equation}
where we have defined $\bar{M}^2=(\frac{1}{N^2}+m^2)$. One can verify that $k^{a}k_{a}=0$ and $\ell^{a}k_{a}=-1$. Then, we have
\begin{equation}
q_{ab}dx^{a}dx^{b} = q_{AB}dx^{A}dx^{B}     
\end{equation}
and
\begin{equation}
q^{ab}dx_{a}dx_{b}=m^{2}dx_{0}^{2}-2\epsilon m^{A}dx_{0}dx_{A}+q^{AB}dx_{A}dx_{B}                        
\end{equation}
where $q^{AB}$ is the inverse of $q_{AB}$. Finally, we have the mixed form we have
\begin{equation}
q^{a}_{b}dx_{a}dx^{b} = -\epsilon m_{B}dx_{0}dx^{B}+\delta ^{A}_{B}dx_{A}dx^{B}
\end{equation}
The non-zero components of $\Theta_{ab}$ are
\begin{equation}
\Theta_{AB}= \frac{\epsilon}{2M}\partial_{t}q_{AB},
\end{equation}
and 
\begin{equation}
\Theta = \left(\frac{\epsilon}{M}\right) \frac{1}{\sqrt{q}}\partial_{t} \sqrt{q},
\end{equation}
where $q$ is the determinant of the $2$-metric $q_{AB}$. We shall now write down the surface term. The surface term, $\sqg\,Q[\ell_{c}]$, can be decomposed using \ref{qgen}. The different terms in this expansion give:
\begin{align}
\sqg \nabla_{a}[\delta \ul^{a}]= -\partial_{t}\left( \epsilon \sqrt{q} ~\frac{\delta M}{M}\right) \label{q1NS} \\
-\delta \left(2 \sqrt{-g} ~\nabla_{a}\ell^{a} \right)= -\delta \left( 2M \sqrt{q}~ \Theta \right)= -\delta \left[2\partial_{t}\left( \epsilon \sqrt{q}\right) \right] \label{q2NS}\\
\sqg ~\nabla_{a}\ell_{b} \delta g^{ab} = M \sqrt{q}~ \Theta_{AB} \delta q^{AB} \label{q3NS} \\
-\sqg g_{ab} \nabla_{c}\ell^{c} \delta g^{ab} = - \sqg \Theta \left[ 2 \epsilon M \delta \left(\frac{\epsilon}{M}\right)+ q_{AB} \delta q^{AB}\right] \label{q4NS}
\end{align}
Our aim was to see if we can fix just $q^{AB}$ on the surface, just as we only need to fix $h^{\alpha \beta}$ on the surface for a spacelike or timelike surface. We have variations of $M$ in \ref{q4NS}. But we have already seen in \ref{NSconstr} that $M$ can be set to unity with a suitable choice of coordinates. Then, the surface term takes the form 
\begin{equation} \label{ST-NS-M1}
\delta \mathcal{A}_{null} = \intlt \left\lbrace - \delta [2 \sqrt{q} (\Theta+ \kappa)] + \sqrt{q} [\Theta_{AB} - (\Theta + \kappa) q_{AB}]\delta q^{AB}\right\rbrace  ~.
\end{equation}
We see that we get back the exact expression we had obtained for GNC in \ref{ST-GNC}, with $\alpha$ and $q_{AB}$ replaced by their corresponding quantities $\kappa$ and $q_{AB}$. 
Note that the variations we have considered sets $\delta \ell^a=0$ and hence we pick up only the remaining terms obtained earlier. To reproduce the terms involving $\delta\ell^a$ we need to consider a more general form of the variation, which we shall now take up.

\subsubsection{Surface Term in NSF for unconstrained variation}

We shall now take the original metric to be the NSF metric but allow the variations to be arbitrary, i.e. the variations are not restricted to those which preserve the original NSF form. We shall start from the expression for the surface term in terms of the variation of the Christoffel symbols given in \ref{surf_term} in \ref{BC-novel}:
\begin{equation}
\sqrt{-g} Q [v_c] = \sqrt{-g} v_{c} (g^{ab} \delta \Gamma^{c}_{ab}-g^{ck} \delta \Gamma^{a}_{ak})~. 
\end{equation}
Substituting the base metric as the GNC metric and $v_{c}=\partial_c x^{3}$, we obtain:
\begin{align}\label{Surf_NSF}
Q[v_{c}]&=g^{tt}\delta \Gamma ^{3}_{tt}+2g^{tA}\delta \Gamma ^{3}_{tA}+2g^{t3}\delta \Gamma ^{3}_{t3}
-g^{3t}\delta \Gamma ^{t}_{tt}-g^{3t}\delta \Gamma ^{A}_{At}-g^{3t}\delta \Gamma ^{3}_{3t}
+q^{AB}\delta \Gamma ^{3}_{AB}
\end{align}
Let us now evaluate each term individually and then put them together. For that, we have the following individual expressions:
\begin{align}
g^{tt}\delta \Gamma ^{3}_{tt}&=\frac{1}{2}\epsilon g^{tt}\partial _{t}\delta g_{tt}
\nonumber
\\
2g^{tA}\delta \Gamma ^{3}_{tA}&=-\epsilon m^{A}\left(\delta g^{33}\partial _{t}m_{A}
+\delta g^{3B}\partial _{t}q_{AB}+\epsilon \partial _{A}\delta g_{tt}\right)
\nonumber
\\
g^{t3}\delta \Gamma ^{3}_{t3}&=\frac{\epsilon}{2} \left[\delta g^{33}
\partial _{t}\left(m^{2}+\frac{1}{N^{2}}\right)+\delta g^{3A}\partial _{t}m_{A}+
\epsilon \partial _{3}\delta g_{tt}\right]
\nonumber
\\
q^{AB}\delta \Gamma ^{3}_{AB}&=\frac{1}{2}q^{AB}\Big[-\epsilon \partial _{t}\delta q_{AB}
+\delta g^{33}\left(-\partial _{3}q_{AB}+\partial _{A}m_{B}+\partial _{B}m_{A}\right)
\nonumber
\\
&+\delta g^{3C}\left(-\partial _{C}q_{AB}+\partial _{A}q_{CB}+\partial _{B}q_{CA}\right)
+\delta g^{3t}\left(-\partial _{t}q_{AB}\right)+ \epsilon \left(\partial _{A}\delta g_{tB}
+\partial _{B}\delta g_{tA}\right)\Big]
\nonumber
\\
\delta \Gamma ^{t}_{tt}&=\frac{1}{2}\left[-\frac{1}{N^{2}}\partial _{t}\delta g_{tt}
-\epsilon \partial _{3} \delta g_{tt}-\epsilon m^{A}
\left(-\partial _{A}\delta g_{tt}+2\partial _{t}\delta g_{At}\right)\right]
\nonumber
\\
\delta \Gamma ^{A}_{At}&=\frac{1}{2}\left[\delta g^{A3}\partial _{t}m_{A}
+\delta q^{AB}\partial _{t}q_{AB}
-\epsilon m^{A}\partial _{A}\delta g_{tt}+q^{AB}\partial _{t}\delta q_{AB}\right]
\nonumber
\end{align}
Then, substitution of these results in \eqref{Surf_NSF} leads to the following expression:
\begin{align}
Q[v_{c}]&=\frac{\epsilon}{2}g^{tt}\partial _{t}\delta g_{tt}
-\epsilon m^{A}\left(\delta g^{33}\partial _{t}m_{A}
+\delta g^{3B}\partial _{t}q_{AB}+\epsilon \partial _{A}\delta g_{tt}\right)
\nonumber
\\
&+\frac{\epsilon}{2} \left[\delta g^{33}
\partial _{t}\left(m^{2}+\frac{1}{N^{2}}\right)+\delta g^{3A}\partial _{t}m_{A}+
\epsilon \partial _{3}\delta g_{tt}\right]
\nonumber
\\
&+\frac{1}{2}q^{AB}\Big[-\epsilon \partial _{t}\delta q_{AB}
+\delta g^{33}\left(-\partial _{3}q_{AB}+\partial _{A}m_{B}+\partial _{B}m_{A}\right)
+\delta g^{3C}\left(-\partial _{C}q_{AB}+\partial _{A}q_{CB}+\partial _{B}q_{CA}\right)
\nonumber
\\
&+\delta g^{3t}\left(-q^{AB}\partial _{t}q_{AB}\right)+ \epsilon \left(\partial _{A}\delta g_{tB}
+\partial _{B}\delta g_{tA}\right)\Big]
\nonumber
\\
&-\frac{\epsilon}{2}\left[-\frac{1}{N^{2}}\partial _{t}\delta g_{tt}
-\epsilon \partial _{3} \delta g_{tt}-\epsilon m^{A}
\left(-\partial _{A}\delta g_{tt}+2\partial _{t}\delta g_{At}\right)\right]
\nonumber
\\
&-\frac{\epsilon}{2}\left[\delta g^{A3}\partial _{t}m_{A}
+\delta q^{AB}\partial _{t}q_{AB}
-\epsilon m^{A}\partial _{A}\delta g_{tt}+q^{AB}\partial _{t}\delta q_{AB}\right]
\end{align}
Now the variation of the expansion scalar can be obtained as
\begin{align}
2\delta \Theta &=2\delta \left(\nabla _{a}\ell ^{a}\right)
\nonumber
\\
&=2\partial _{t}\delta g^{t3}+2\partial _{3}\delta g^{33}+2\partial _{A}\delta g^{A3}
+2\delta g^{a3}\partial _{a}\ln \sqrt{-g}+2\partial _{t}\left(\delta \ln \sqrt{-g}\right),
\end{align}
which can be used subsequently to obtain the following result:
\begin{align}
2\delta \left(\sqrt{-g}\Theta \right)&=2\sqrt{-g}\partial _{t}\delta g^{t3}+2\sqrt{-g}\partial _{3}\delta g^{33}+2\sqrt{-g}\partial _{A}\delta g^{A3}
+2\delta g^{a3}\partial _{a}\sqrt{-g}
\nonumber
\\
&+2\sqrt{-g}\partial _{t}\left(\delta \ln \sqrt{-g}\right)-2\sqrt{-g}\Theta \delta g^{t3}-\sqrt{-g}\Theta q_{AB}\delta q^{AB}~.
\end{align}
Also, we have the results
\begin{align}
\delta g^{tt}=-\delta g^{33}\qquad \delta g_{tA}=-\epsilon g_{tA}\delta g^{3t}-\epsilon q_{AB}\delta g^{3B}-\epsilon g_{A3}\delta g^{33}~.
\end{align}
Thus, we arrive at the following result:
\begin{align}
\sqrt{-g}Q&+2\delta \left(\sqrt{-g}\Theta \right)=\frac{\epsilon}{2}\partial _{t}\left(\sqrt{-g}g^{tt}\delta g_{tt}\right)-\partial _{A}\left(\sqrt{-g}m^{A}\delta g_{tt}\right)
\nonumber
\\
&+\epsilon \partial _{A}\left(\sqrt{-g}q^{AB}\delta g_{tB}\right)
+\left(\Theta _{AB}-\Theta q_{AB}\right)\delta q^{AB}+\sqrt{-g}Q_{extra}
\end{align}
As usual, we have the standard counter-term and variation proportional to $\delta q^{AB}$ which are by now familiar. In addition to boundary terms and counter-terms, we have the following extra terms:
\begin{align}
\sqrt{-g}Q_{extra}&=-\sqrt{-g}\delta g^{3t}q^{AB}\partial _{t}q_{AB}
-\epsilon \sqrt{-g}\delta g^{3A}\left[m_{A}q^{CD}\partial _{t}q_{CD}+\partial _{t}m_{A}\right]
\nonumber
\\
&+\sqrt{-g}\delta g^{33}\left[-\partial _{t}\left(m^{2}+\frac{1}{N^{2}}\right)-\frac{\epsilon}{2}\left(m^{2}+\frac{1}{N^{2}}\right)q^{AB}\partial _{t}q_{AB}\right]
\end{align}
These terms are all proportional to $\delta g^{3a}$  i.e. to $\delta \ell ^{a}$, as to be expected, with the proper coefficients dictated by \ref{NewLabel03}. This matches with our earlier conclusion. If we restrict to variations of NSF parameters,  then it is clear that $\delta \ell ^{a}=0$. Then, in accordance with the general result presented in \ref{NewLabel03}, only variations of $q^{AB}$ need to be fixed on the boundary, as we have shown explicitly in the previous section.

\section{More on GNC and NSF}
\subsection{Inverse Metric and Christoffel Symbols in GNC}\label{GNCMetric}

With coordinates ordered as $\left(u,r,x^A\right)$, the metric as presented in \ref{GNCline} and its inverse may be written in matrix notation as
\begin{equation}\label{GNCmet}
g_{ab}=
\left(\begin{array}{lll}
-2r\alpha & 1 & -r\beta _{A}\\
1 & 0 & 0\\
-r\beta _{A} & 0 & q_{AB}
\end{array}\right),~~~~~~~
g^{ab}=
\left(\begin{array}{lll}
0 & 1 & 0\\
1 & 2r\alpha +r^{2}\beta ^{2} & r\beta ^{A}\\
0 & r\beta ^{A} & q^{AB}
\end{array}\right)
\end{equation}
where $q ^{AB}=(q ^{-1})^{AB}$ is the inverse matrix of $q _{AB}$, $\beta ^{A}=q ^{AB}\beta _{B}$ and $\beta ^{2}=\beta _{A}\beta ^{A}$. The determinant of the metric is $g=q$, where $q$ is the determinant of the $2$-metric $q_{AB}$. Let $\hat{D}_{A}$ be the covariant derivative operator associated with the two-dimensional metric $q_{AB}$. For example,
\begin{equation}
\hat{D}_{A} \beta_{B} = \partial_{A} \beta_{B}- \hat{\Gamma}^{C}_{AB}\beta_{C},
\end{equation}
where 
\begin{equation}
\hat{\Gamma}^{C}_{AB} = \frac{q^{CD}}{2}\left(-\partial_D q_{AB}+\partial_A q_{BD}+\partial_B q_{DA} \right)~,
\end{equation}
and  $\hat{D}_{A} \beta=\partial_A \beta$ for the two-dimensional scalar $\beta$. Introducing $\bar{\alpha}=-2r\alpha$ and $\bar{\beta} _{A}=-r\beta _{A}$, the Christoffel symbols have the following expressions \cite{Morales}:
\begingroup
\allowdisplaybreaks
\begin{eqnarray}\label{GNCchr}
\Gamma ^{u}_{uu}&=&-\frac{1}{2}\partial _{r}\bar{\alpha}\\
\Gamma ^{u}_{uA}&=&-\frac{1}{2}\partial _{r}\bar{\beta} _{A}\\
\Gamma ^{u}_{AB}&=&-\frac{1}{2}\partial _{r}q _{AB}\\
\Gamma ^{u}_{ur}&=&\Gamma ^{u}_{rr}=\Gamma ^{u}_{rA}=0\\
\Gamma ^{r}_{ur}&=&\frac{1}{2}\left(\partial _{r}\bar{\alpha}-\bar{\beta}^{C}\partial _{r}\bar{\beta}_{C} \right)\\
\Gamma ^{r}_{rA}&=&\frac{1}{2}\left(\partial _{r}\bar{\beta}_{A}-\bar{\beta}^{C}\partial _{r}q _{CA} \right)\\
\Gamma ^{r}_{uu}&=&-\frac{1}{2}\left(\bar{\beta}^{C}\bar{\beta}_{C}-\bar{\alpha} \right)\partial _{r}\bar{\alpha}+\frac{1}{2}\partial _{u}\bar{\alpha}+\frac{1}{2}\bar{\beta}^{C}\partial_{C}\bar{\alpha}- \bar{\beta}^{C}\partial _{u}\bar{\beta}_{C}\\
\Gamma ^{r}_{AB}&=&-\frac{1}{2}\left\lbrace \partial _{u}q _{AB}+\left(\bar{\beta}^{C}\bar{\beta}_{C}-\bar{\alpha} \right)\partial _{r}q _{AB}\right\rbrace +\frac{1}{2}\left(\hat{D}_{A}\bar{\beta} _{B}+\hat{D}_{B}\bar{\beta} _{A} \right)\\
\Gamma ^{r}_{uA}&=&-\frac{1}{2}\left(\bar{\beta}^{C}\bar{\beta}_{C}-\bar{\alpha} \right)\partial _{r}\bar{\beta}_{A}+\frac{1}{2}\partial_{A}\bar{\alpha}-\frac{1}{2}\bar{\beta}^{B}\left(\partial _{u}q _{AB}+\partial_{A}\bar{\beta}_{B}-\partial_{B}\bar{\beta} _{A} \right)\\
\Gamma ^{r}_{rr}&=&0\\
\Gamma ^{A}_{BC}&=&\frac{1}{2}\bar{\beta}^{A}\partial _{r}q _{BC}+\hat{\Gamma}^{A}_{BC}\\
\Gamma ^{A}_{Bu}&=&\frac{1}{2}\bar{\beta}^{A}\partial _{r}\bar{\beta}_{B}+\frac{1}{2}q ^{CA}\partial _{u}q _{BC}+ \frac{1}{2}q ^{CA}\left(\partial_{B}\bar{\beta}_{C}-\partial_{C}\bar{\beta} _{B}\right)\\
\Gamma ^{A}_{Br}&=&\frac{1}{2}q ^{CA}\partial _{r}q _{BC}\\
\Gamma ^{A}_{uu}&=&\frac{1}{2}\bar{\beta}^{A}\partial _{r}\bar{\alpha}-\frac{1}{2}q ^{CA}\partial_{C}\bar{\alpha}+q ^{AC}\partial _{u}\bar{\beta}_{C}\\
\Gamma ^{A}_{ur}&=&\frac{1}{2}q ^{CA}\partial _{r}\bar{\beta}_{C}\\
\Gamma ^{A}_{rr}&=&0
\end{eqnarray}
\endgroup

\subsection{Eliminating the Extra Parameter M in the NSF Metric} \label{KillM}

The NSF metric was given in \ref{NSline-short} and is of the form
\begin{equation} \label{NSline-short1}
 ds^2 = \widetilde{M}^2 (dx^{3})^2 + 2 M \epsilon dt dx^3 + 2 q_{AB} m^A dx^B dx^3 + q_{AB} dx^A dx^B,
\end{equation}
The condition that the null vector $\ell^{a}= \partial/\partial t$ is affinely parametrized leads to $\partial_t M =0$. So, we have $M=M(x^1,x^2,x^3)$.

Let us make the coordinate transformation to a new coordinate $\bar{t} = t M(x^1,x^2,x^3)$ so that
\begin{equation}
M dt = d \bar{t} - t \frac{\partial M}{\partial x^1}  dx^1 - t \frac{\partial M}{\partial x^2}  dx^2 - t \frac{\partial M}{\partial x^3} dx^3~.
\end{equation}
The metric in the new coordinates would take the form
\begin{equation} \label{NSline-short-transf1}
ds^2 = (\widetilde{M}^2- 2 \epsilon \bar{t}~ \frac{\partial \ln M}{\partial x^3}) (dx^{3})^2 + 2 \epsilon d \bar{t} dx^3 + 2 (q_{AB} m^A- \epsilon \bar{t}~ \frac{\partial \ln M}{\partial x^B}) dx^B dx^3 + q_{AB} dx^A dx^B~.
\end{equation}
We can see that this metric is of the same form as we would have obtained by putting $M=1$ in \ref{NSline-short1}. $\partial / \partial \bar{t} $ is a null geodesic with affine parametrization, as we would have expected since $\bar{t}$ is a linear function of $t$ and hence also an affine parameter. Thus, we have managed to encode the same information as in \ref{NSline-short1} with one less parameter. But one difference with \ref{NSline-short1} is that while $\partial / \partial x_3$ was taken to be a spacelike vector in \ref{NSline-short1}, the vector $\partial / \partial x_3$ in \ref{NSline-short-transf} (which is a different vector) is not necessarily spacelike.

\subsection{Inverse Metric and Christoffel Symbols for the NSF Metric}\label{NSapp}

The NSF metric corresponding to the line element in \ref{NSline-short} and its inverse in matrix form are given below in the coordinates $(t,x^1,x^2,x^3)$:
\begin{equation}\label{NSmet}
g_{ab}=
\left(\begin{array}{lll}
 0 & 0 & \epsilon M\\
 0 & q_{AB} & m_{A}\\
 \epsilon M & m_{A} & m^{2}+\frac{M^{2}}{N^{2}}
\end{array}\right),~~~~~~~
g^{ab}=
\left(\begin{array}{lll}
 -\frac{1}{N^{2}} & -\frac{\epsilon m^{A}}{M} & \frac{\epsilon} {M}\\
 -\frac{\epsilon m^{A}}{M} & q^{AB} & 0\\
 \frac{\epsilon} {M} & 0 & 0
\end{array}\right),
\end{equation}
where $q^{AB}=(q^{-1})^{AB}$ is the inverse matrix of $q_{AB}$, $m^{A}=q^{AB}m_{B}$ and $m^{2}=m_{A}m^{A}$. The determinant of the metric is $g=M^2 q$, where $q$ is the determinant of the $2$-metric $q_{AB}$.

Introducing $\hat{D}_{A}$ as the covariant derivative operator corresponding to the two-dimensional metric $q_{AB}$, as in \ref{GNCMetric}, the Christoffel symbols for the above metric are
\begingroup
\allowdisplaybreaks
\begin{eqnarray}\label{NSchris}
\Gamma ^{t}_{tt}&=&\partial _{t}\ln M\\
\Gamma ^{t}_{At}&=&\frac{1}{2}\partial _{A}\ln M+\frac{\epsilon}{2M}q_{AB}\partial _{t}m^{B}\\
\Gamma ^{t}_{AB}&=&\frac{1}{2N^{2}}\partial _{t}q_{AB}+\frac{\epsilon}{2M}\left(\hat{D}_{A}m_{B}+\hat{D}_{B}m_{A} \right) -\frac{\epsilon}{2M}\partial _{3}q_{AB}\\
\Gamma ^{t}_{x_{3}t}&=&\frac{1}{2}m^{A}\partial _{A}\ln M +\frac{\epsilon}{2M}\partial _{t}\left(\frac{M^{2}}{N^{2}} \right)+\frac{\epsilon}{2M}m_{A}\partial _{t}m^{A}\\
\Gamma ^{t}_{x_{3}A}&=&\frac{\epsilon}{2M}\partial_{A}\left(m^{2}+\frac{M^{2}}{N^{2}} \right) -\frac{\epsilon}{2N^{2}}\left(\partial_{A}M-\epsilon \partial _{t}m_{A}\right)
\nonumber
\\
&-&\frac{\epsilon m^{B}}{2M}\left(\partial_{A}m_{B}-\partial_{B}m_{A}\right) -\frac{\epsilon m^{B}}{2M}\partial _{3}q_{AB}\\
\Gamma ^{t}_{x_{3}x_{3}}&=&\frac{1}{2N^{2}}\partial _{t}\left(m^{2}+\frac{M^{2}}{N^{2}}\right) +\frac{\epsilon m^{A}}{2M}\partial_{A}\left(m^{2}+\frac{M^{2}}{N^{2}} \right)-\frac{\epsilon M}{N^{3}}\partial _{3}N-\frac{\epsilon}{2M} m^A m^B \partial_3 q_{AB}\\
\Gamma ^{A}_{Bt}&=&\frac{1}{2}q^{AC}\partial _{t}q_{BC}\\
\Gamma ^{A}_{BC}&=&\frac{\epsilon m^{A}}{2M}\partial _{t}q_{BC}+\hat{\Gamma}^{A}_{BC}\\
\Gamma ^{A}_{tx_{3}}&=&-\frac{\epsilon}{2}q^{AB}\partial _{B}M +\frac{q^{AB}}{2}\partial _{t}m_{B}\\
\Gamma ^{A}_{Bx_{3}}&=& -\frac{m^{A}}{2M}\partial_{B}M +\frac{\epsilon m^{A}}{2M}\partial _{t}m_{B} +\frac{1}{2}q^{AC}\left(\partial_{B}m_{C}-\partial_{C}m_{B}\right) +\frac{1}{2}q^{AC}\partial _{3}q_{BC}\\
\Gamma ^{A}_{tt}&=&0\\
\Gamma ^{A}_{x_{3}x_{3}}&=& \frac{\epsilon m^{A}}{2M}\partial _{t}\left(m^{2}+\frac{M^{2}}{N^{2}}\right) - m^{A} \partial _{3}\ln M -\frac{1}{2}q^{AB}\partial_{B}\left(m^{2}+\frac{M^{2}}{N^{2}}\right) +q^{AB}\partial _{3}m_{B}\\
\Gamma ^{x_{3}}_{x_{3}x_{3}}&=&-\frac{\epsilon}{2M}\partial _{t}\left(m^{2}+\frac{M^{2}}{N^{2}}\right)+\epsilon \partial _{3}\ln M\\
\Gamma ^{x_{3}}_{Ax_{3}}&=&-\frac{\epsilon}{2M}\left(\partial _{t}m_{A}-\epsilon \partial _{A}M\right)\\
\Gamma ^{x_{3}}_{AB}&=&-\frac{\epsilon}{2M}\partial _{t}q_{AB}\\
\Gamma ^{x_{3}}_{tx_{3}}&=&\Gamma ^{x_{3}}_{tt}=\Gamma ^{x_{3}}_{tA}=0~.
\end{eqnarray}
\endgroup

\section{Interpretation of the Counter-term in the Non-null Case}\label{nonnull_CT_interp}

We have seen that the counter-term in the null case can be interpreted as the difference in the $2$-surface areas orthogonal to $\boldsymbol{\ell}$ and $\boldsymbol{k}$ at the boundaries of the null congruence on the null surface (see \ref{CT-diffA}) when $\boldsymbol{\ell}^2=0$ everywhere. The counter-term to be added to the boundary in the non-null case is the integral over the boundary of the expression $2 \sqh \nabla_a n^a$ (see \ref{S.T-nonnull-1}). We shall work with the case where the boundary is a spacelike surface, a $t$-constant surface in coordinates $(t,y_1,y_2,y_3)$, and provide an interpretation for the counter-term. Working along the same lines, a similar interpretation can be given for the counter-term on a general non-null surface. The metric has the ADM form \cite{Arnowitt:1962hi,gravitation}
\begin{equation}
ds^2 = -N^2 dt^2 + h_{\alpha \beta} \left(dx^{\alpha}+N^{\alpha}dt\right)\left(dx^{\beta}+N^{\beta}dt\right)  ~,
\end{equation}
with the inverse metric components given by
\begin{equation}
g^{tt}= -\frac{1}{N^2}; \quad~g^{t \alpha}=\frac{N^{\alpha}}{N^2};~\quad g^{\alpha \beta} = h^{\alpha \beta}-\frac{N^{\alpha} N^{\beta}}{N^2}~.
\end{equation}
The normal according to our conventions (see \ref{BC-novel}) is $n_a=\partial_a t$. The counter-term integrand $\sqrt{h} \nabla_a n^a$ can be manipulated as follows:
\begin{equation}\label{exp_CT}
\sqrt{h} \nabla_a n^a = \frac{\sqrt{h}}{\sqg} \partial_a \left(\sqg n^a \right)=\frac{1}{N} \partial_a \left(N \sqrt{h} n^a \right) = n^a \partial_a \left( \sqrt{h}  \right)+\frac{\sqrt{h}}{N} \partial_a \left(N  n^a \right)~.
\end{equation}
We have $n^a = g^{at}= \left(1/N,  N^{\alpha}/N \right)$, the second term in \ref{exp_CT} can be written in the following form:
\begin{align}\label{second_term}
 \frac{\sqrt{h}}{N} \partial_a \left(N  n^a \right)= \frac{ \sqrt{h}}{N} \partial_{\alpha} N^{\alpha}~.
\end{align}
Let $e^{a}_{\alpha}$, $\alpha=1,2,3$, represent the coordinate basis vectors on the $3$-surfaces normal to $n^a$. They satisfy $e^a_{\alpha}n_a=0$. We shall consider the case where these basis vectors are Lie transported along $n^a$. So we have
\begin{align}
\left[ n^a, e^a_{\alpha}\right]=0 \Longrightarrow  n^b \partial_b e^{a}_{\alpha}-e^{b}_{\alpha} \partial_b n^a =0~.
\end{align}
In the coordinates where $e^{a}_{\alpha}$ are basis vectors, we would have $e^{a}_{\alpha}=\df^{a}_{\alpha}$. Hence, the above condition reduces to
\begin{align}\label{LieDcond}
\partial_{\alpha} n^a =0 \Longrightarrow \partial_\alpha N =0 \textrm{ and } \partial_\alpha N^\alpha =0~.
\end{align}
These conditions also imply that $n^a$ are tangent vectors to affinely parametrized geodesics. Using $n_a=N \nabla_a t$ and $\nabla_a \nabla_b t= \nabla_b \nabla_a t$, we can reduce the geodesic equation to the condition
\begin{align}
 n^a \nabla_a n_b=0 \Longrightarrow  h^{ab} \partial_a \ln N =0~.
\end{align}
Since $h^{ab}$ has only the spatial $h^{\alpha \beta}$ components, \ref{LieDcond} implies the RHS is zero. Hence, $n^a$ should satisfy the geodesic condition.
Using \ref{second_term} and \ref{LieDcond}, \ref{exp_CT} reduces to
\begin{align}
\sqrt{h} \nabla_a n^a = n^a \partial_a \left( \sqrt{h}  \right)~,
\end{align}
which is the change in the volume of a $3$-surface element along $\ell^a$. Taking $\tau$ to be the parameter along the integral curves of $n^a$, we may also write the above equation as
\begin{align}
\sqrt{h} \nabla_a n^a = \frac{ \partial \left( \sqrt{h}  \right)}{\partial \tau}~.
\end{align}
Unlike the null case in \ref{CT-diffA}, where the derivative was on the surface, this derivative is off the boundary surface and hence will not get integrated in the boundary integration to be interpreted as the difference in the volume of the $3$-surface element at two different points.
\section{Details of Various Calculations}

\subsection{Derivation of \ref{ST2}}
\label{APPdervST2}

The boundary term in the non-null case has the following expression:
\begin{eqnarray}
\sqrt{|h|}D_{a}\left(\delta \un^{a}\right)&=& \partial _{\alpha}\left(\sqrt{|h|}\delta \un^{\alpha}\right)
\nonumber
\\
&=& \partial _{\alpha}\left[\sqrt{|h|}\left(N\delta \ell ^{\alpha}+2\ell ^{\alpha}\delta N\right)\right]
\nonumber
\\
&=&\partial _{\alpha}\left[\sqrt{-g}\left(\delta \ell ^{\alpha}+2\ell ^{\alpha}\delta \ln N\right)\right], \label{ST1}
\end{eqnarray}
where we have used $\df \ell_a=0$ to get to the second line and \ref{g_decomp_Nh} in the last line. On the other hand, the surface term on the null surface in \ref{NewLabel03} can be manipulated as follows:
\begin{align}
\partial_{a}\left[\sqrt{-g}\Pi ^{a}_{\phantom{a}b}\delta \ul^{b}\right]=\partial_{\alpha}\left[\sqrt{-g}\Pi ^{\alpha}_{\phantom{a}b}\delta \ul^{b}\right] &= \partial_{\alpha}\left[\sqrt{-g}\delta \ul^{\alpha}\right] + \partial_{\alpha}\left[\sqrt{-g}k^{\alpha} \ell_{b}\delta \ul^{b}\right] \nn \\
&= \partial_{\alpha}\left[\sqrt{-g}\delta \ell^{\alpha}\right] + \partial_{\alpha}\left[\sqrt{-g}k^{\alpha} \ell_{b}\delta \ell^{b}\right] \nn\\
&= \partial_{\alpha}\left[\sqrt{-g}\delta \ell^{\alpha}\right] \label{ul}, 
\end{align}
where we have used $\delta \ell_{a}=0$ to write $\delta \ul^{\alpha} = \df \ell^{a} + g^{ab} \df \ell_{a} = \df \ell^{a}$ and used $\df (\ell^2) = \ell_{a} \df \ell^{a} =0$ on the null surface.

\subsection{Derivation of \ref{K2}}
\label{APPdervK2}

The variation of the counter-term can be written as:
\begin{align}
\delta \left(2K\sqrt{|h|}\right)&=-\delta \left(2\sqrt{|h|}\nabla _{a}n^{a}\right)
\nonumber
\\
&=-2 \delta \left[\frac{1}{N}\partial _{a}\left(\sqrt{-g}g^{ab}n_{b}\right) \right]\nn\\
&=-2\delta \left[\frac{1}{N}\partial _{a}\left(\sqrt{-g}g^{a\phi}N\right) \right]
\nonumber
\\
&=-2\delta \left[\partial _{a}\left(\sqrt{-g}g^{a\phi}\right) \right]
-2\delta \left[\partial _{a}\left(\ln N \right)\sqrt{-g}g^{a\phi} \right]
\nonumber
\\
&=-2\delta \left[\partial _{a}\left(\sqrt{-g}\ell^{a} \right) \right]
-2\delta \left[\sqrt{-g}\ell ^{a}\partial _{a}\left(\ln N\right) \right] \nn\\
&=-2\delta \left[\partial _{a}\left(\sqrt{-g}\ell^{a} \right) \right]+\delta \left[\sqrt{-g}\left(\partial _{\phi}g^{\phi \phi}
+\frac{g^{\alpha \phi}}{g^{\phi \phi}}\partial _{\alpha}g^{\phi \phi}\right) \right]~, \label{K1}
\end{align}
where the last line has been obtained using the relation $g^{\phi \phi}=\epsilon / N^2$. In the limit $N\rightarrow \infty$, it is clear that all terms in \ref{K1} are finite under our assumptions. 
For a null surface, we have the following result using \ref{exp_nabla_l}:
\begin{equation}
\partial _{a}\left(\sqrt{-g}\ell ^{a}\right)=\sqrt{-g}\nabla _{a}\ell ^{a}
=\sqrt{-g}\left\lbrace \left(\Theta +\kappa\right)-\frac{1}{2}k^{a}\nabla _{a}\ell ^{2}\right\rbrace
=\sqrt{-g}\left\lbrace \left(\Theta +\kappa\right)-\frac{1}{2}k^{a}\partial _{a}g^{\phi \phi}\right\rbrace \label{rand1}
\end{equation} 
As we are interested in considering variations such that $g^{\phi \phi}$ is kept zero on the surface (since we demanded $\df(\ell_{a} \ell^{a})=0$ in obtaining \ref{NewLabel03}), the last term in \ref{rand1} can be manipulated as follows in the null limit:
\begin{align}
 -\frac{1}{2}k^{a}\partial _{a}g^{\phi \phi} =  -\frac{1}{2}k^{\phi}\partial _{\phi}g^{\phi \phi} = \frac{1}{2}\partial _{\phi}g^{\phi \phi}, \label{rand2}
\end{align}
where we have used $k^{a}\ell_{a}=k^{\phi} = -1$ in the last line. Thus, we can write the null limit of \ref{K1} as
\begin{align}
\delta \left(2\sqrt{|h|}K\right)&\overset{r\rightarrow0}=-2\delta \left[\sqrt{-g}\left(\Theta +\kappa \right) \right]
-\delta \left[\sqrt{-g}\partial _{\phi}g^{\phi \phi}\right]
+\delta \left[\sqrt{-g}\left(\partial _{\phi}g^{\phi \phi}
+\frac{g^{\alpha \phi}}{g^{\phi \phi}}\partial _{\alpha}g^{\phi \phi}\right) \right]
\nonumber
\\
&=-2\delta \left[\sqrt{-g}\left(\Theta +\kappa \right) \right]
+\delta \left[ \sqg \frac{g^{\alpha \phi}}{g^{\phi \phi}}\partial _{\alpha}g^{\phi \phi} \right]  \nn\\
&= -2\delta \left[\sqrt{-g}\left(\Theta +\kappa \right) \right]
-2\delta \left[ \sqg~ \ell^{\alpha} \partial_{\alpha} \ln N \right]  \label{K21}
\end{align}

\subsection{Derivation of \ref{Khh}}
\label{APPdervKhh}

We will first calculate the quantity $h_{ij}\delta h^{ij}$:
\begin{eqnarray}
h_{ij}\delta h^{ij}&=&  \left(h_{ij}\delta g^{ij} - 2 \epsilon h_{ij} n^{i}\delta n^{j}\right) \nn \\
&=&  h_{ij} \df g^{ij} \nn\\
&=&\left(g_{ij}\delta g^{ij}-\epsilon n_{i}n_{j}\delta g^{ij}\right)
\nonumber
\\
&=&\left(g_{ij}\delta g^{ij}-\epsilon N^{2}\ell_{i}\ell _{j}\delta g^{ij}\right)
\nonumber
\\
&=&\left(g_{ij}\delta g^{ij}-\epsilon N^{2}\delta g^{\phi \phi}\right)
\nonumber
\\
&=&\left(g_{ij}\delta g^{ij}+2 \delta  \ln N\right)
\end{eqnarray}
On the null side, we have the following result:
\begin{eqnarray}
q_{ab}\delta q^{ab}&=&q_{ab}\delta g^{ab}=g_{ab}\delta g^{ab}+2\ell_{a}k_{b}\delta g^{ab}
\nonumber
\\
&=&g_{ab}\delta g^{ab}+2k_{b}\delta g^{b\phi}=g_{ab}\delta g^{ab}+2k_{b}\delta \ell ^{b}
\end{eqnarray}
Now, if we look at the derivation of \ref{K2}, it is easy to see that the expression is valid without the $\df$ too. Thus, we have
\begin{equation}
 \sqrt{|h|}K\overset{r\rightarrow0}=- \sqrt{-g}\left[\left(\Theta +\kappa \right)+ \ell^{\alpha} \partial_{\alpha} \ln N \right]~.
\end{equation}
We are ready now to attack the last term in \ref{S.T-nonnull-3}:
\begin{eqnarray}
\sqrt{|h|}Kh_{ij}\delta h^{ij}
&\overset{r\rightarrow0}=&-\sqrt{-g}\left[\left(\Theta +\kappa \right)
+\ell^{\alpha} \partial_{\alpha} \ln N\right]
\times \left[g_{ij}\delta g^{ij}+2{\delta \ln N} \right]
\nonumber
\\
&=&-\sqrt{-g}\left(\Theta +\kappa \right)q_{ab}\delta q^{ab} +2\sqrt{-g} \left(\Theta +\kappa \right)k_{b}\delta \ell ^{b}
-2\sqrt{-g}\left(\Theta +\kappa \right){\delta \ln N}
\nonumber
\\
&&-\sqg \ell^{\alpha} \partial_{\alpha} \ln N
g_{ij}\delta g^{ij}
-2\sqrt{-g}\left(\ell^{\alpha}\partial _{\alpha}\ln N\right)~ \delta \ln N \label{Khh2}
\end{eqnarray}
\subsection{Derivation of \ref{Kij2}}
\label{APPdervKij2}

We have:
\begin{eqnarray}
- \sqrt{|h|}K_{ij}\delta h^{ij}&=&- \sqrt{|h|}K_{ij}\delta g^{ij} + 2\epsilon \sqrt{|h|}K_{ij} n^{i}\delta n^{j}  \nn\\
&=&- \sqrt{|h|}K_{ij}\delta g^{ij} \nn\\
&=& - \sqrt{|h|}\delta g^{ij} \left[-\nabla _{i}n_{j}+\epsilon n_{i}n^{m}\nabla _{m}n_{j}\right]
\nonumber
\\
&=&\sqrt{-g}\left[\nabla _{i}\ell _{j}+\ell _{j}\partial _{i}\ln N
-\epsilon N^{2}\ell _{i}\ell ^{m}\nabla _{m}\ell _{j}-\epsilon N^{2}
\ell _{i}\ell _{j}\ell ^{m}\nabla _{m} \ln N \right]\delta g^{ij}~. \label{Kij1}
\end{eqnarray}
The third term in \ref{Kij1} is
\begin{eqnarray}
-\epsilon N^{2}\ell _{i}\ell ^{m}\nabla _{m}\ell _{j}\delta g^{ij}
&=&-\frac{1}{2}\epsilon N^{2}\ell _{i}\nabla _{j}\left(\ell ^{2}\right)\delta g^{ij}
\nonumber
\\
&=&-\frac{1}{2g^{\phi \phi}}\left(\partial _{j}g^{\phi \phi}\right)\delta \ell^{j}\nn\\
&=& \df \ell^{j} \partial_j \ln N~,
\end{eqnarray}
where we have used the symmetry of $\nabla_a \ell_b$ in the second step. This term adds with the second term in \ref{Kij1}, while the last term in \ref{Kij1} gives
\begin{eqnarray}
-\epsilon N^2 \ell _{i}\ell _{j}\ell ^{m}\left(\nabla _{m}\ln N \right)\delta g^{ij}
&=& -\epsilon N^2 \ell _{j}\ell ^{m}\left(\nabla _{m}\ln N \right)\delta \ell^{j} \nn \\
&=& - N^2 \ell ^{m}\left(\nabla _{m}\ln N \right)\delta \left(\frac{1}{N^2} \right) \nn \\
&=&  2 \ell^{m} \left(\partial_{m} \ln N \right)\df \ln N~,
\end{eqnarray}
where we have used $\ell_a \df \ell^a=\df \ell^\phi =\df g^{\phi \phi}$ in the first step. Adding everything up, we obtain
\begin{align}\label{Kij11}
 - \sqrt{|h|}K_{ij}\delta h^{ij} = \sqrt{-g}\left[\nabla _{i}\ell _{j} \delta g^{ij}+2 \left(\partial _{i}\ln N\right) \df \ell^{i} + 2 \ell^{i} \left(\partial_{i} \ln N \right)\df \ln N \right]~.
\end{align}
All terms here are finite in the null limit. In the last term, $\df \ln N$ is finite under our assumptions and $\ell^{i} \left(\partial_{i} \ln N \right)$ has been shown to be finite in \ref{K1}. $\left(\partial _{i}\ln N\right) \df \ell^{i}$ can also be easily shown to be finite. The first term in \ref{Kij11} needs to be decomposed in the null limit. This is done as follows: 
\begin{eqnarray}
\nabla _{i}\ell _{j}\delta g^{ij}&\overset{r\rightarrow0}=&\nabla _{i}\ell _{j}\left[\delta q^{ij}-2\delta \left(\ell ^{i}k^{j}\right)\right]
\nonumber
\\
&=&\nabla _{i}\ell _{j}\delta q^{ij}-\delta k^{i}\partial _{i}\ell ^{2}
-2\delta \ell ^{i}k^{j}\nabla _{i}\ell _{j}
\nonumber
\\
&=&\Theta _{ij}\delta q^{ij}+2\delta \ell ^{i}\ell ^{j}\nabla _{i}k_{j}~.
\end{eqnarray}
In the last step, we have used \ref{NewLabel10} for 
the first term and $\df \ell_{a}=0$ was used to put 
$\delta k^{i}\partial _{i}\ell ^{2}=0$.
Putting it all together, we arrive at
\begin{eqnarray}
-\sqrt{|h|}K_{ij}\delta h^{ij}\overset{r\rightarrow0}=\sqrt{-g}\Big[\Theta _{ij}\delta q^{ij}+2\delta \ell ^{i}\ell ^{j}\nabla _{i}k_{j}
+2 \left(\partial _{i}\ln N\right) \df \ell^{i} + 2 \ell^{i} \left(\partial_{i} \ln N \right)\df \ln N \Big]~. \label{Kij21}
\end{eqnarray}

\subsection{Derivation of \ref{ADM-2-Null}}
\label{APPdervADM-2-Null} 
The extra terms in \ref{result1} are
\begin{align}
 \textrm{Extra Terms} &=2\partial _{\alpha}\left[\sqg \ell ^{\alpha}\delta \ln N \right]-2\delta \left[ \sqg~ \ell^{\alpha} \partial_{\alpha} \ln N \right] -2\sqrt{-g}\left(\Theta +\kappa \right){\delta \ln N}
\nonumber
\\
&-\sqg \left( \ell^{\alpha} \partial_{\alpha} \ln N \right)
g_{ij}\delta g^{ij}
-2\sqrt{-g}\left(\ell^{\alpha}\partial _{\alpha}\ln N \right)\delta \ln N  \nn \\
&+\sqrt{-g}\Big[2 \left(\partial _{i}\ln N\right) \df \ell^{i} + 2 \ell^{i} \left(\partial_{i} \ln N \right)\df \ln N \Big]\label{result2}
\end{align}
We shall start by working on the first, second and the fourth of these extra terms. These are
\begin{align}
 &2\partial _{\alpha}\left[\sqg \ell ^{\alpha}\delta \ln N\right]-2\delta \left[ \sqg~ \ell^{\alpha} \partial_{\alpha} \ln N \right]-\sqg \left(\ell^{\alpha} \partial_{\alpha} \ln N \right)
g_{ij}\delta g^{ij}\nn\\
= ~&2  \left[\partial_{\alpha}\left( \sqg \ell^{\alpha}\right) \df \ln N - \sqg\df \ell^{\alpha}\partial_{\alpha} \ln N \right]
\end{align}
Substituting back, we have the extra terms as
\begin{align}\label{result3}
\textrm{Extra Terms}= 2  \left\{\left[\partial_{\alpha} \left(\sqg \ell^{\alpha}\right)-\sqg \left(\Theta +\kappa \right)\right]\df \ln N+ \sqg \df \ell^{\phi} \partial_{\phi} \ln N + \sqg \left(\ell^{\phi}\partial_{\phi} \ln N \right) \df \ln N \right\}~.
\end{align}
Now, from \ref{rand1} and \ref{rand2}, we know that the first term in \ref{result3} can be manipulated as
\begin{align}
 \left[\partial_{\alpha}\left(\sqg \ell^{\alpha}\right)- \sqg \left(\Theta +\kappa \right)\right]\df \ln N&=\left[\sqg \left(\nabla_{a}\ell^{a}-\Theta -\kappa \right)-\partial_{\phi}\left(\sqg \ell^{\phi}\right)\right]\df \ln N \nn \\
 &= \left(\frac{\sqg}{2} \partial_{\phi} g^{\phi \phi}- \sqg \partial_{\phi} \ell^{\phi} \right) \df \ln N \nn \\
 &= \left(\frac{-\sqg}{2} \partial_{\phi} g^{\phi \phi}\right) \df \ln N \nn \\
 &= \left(\sqg \frac{\epsilon}{N^2} \partial_{\phi} \ln N \right)\df \ln N \nn\\
 &= \left(\sqg \ell^{\phi} \partial_{\phi} \ln N \right)\df \ln N,
\end{align}
where we have made use of \ref{eq:gphiphi} and $\ell^{a}=g^{a \phi}$. In the second step, we have put the term $-\left(\sqg \ell^{\phi} \partial_{\phi}\sqg \right) \df \ln N$ to zero as we know that $\ell^{\phi}=\ell^{a}\ell_{a}=0$ on the null surface while $\sqg$ and its derivatives are assumed to be finite everywhere and $\df \ln N$ is finite under our assumption about the behaviour of $g^{\phi \phi}$ near the surface.

Thus, the first and the last terms in \ref{result3} are identical. The middle term should also be then of the same form with an extra factor of $-2$. Indeed, we have
\begin{align}
\df \ell^{\phi} \partial_{\phi} \ln N = \frac{-2 \epsilon}{N^2} \left(\partial_{\phi} \ln N\right)\df \ln N   =-2 \ell^{\phi}\left(\partial_{\phi} \ln N\right) \df \ln N ~,
\end{align}
so that \ref{result3} reduces to
\begin{equation*}
\textrm{Extra Terms}= 0!
\end{equation*}

\section{Working with a General $\ell_a=A\nabla_a \phi$}\label{app:gen-normal}
In this appendix, we shall consider the general case of our null normal being of the form $\ell_a = A \partial_a \phi=A v_{a}$, for an arbitrary scalar $A$ which may depend on the metric. We no longer have the results $\nabla_a \ell_b=\nabla_b \ell_a$ everywhere and $\df \ell_a=0$ everywhere that we had used profusely in \ref{paddy_deriv-null_surf}. But we do have the result
\begin{equation}\label{var_l}
\df \ell_a = \df A \nabla_a \phi=\df \ln A ~\ell_a
\end{equation}
From \ref{null-ST-1}, the boundary term on the null surface is
\begin{equation}\label{null-ST-11}
\sqg Q[v_{a}]=\frac{\sqg}{A}Q[\ell_{a}]= \frac{1}{A}\left\{ \sqrt{-g} \nabla_c [\delta \ul^{c}] - 2 \delta (\sqrt{-g} \nabla_a \ell^{a}) + \sqrt{-g} (\nabla_a \ell_b -g_{ab} \nabla_c \ell^{c}) \delta g^{ab} \right\}~.
\end{equation}
Labelling the first term as $\sqg R_1$, we have
\begin{equation}
\sqg R_1\equiv\frac{\sqrt{-g}}{A} \nabla_a [\delta \ul^{a}] =\frac{1}{A} \partial_{a}[\sqg \delta \ul^{a}]= \partial_{a}[\frac{\sqg}{A} \delta \ul^{a}]-\sqg \delta \ul^{a}\partial_{a}\left(\frac{1}{A}\right)
\end{equation}
We shall now use the projector $\Pi^a_{\phantom{b}b}$ to separate out the surface derivatives in the first term.
\begin{align}
\partial_{a}[\frac{\sqg}{A} \delta \ul^{a}] &=\partial_{a}[\frac{\sqg}{A} \Pi^a_{\phantom{b}b}\delta \ul^{b}]-\partial_{a}[\frac{\sqg}{A} k^a \ell_b \delta \ul^{b}]  \nn\\
&= \partial_{a}[\frac{\sqg}{A} \Pi^a_{\phantom{b}b}\delta \ul^{b}]- \df\left(\ell_a \ell^a\right)\partial_b\left(\frac{\sqg}{A} k^b\right) -\frac{\sqg}{A} k^b \partial_b \left[\df\left(\ell_a \ell^a\right) \right] \nn\\ 
&= \partial_{a}[\frac{\sqg}{A} \Pi^a_{\phantom{b}b}\delta \ul^{b}]-\frac{\sqg}{A} k^b \partial_b \left[\df\left(\ell_a \ell^a\right) \right]\label{null-ST-41}, 
\end{align}
where the last step was obtained by using our assumption $\df\left(\ell_a \ell^a\right)=0$ on the null surface. Using this expression, we have
\begin{align}
\sqg R_1=\partial_{a}[\frac{\sqg}{A} \Pi^a_{\phantom{b}b}\delta \ul^{b}]-\frac{\sqg}{A} k^b \partial_b \left[\df\left(\ell_a \ell^a\right) \right]-\sqg\delta \ul^{a} \partial_{a}\left(\frac{1}{A}\right) \label{null-ST-42}
\end{align}
The first term in \ref{null-ST-42} is a surface derivative on the null surface as $\Pi^a_{\phantom{b}b}\ell_a=0$. The second term in \ref{null-ST-42} has variations of the derivatives of the metric. We shall take out the $\df$ to obtain
\begin{equation}\label{null-ST-51}
-k^b \partial_b \left[\df\left(\ell_a \ell^a\right) \right] = -\df \left[k^b \partial_b \left(\ell_a \ell^a\right) \right]+ \df k^b \partial_b \left(\ell_a \ell^a\right) ~.
\end{equation}    
Substituting in \ref{null-ST-42}, we obtain
\begin{align}
\sqg R_1 =&\partial_{a}[\frac{\sqg}{A} \Pi^a_{\phantom{b}b}\delta \ul^{b}]-\frac{\sqg}{A} \df \left[k^b \partial_b \left(\ell_a \ell^a\right) \right]+\frac{\sqg}{A}\df k^{b} \partial_b \left(\ell_a \ell^a\right)-\sqg\delta \ul^{a}\partial_{a}\left(\frac{1}{A}\right)  \nn\\
=&\partial_{a}[\frac{\sqg}{A} \Pi^a_{\phantom{b}b}\delta \ul^{b}]- \df \left[ \frac{\sqg}{A} k^b \partial_b \left(\ell_a \ell^a\right) \right] - \frac{\sqg}{2A} \left[ k^b \partial_b \left(\ell_a \ell^a\right) \right] g_{ij} \df g^{ij}+\sqg k^b \partial_b \left(\ell_a \ell^a\right) \df \left(\frac{1}{A}\right)\nn\\&+\frac{\sqg}{A}\df k^{b} \partial_b \left(\ell_a \ell^a\right)-\sqg\delta \ul^{a}\partial_{a}\left(\frac{1}{A}\right) \label{Q1-simplified-1}
\end{align}
Here, all the variations of the derivatives of the metric are in the first two terms, assuming $A$ does not depend on the derivatives of the metric. The second term in \ref{null-ST-11} is
\begin{align}\label{Q2-simplified-1}
\sqg R_2\equiv - \frac{2}{A} \delta (\sqrt{-g} \nabla_a \ell^{a})=-2 \delta (\frac{\sqrt{-g}}{A} \nabla_a \ell^{a})+2\sqrt{-g} \nabla_a \ell^{a}\delta \left(\frac{1}{A}\right)
\end{align}
Substituting \ref{Q1-simplified-1} and \ref{Q2-simplified-1}  back in \ref{null-ST-11}, and using the relation
\begin{equation}
\nabla_a \ell^a+\frac{k^a}{2}  \partial_a \left(\ell_b \ell^b\right)= \df^{a}_{b} \nabla_a \ell^b + k^a \ell_b \nabla_a \ell^b = \Pi^a_{\phantom{b}b}\nabla_a \ell^b~,
\end{equation}
in three places, the boundary term on the null surface reduces to
\begin{align} \label{null-ST-intermediate-1}
\sqg Q[\ell_{a}]=&\partial_{a}[\frac{\sqg}{A} \Pi^a_{\phantom{b}b}\delta \ul^{b}] - 2 \delta (\frac{\sqg}{A} \Pi^a_{\phantom{b}b}\nabla_a \ell^b) + \frac{\sqg}{A} (\nabla_a \ell_b -g_{ab} \Pi^c_{\phantom{b}d}\nabla_c \ell^d) \delta g^{ab}\nn \\&+2\sqrt{-g} \Pi^a_{\phantom{b}b}\nabla_a \ell^b \delta \left(\frac{1}{A}\right)+\frac{\sqg}{A}\df k^{b} \partial_b \left(\ell_a \ell^a\right)-\sqg\delta \ul^{a}\partial_{a}\left(\frac{1}{A}\right)~,
\end{align}
When $A=1$, the last three terms vanish and this result reduces to the result in \ref{null-ST-intermediate}. To see why the second-to-last term did not appear for $A=1$, note that this term will only have normal derivatives as $\ell_a \ell^a$ is fixed to zero everywhere on the null surface. Hence, only $\df k^{\phi}$ contributes in our $\left(\phi,y^1,y^2,y^3\right)$ coordinate system. When $A=1$, $\df k^{\phi}=\df\left(\ell_ak^{a}\right)=0$. In the general case, $k^a\ell_a=-1$ means $k^{\phi}A=-1$, which gives
\begin{equation}\label{kphi}
k^{\phi}=-1/A; \quad \quad \df k^{\phi}=\frac{\df \ln A}{A}~.
\end{equation}
In \ref{null-ST-intermediate-1}, we have succeeded in separating out a total derivative on the surface and a total variation to remove all derivatives of the metric. The counter-term to be added in this case is the integral over the null surface of
\begin{equation}
2 \frac{\sqg}{A} \Pi^a_{\phantom{b}b}\nabla_a \ell^b =2 \frac{\sqg}{A} \left(q^a_{\phantom{b}b}\nabla_a \ell^b-\ell^a k_{b}\nabla_a \ell^b\right)=2 \frac{\sqg}{A} \left(\Theta+\kappa\right)~. 
\end{equation}
We shall now do the analysis of what is to be fixed on the null boundary in this case. Since \ref{l_k_conditions} are taken to be valid even on variation, the relations $q^{ab}\ell_a=0$ and $q^{ab}k_a=0$ are respected by the variations and terms of the form $\ell_a \ell_b\df q^{ab} $, $\ell_a k_b\df q^{ab}$ and $k_a k_b\df q^{ab}$ would reduce to zero, just as in \ref{paddy_deriv-null_surf}. Thus, we can simplify $g_{ab}\df g^{ab}$ as follows:
\begin{align}
g_{ab} \df g^{ab}&=g_{ab}\left[\df q^{ab}-\df \left(\ell^{a}  k^{b}\right)-\df \left(\ell^{b} k^{a}\right) \right]= q_{ab}\df q^{ab}+2 \left(\ell_{a}k_{b} + \ell_{b}k_{a}\right) \df \left(\ell^{a}  k^{b}\right)= q_{ab}\df q^{ab}+2 \ell_{b}k_{a} \df \left(\ell^{a}  k^{b}\right) \nn\\ &= q_{ab}\df q^{ab} - 2 k_{a}\df \ell^{a}- 2 \ell_{a}\df k^{a}=q_{ab}\df q^{ab} - 2 k_{a}\df \ell^{a}- 2 \df \ln A,  \label{gdfg-1}
\end{align}
where we have used \ref{kphi} in the last step.

Next, we shall simplify $\left(\nabla_{a} \ell_b \right) \df g^{ab}$. We have
\begin{align}
\left(\nabla_{a} \ell_b \right) \df g^{ab} &= \left(\nabla_{a} \ell_b \right) \df q^{ab} - \df\left(\ell^a k^b \right) \nabla_a \ell_b- \df\left(\ell^b k^a \right) \nabla_a \ell_b \nn\\
&= \left(\nabla_{a} \ell_b \right) \df q^{ab} -\df{\ell^a} k^b \nabla_a \ell_b - \df{k^b }\ell^a \nabla_a \ell_b-\df{\ell^b} k^a \nabla_a \ell_b - \df{k^a }\ell^b \nabla_a \ell_b\nn\\
&= \left(\nabla_{a} \ell_b \right) \df q^{ab} - \df{\ell^a} k^b \left(\nabla_a \ell_b+\nabla_b \ell_a\right) - \kappa \df{k^b } \ell_b-\frac{\df k^a}{2}\partial_a \ell^2 \nn\\
&= \left(\nabla_{a} \ell_b \right) \df q^{ab} - \df{\ell^a} k^b \left(\nabla_a \ell_b+\nabla_b \ell_a\right)- \kappa \df \ln A-\frac{\df k^a}{2}\partial_a \ell^2 , \label{dldg-11}
\end{align}
where we have used $\ell^a \nabla_a \ell_b=\kappa \ell_b$ (see \ref{kappa_eq_1}) and $\ell_a\df k^a=\df \ln A$ (see \ref{kphi}). The expression $\left(\nabla_{a} \ell_b \right) \df q^{ab}$ can be simplified as follows:
\begin{align}\label{NewLabel101}
\left(\nabla_{a} \ell_b \right) \df q^{ab} = \df^m_a \df^n_b \left(\nabla_{m} \ell_n \right) \df q^{ab} &= \left(q^m_a -\ell^m k_a-k^m \ell_a\right) \left(q^n_b-\ell^n k_b-k^n \ell_b\right) \left(\nabla_{m} \ell_n \right) \df q^{ab}   \nn\\
&= \left(q^m_a -\ell^m k_a \right) \left(q^n_b-\ell^n k_b \right) \left(\nabla_m \ell_n \right) \df q^{ab} \nn \\
&=\left(q^m_a q^n_b \nabla_m \ell_n - q^m_a \ell^n k_b \nabla_m \ell_n - q^n_b \ell^m k_a \nabla_m \ell_n \right) \df q^{ab} \nn \\
&=\left(\Theta_{ab}- \frac{k_b q^m_a\partial_m \ell^2}{2}-\kappa q^n_b  k_a  \ell_n \right) \df q^{ab} \nn \\
&=\Theta_{ab} \df q^{ab}
\end{align}
where we used $\ell_a\df q^{ab}=\df \left(q^{ab} \ell_a \right)-q^{ab}\df \ell_a=- q^{ab}\ell_a \df \ln A=0$ to get to the second line, $k_a k_b \df q^{ab}=0$ to get to the third line and the definition of $\Theta_{ab}$ and $\kappa$ (see \ref{app:SFF+exp} and \ref{kappa_eq_1}) to get to the fourth line. The final result is obtained using $\ell_a q^{ab}=0$ and the fact that $q^m_a\partial_m \ell^2$, a derivative on the null surface, is zero since $\ell^2$ is zero all over the null surface. Next, we shall simplify the last two terms in \ref{dldg-11} as follows:
\begin{align}
- \kappa \df \ln A-\frac{\df k^a}{2}\partial_a \ell^2 =&- \kappa \df \ln A-\frac{\df k^\phi}{2}\partial_\phi \ell^2
=- \kappa \df \ln A-\frac{\df \ln A}{2A}\partial_\phi \ell^2 \nn \\
=& -\df \ln A \left(\kappa+ \frac{\partial_{\phi}\ell^2}{2A}\right) 
=-\df \ln A \left(\kappa- \frac{k^a \partial_{a}\ell^2}{2}\right)=-\df \ln A \left(\kappa+ \tilde{\kappa}\right)~. \label{int}
\end{align}
Here, the first step made use of the fact that $\ell^2=0$ all over the null surface and hence only the derivative along $\phi$ contributes and the second step as well as the second-to-last step used \ref{kphi}. The last line used the definition $2\tilde{\kappa}=-k^{a}\nabla _{a}\ell ^{2}$ (see \ref{app:kappatheta}). 
The remaining terms in \ref{null-ST-intermediate-1} are
\begin{align}
2\sqrt{-g} \Pi^a_{\phantom{b}b}\nabla_a \ell^b \delta \left(\frac{1}{A}\right)&=-2 \frac{\sqg}{A}\left(\Theta+\kappa\right) \df \ln A \\
\frac{\sqg}{A}\df k^{b} \partial_b \left(\ell_a \ell^a\right)&=2\frac{\sqg}{A} \tilde{\kappa} \df \ln A \label{70}\\
-\sqg\delta \ul^{a}\partial_{a}\left(\frac{1}{A}\right)&=\frac{\sqg}{A} \left(\df \ell^a\partial_a \ln A+g^{ab}\partial_a \ln A\df \ell_b\right)\nn\\
&=\frac{\sqg}{A} \left(\df \ell^a\partial_a \ln A+\ell^{a}\partial_a \ln A\df \ln A\right) \nn \\
&=\frac{\sqg}{A} \left(\df \ell^a\partial_a \ln A+(\kappa-\tilde{\kappa})\df \ln A\right)
\end{align}
\ref{70} can be obtained using the same manipulations that we performed in \ref{int}. Combining all the above results, the boundary term in \ref{null-ST-intermediate-1} becomes
\begin{align}
\sqg Q[\ell_{a}]=&\partial_{a}[\frac{\sqg}{A} \Pi^a_{\phantom{b}b}\delta \ul^{b}] - 2 \delta \left[ \frac{\sqg}{A} \left(\Theta+\kappa\right)\right] + \frac{\sqg}{A} \left[\left(\Theta_{ab}-\left(\Theta+\kappa\right) q_{ab}\right)\df q^{ab}\right]\nn\\& +\frac{\sqg}{A}  \left(2 k_{a}\left(\Theta+\kappa\right)-  k^b \left(\nabla_a \ell_b+\nabla_b \ell_a\right)+\partial_a \ln A\right)\df \ell^{a}~.\label{null-ST-general-final}
\end{align}
The changes from \ref{q_with_A_1} are as follows: $\sqg$ has been replaced with $\sqg/A$ everywhere. The $\partial_a \ln A$ is extra and $2 \ell^{b}\nabla _{c}k_{b} \df \ell^{c}$ has been replaced with $- k^b \left(\nabla_a \ell_b+\nabla_b \ell_a\right) \df \ell^{a}$.


\bibliography{mybibliography-gravity}

\bibliographystyle{./utphys}

\end{document}